%% file: 0_PAPER_MAIN.tex
\documentclass[
    a4paper,
    colorlinks=true,
    linkcolor=medblau,
    citecolor=medblau,
    urlcolor=medblau,
]{article}

\usepackage{arxiv}

\usepackage[utf8]{inputenc} % allow utf-8 input
\usepackage[T1]{fontenc}    % use 8-bit T1 fonts
\usepackage{url}            % simple URL typesetting
\usepackage{booktabs}       % professional-quality tables
\usepackage{amsfonts}       % blackboard math symbols
\usepackage{amsmath}
\usepackage{nicefrac}       % compact symbols for 1/2, etc.

\usepackage{microtype}      % microtypography
\usepackage{lipsum}		   
\usepackage[numbers,sort&compress]{natbib}
\usepackage{doi}
\usepackage{siunitx}
\usepackage{enumitem}
\usepackage[version=4]{mhchem}
\usepackage{titlesec}
\usepackage[all]{nowidow}

\usepackage{graphicx}
\usepackage{xcolor} % Falls noch nicht geladen
\input{fau-colors}

\usepackage{transparent}
\usepackage{adjustbox}
\usepackage{subcaption}
\usepackage[labelfont=bf]{caption}
\usepackage{graphicx}
\usepackage{tikz}
\newcommand{\field}{0.145}
\newcommand{\fields}{0.14}

\usepackage{tabularx}
\usepackage{multirow}
\usepackage{booktabs,makecell} 

\usepackage{pgfplots}
\pgfplotsset{compat=1.18}
\usepgfplotslibrary{statistics}
\usepackage{tikz}
\usetikzlibrary{tikzmark,patterns}
\usetikzlibrary{arrows,arrows.meta,shapes.arrows}

\usepackage[nogroupskip,nonumberlist,nopostdot]{glossaries-extra}
\usepackage{glossary-mcols}

\newglossarystyle{mycoltreeheader}{%
  \setglossarystyle{mcoltree*}%
  \setlength{\columnsep}{8pt} % 
  \newcolumntype{Y}{>{\raggedright\arraybackslash}p{1.15cm}}
  \newcolumntype{U}{>{\raggedright\arraybackslash}p{.7cm}}
}

\setabbreviationstyle[acronym]{long-short}
\setglossarystyle{mycoltreeheader}
\newglossary[slg]{symbolslist}{syi}{syg}{Nomenclature}
\makeglossaries

\input{2_Glossary}

\usepackage{flafter}
\usepackage{placeins,etoolbox} % Auto Floatbarrier
\usepackage{comment}

\titlespacing{\section}{0pt}{6pt}{3pt} % Section 
\titlespacing{\subsection}{0pt}{4pt}{1pt} % Section
\usepackage{hyperref}

\title{Fourier Neural Operators for Two-Phase, 2D Mold-Filling Problems Related to Metal Casting}

\addauthor{Edgard M.~Minete}{1}{edgard.minete@fau.de}{0000-0003-0238-4661}
\addauthor{Mathis Immertreu}{1}{mathis.immertreu@fau.de}{0009-0006-0327-7461}
\addauthor{Fabian Teichmann}{1}{fabian.teichmann@fau.de}{0000-0001-9353-9168}
\addauthor{Sebastian Müller}{1}{seb.mueller@fau.de}{0000-0001-8693-6341}

\addaffil{1}{Friedrich-Alexander-Universität Erlangen-Nürnberg, Chair of Casting Technology, Fürth, 90762, Germany}

\hypersetup{
pdftitle={Fast two-phase mold filling via Fourier–graph neural operators},
pdfauthor={Edgard M.~Minete, Mathis ~Immertreu, Fabian ~Teichmann, Sebastian ~Müller},
pdfkeywords={Neural operator, Fourier neural operator, Surrogate modeling, Filling problems, Mold filling, Metal Casting, Cahn–Hilliard–Navier–Stokes}
}

\begin{document}
\maketitle
\begin{abstract}
	\input{1_Abstract}
\end{abstract}

% Keywords
\keywords{Neural operator \and Fourier neural operator \and Surrogate modeling \and Filling problems \and Mold filling \and Metal casting \and Cahn–Hilliard–Navier–Stokes}

% Glossary
\printglossary[type=symbolslist,title={Nomenclature}]
%\printnoidxglossary[type=symbolslist,title={Nomenclature}]
%\printglossaries % If glossary is not working: compile one time with this comand and then again with the command above.

% Introduction Section
\input{3_Introduction}

% Methodology
\input{4_Methodology}

% Results
\input{5_Results_and_Discussion}

% Conclusion
\input{6_Conclusion}

% Akowledgement
\input{7_Acknowledgement}

% Contributions
\input{8_Contributions}

\FloatBarrier

% Bibliography
\bibliography{9_References}

\end{document}

%% file: fau-colors.tex
\definecolor{faublau}{RGB}{4,49,106}
\definecolor{faudunkelblau}{RGB}{4,30,66}
\definecolor{philgelb}{RGB}{253,183,53}
\definecolor{philorange}{RGB}{232,119,34}
\definecolor{rwrot}{RGB}{197,15,60}
\definecolor{rwdunkelrot}{RGB}{151,27,47}
\definecolor{medblau}{RGB}{24,180,241}
\definecolor{meddunkelblau}{RGB}{0,82,135}
\definecolor{natgrün}{RGB}{123,183,37}
\definecolor{natdunkelgrün}{RGB}{38,97,65}
\definecolor{tfmetallic}{RGB}{140,159,177}
\definecolor{tfdunkelmetallic}{RGB}{47,88,110}
\definecolor{schwarz}{RGB}{0,0,0}

\definecolor{faublau062}{RGB}{97,125,161}
\definecolor{faublau037}{RGB}{160,177,198}
\definecolor{faublau025}{RGB}{192,203,218}
\definecolor{faublau012}{RGB}{211,220,242}

\definecolor{faudunkelblau062}{RGB}{97,113,136}
\definecolor{faudunkelblau037}{RGB}{160,169,183}
\definecolor{faudunkelblau025}{RGB}{192,205,208}
\definecolor{faudunkelblau012}{RGB}{223,226,241}

\definecolor{philgelb062}{RGB}{254,206,118}
\definecolor{philgelb037}{RGB}{254,228,178}
\definecolor{philgelb025}{RGB}{254,237,204}
\definecolor{philgelb012}{RGB}{255,245,224}

\definecolor{philorange062}{RGB}{239,163,105}
\definecolor{philorange037}{RGB}{246,203,171}
\definecolor{philorange025}{RGB}{249,221,200}
\definecolor{philorange012}{RGB}{252,237,226}

\definecolor{rwrot062}{RGB}{221,115,124}
\definecolor{rwrot037}{RGB}{235,171,174}
\definecolor{rwrot025}{RGB}{241,200,201}
\definecolor{rwrot012}{RGB}{252,220,227}

\definecolor{rwdunkelrot062}{RGB}{190,113,125}
\definecolor{rwdunkelrot037}{RGB}{216,167,198}
\definecolor{rwdunkelrot025}{RGB}{230,198,205}
\definecolor{rwdunkelrot012}{RGB}{244,226,229}

\definecolor{medblau062}{RGB}{109,208,246}
\definecolor{medblau037}{RGB}{167,236,250}
\definecolor{medblau025}{RGB}{197,236,251}
\definecolor{medblau012}{RGB}{227,250,255}

\definecolor{meddunkelblau062}{RGB}{90,164,179}
\definecolor{meddunkelblau037}{RGB}{158,189,209}
\definecolor{meddunkelblau025}{RGB}{191,218,245}
\definecolor{meddunkelblau012}{RGB}{222,241,252}

\definecolor{natgrün062}{RGB}{172,210,117}
\definecolor{natgrün037}{RGB}{205,228,172}
\definecolor{natgrün025}{RGB}{222,237,200}
\definecolor{natgrün012}{RGB}{232,252,220}

\definecolor{natdunkelgrün062}{RGB}{118,155,135}
\definecolor{natdunkelgrün037}{RGB}{172,195,174}
\definecolor{natdunkelgrün025}{RGB}{201,215,207}
\definecolor{natdunkelgrün012}{RGB}{222,235,220}

\definecolor{tfmetallic062}{RGB}{182,194,206}
\definecolor{tfmetallic037}{RGB}{211,218,225}
\definecolor{tfmetallic025}{RGB}{235,240,247}
\definecolor{tfmetallic012}{RGB}{243,245,247}

\definecolor{tfdunkelmetallic062}{RGB}{124,150,163}
\definecolor{tfdunkelmetallic037}{RGB}{176,191,200}
\definecolor{tfdunkelmetallic025}{RGB}{228,233,236}
\definecolor{tfdunkelmetallic012}{RGB}{229,233,236}

\definecolor{schwarz062}{RGB}{94,94,94}
\definecolor{schwarz037}{RGB}{158,158,158}
\definecolor{schwarz025}{RGB}{191,191,191}
\definecolor{schwarz012}{RGB}{222,222,222}

\colorlet{col1}{faublau}           % blue
\colorlet{col2}{tfdunkelmetallic}  % dark metallic blue
\colorlet{col3}{rwrot}             % red
\colorlet{col4}{natgrün}           % green
\colorlet{col5}{philgelb}          % yellow
\colorlet{col6}{philorange}        % orange
\colorlet{col7}{rwdunkelrot}       % dark red
\colorlet{col8}{natdunkelgrün}     % dark green
\colorlet{col9}{medblau}           % medium blue
\colorlet{col10}{meddunkelblau}    % dark medium blue
\colorlet{col11}{tfmetallic}       % metallic blue
\colorlet{col12}{faudunkelblau}    % dark blue
\colorlet{col13}{philgelb062}      % lighter yellow tint (for variety)
\colorlet{col14}{tfdunkelmetallic012} % very light metallic tint
\colorlet{col15}{schwarz}          % black

%% file: 2_Glossary.tex
\newacronym{CFD}{CFD}{Computational Fluid Dynamics}
\newacronym{cfd}{CFD}{Computional Fluid Dynamics}
\newacronym{CNNs}{CNNs}{Convolutional Neural Networks}
\newacronym{CHNS}{CHNS}{Cahn–Hilliard–Navier–Stokes}
\newacronym{FNS}{FNS}{Fourier Neural Solver}
\newacronym{FNOs}{FNOs}{Fourier Neural Operators}
\newacronym{FNO}{FNO}{Fourier Neural Operator}
\newacronym{GNO}{GNO}{Graph Neural Operator}
\newacronym{GNNs}{GNNs}{Graph Neural Networks}
\newacronym{PINNs}{PINNs}{Physics-informed neural networks}
\newacronym{PDE}{PDE}{Partial Differential Equation}
\newacronym{GINO}{GINO}{Geometry-Informed Neural Operator}
\newacronym{CMAE}{CMAE}{Computer Methods in Applied Mechanics and Engineering}
\newacronym{HPC}{HPC}{High Performance Computer}
\newacronym{BC}{BC}{Boundary Conditions}
\newacronym{IC}{IC}{Initial Conditions}
\newacronym{AdaIN}{AdaIN}{Adaptative Instance Normalization}
\newacronym{FFT}{FFT}{Fast Fourier Transform}
\newacronym{RMSE}{RMSE}{Root Mean Squared Error}
\newacronym{MSE}{MSE}{Mean Squared Error}
\newacronym{rk}{$r_k$}{relative $L_2$ percentage error}
\newacronym{hpdc}{HPDC}{High Pressure Die Casting}
\newacronym{vof}{VOF}{Volume-of-Fluid}
\newacronym{cae}{CAE}{Computer-Aided Engineering}

\newglossaryentry{u_vec}{
  type=symbolslist,
  name={\ensuremath{\mathbf{u}}},
  description={Velocity vector field},
  user1={m/s}
}
\newglossaryentry{u_vec_hat}{
  type=symbolslist,
  name={\ensuremath{\mathbf{\hat{u}}}},
  description={Predicted velocity vector field},
  user1={m/s}
}
\newglossaryentry{u}{
  type=symbolslist,
  name={\ensuremath{u}},
  description={Velocity in $x$-direction},
  user1={m/s}
}
\newglossaryentry{u_hat}{
  type=symbolslist,
  name={\ensuremath{\hat{u}}},
  description={Predicted velocity in $x$-direction},
  user1={m/s}
}
\newglossaryentry{v}{
  type=symbolslist,
  name={\ensuremath{v}},
  description={Velocity in $y$-direction},
  user1={m/s}
}
\newglossaryentry{v_hat}{
  type=symbolslist,
  name={\ensuremath{\hat{v}}},
  description={Predicted velocity in $y$-direction},
  user1={m/s}
}
\newglossaryentry{p}{
  type=symbolslist,
  name={\ensuremath{p}},
  description={Pressure},
  user1={Pa}
}
\newglossaryentry{p_hat}{
  type=symbolslist,
  name={\ensuremath{\hat{p}}},
  description={Predicted pressure},
  user1={Pa}
}
\newglossaryentry{alpha}{
  type=symbolslist,
  name={\ensuremath{\alpha}},
  description={Volume fraction of fluid},
  user1={--}
}
\newglossaryentry{alpha_hat}{
  type=symbolslist,
  name={\ensuremath{\hat{\alpha}}},
  description={Predicted volume fraction of fluid},
  user1={--}
}
\newglossaryentry{alpha1}{
  type=symbolslist,
  name={\ensuremath{\alpha_{1}}},
  description={Volume fraction of fluid 1},
  user1={--}
}
\newglossaryentry{alpha2}{
  type=symbolslist,
  name={\ensuremath{\alpha_{2}}},
  description={Volume fraction of fluid 2},
  user1={--}
}

\newglossaryentry{phi}{
  type=symbolslist,
  name={\ensuremath{\phi}},
  description={Phase-field variable},
  user1={--}
}
\newglossaryentry{psi}{
  type=symbolslist,
  name={\ensuremath{\Psi}},
  description={Chemical potential},
  user1={}
}
\newglossaryentry{lambda_mix}{
  type=symbolslist,
  name={\ensuremath{\lambda}},
  description={Mixing energy density parameter},
  user1={}
}
\newglossaryentry{epsilon}{
  type=symbolslist,
  name={\ensuremath{\varepsilon}},
  description={Interface thickness},
  user1={mm}
}
\newglossaryentry{gamma}{
  type=symbolslist,
  name={\ensuremath{\gamma}},
  description={Mobility parameter},
  user1={}
}
\newglossaryentry{chi}{
  type=symbolslist,
  name={\ensuremath{\chi}},
  description={Mobility tuning parameter},
  user1={--}
}

\newglossaryentry{Fx}{
  type=symbolslist,
  name={\ensuremath{F_{x}}},
  description={Surface tension force in x},
  user1={N/m$^{3}$}
}
\newglossaryentry{Fy}{
  type=symbolslist,
  name={\ensuremath{F_{y}}},
  description={Surface tension force in y},
  user1={N/m$^{3}$}
}
\newglossaryentry{rho}{
  type=symbolslist,
  name={\ensuremath{\rho}},
  description={Mixture density},
  user1={kg/m$^{3}$}
}
\newglossaryentry{rho1}{
  type=symbolslist,
  name={\ensuremath{\rho_{1}}},
  description={Density of fluid $1$},
  user1={kg/m$^{3}$}
}
\newglossaryentry{rho2}{
  type=symbolslist,
  name={\ensuremath{\rho_{2}}},
  description={Density of fluid $2$},
  user1={kg/m$^{3}$}
}
\newglossaryentry{mu}{
  type=symbolslist,
  name={\ensuremath{\mu}},
  description={Mixture dynamic viscosity ($\mu=\mu_{1}\alpha_{1}+\mu_{2}\alpha_{2}$)},
  user1={Pa$\cdot$s}
}
\newglossaryentry{mu1}{
  type=symbolslist,
  name={\ensuremath{\mu_{1}}},
  description={Dynamic viscosity of fluid $1$},
  user1={Pa$\cdot$s}
}
\newglossaryentry{mu2}{
  type=symbolslist,
  name={\ensuremath{\mu_{2}}},
  description={Dynamic viscosity of fluid $2$},
  user1={Pa$\cdot$s}
}
\newglossaryentry{g}{
  type=symbolslist,
  name={\ensuremath{g}},
  description={Gravitational acceleration},
  user1={m/s$^{2}$}
}

\newglossaryentry{xcoord}{
  type=symbolslist,
  name={\ensuremath{x}},
  description={Spatial coordinate in $x$},
  user1={mm}
}
\newglossaryentry{ycoord}{
  type=symbolslist,
  name={\ensuremath{y}},
  description={Spatial coordinate in $y$},
  user1={mm}
}
\newglossaryentry{x}{
  type=symbolslist,
  name={\ensuremath{\mathbf{x}}},
  description={Coordinate vector},
  user1={mm}
}
\newglossaryentry{xi}{
  type=symbolslist,
  name={\ensuremath{\mathbf{x_i}}},
  description={Query vertice},
  user1={mm}
}
\newglossaryentry{time}{
  type=symbolslist,
  name={\ensuremath{t}},
  description={Time},
  user1={s}
}
\newglossaryentry{Thorizon}{
  type=symbolslist,
  name={\ensuremath{T}},
  description={Temporal horizon of the NO},
  user1={s}
}
\newglossaryentry{domain}{
  type=symbolslist,
  name={\ensuremath{\Omega}},
  description={Mold / ingate computational domain},
  user1={--}
}
\newglossaryentry{inletBoundary}{
  type=symbolslist,
  name={\ensuremath{\Gamma_{\mathrm{in}}}},
  description={Inlet boundary},
  user1={--}
}
\newglossaryentry{outletBoundary}{
  type=symbolslist,
  name={\ensuremath{\Gamma_{\mathrm{out}}}},
  description={Outlet boundary},
  user1={--}
}
\newglossaryentry{wallBoundary}{
  type=symbolslist,
  name={\ensuremath{\Gamma_{\mathrm{w}}}},
  description={No-slip wall boundary},
  user1={--}
}
\newglossaryentry{operator}{
  type=symbolslist,
  name={\ensuremath{\mathcal{N}_{\vartheta}}},
  description={Neural operator},
  user1={--}
}
\newglossaryentry{fman}{
  type=symbolslist,
  name={\ensuremath{f_{man}}},
  description={Feature function},
  user1={--}
}
\newglossaryentry{kappa}{
  type=symbolslist,
  name={\ensuremath{\kappa}},
  description={Learnable neural network kernel},
  user1={--}
}
\newglossaryentry{theta_hat}{
  type=symbolslist,
  name={\ensuremath{\hat{\theta}}},
  description={Neural network kernel parameters},
  user1={--}
}
\newglossaryentry{solution_map}{
  type=symbolslist,
  name={\ensuremath{\mathcal{S}}},
  description={Solution map},
  user1={--}
}
\newglossaryentry{inlet_setup}{
  type=symbolslist,
  name={\ensuremath{I}},
    description={Inlet setup},
  user1={--}
}
\newglossaryentry{input_a}{
  type=symbolslist,
  name={\ensuremath{a}},
    description={Input Bundle on \(\Omega\)},
  user1={--}
}
\newglossaryentry{latent}{
  type=symbolslist,
  name={\ensuremath{U}},
  description={Latent signal},
  user1={--}
}
\newglossaryentry{flayer}{
  type=symbolslist,
  name={\ensuremath{\mathcal{L}^l}},
  description={Fourier layer},
  user1={--}
}
\newglossaryentry{FFF}{
  type=symbolslist,
  name={\ensuremath{\mathcal{F}}},
  description={Fast Fourier Transform},
  user1={--}
}
\newglossaryentry{integral}{
  type=symbolslist,
  name={\ensuremath{K}},
  description={GNO local aggregation operator},
  user1={--}
}
\newglossaryentry{steps}{
  type=symbolslist,
  name={\ensuremath{k}},
  description={Prediction steps},
  user1={--}
}
\newglossaryentry{closs}{
  type=symbolslist,
  name={\ensuremath{c}},
  description={Cumulative past loss},
  user1={--}
}
\newglossaryentry{lloss}{
  type=symbolslist,
  name={\ensuremath{\ell}},
  description={Per–step, per–field relative \(L_2\) error},
  user1={--}
}
\newglossaryentry{tweight}{
  type=symbolslist,
  name={\ensuremath{\gamma_t}},
  description={Temporal weight},
  user1={--}
}
\newglossaryentry{tau}{
  type=symbolslist,
  name={\ensuremath{\tau}},
  description={Causality parameter},
  user1={--}
}
\newglossaryentry{trainingloss}{
  type=symbolslist,
  name={\ensuremath{L_{opt}}},
  description={Training/optimization loss},
  user1={--}
}
\newglossaryentry{stepsT}{
  type=symbolslist,
  name={\ensuremath{H}},
  description={Total number of prediction steps},
  user1={--}
}
\newglossaryentry{fields}{
  type=symbolslist,
  name={\ensuremath{q}},
  description={Target field $\in\{u,v,p,\alpha\}$},
  user1={--}
}
\newglossaryentry{fieldsp}{
  type=symbolslist,
  name={\ensuremath{\hat{q}}},
  description={Prediction field $\in\{\hat{u},\hat{v},\hat{p},\hat{\alpha}\}$},
  user1={--}
}
\newglossaryentry{verticesN}{
  type=symbolslist,
  name={\ensuremath{N}},
  description={Number of mesh vertices},
  user1={--}
}
\newglossaryentry{rweights}{
  type=symbolslist,
  name={\ensuremath{R^\ell}},
  description={Learnable spectral weights},
  user1={--}
}
\newglossaryentry{plinearmap}{
  type=symbolslist,
  name={\ensuremath{W}},
  description={Point-wise linear map},
  user1={--}
}
\newglossaryentry{rerror}{
  type=symbolslist,
  name={\ensuremath{r}},
  description={Relative $L_2$ percentage error},
  user1={--}
}

\newglossaryentry{Vmag}{
  type=symbolslist,
  name={\ensuremath{V}},
  description={Inlet velocity magnitude on \ensuremath{\Gamma_{\mathrm{in}}}},
  user1={\%}
}
\newglossaryentry{zeta}{
  type=symbolslist,
  name={\ensuremath{\zeta}},
  description={Inlet flow direction at $\Gamma_{\mathrm{in}}$},
  user1={--}
}
\newglossaryentry{beta_n}{
  type=symbolslist,
  name={\ensuremath{\beta_{n}}},
  description={Angle of wall normal},
  user1={rad}
}
\newglossaryentry{theta_w}{
  type=symbolslist,
  name={\ensuremath{\theta_{w}}},
  description={Static contact angle at the wall},
  user1={deg}
}
\newglossaryentry{phi0}{
  type=symbolslist,
  name={\ensuremath{\phi_{0}}},
  description={Initial phase-field variable},
  user1={--}
}
\newglossaryentry{Dwi}{
  type=symbolslist,
  name={\ensuremath{D_{\mathrm{wi}}}},
  description={Distance to the initial interface},
  user1={mm}
}
\newglossaryentry{Process_param}{
  type=symbolslist,
  name={\ensuremath{\theta}},
  description={Process parameters},
}

\newglossaryentry{A_inlet_size}{
  type=symbolslist,
  name={\ensuremath{A}},
  description={Inlet size},
  user1={mm}
}
\newglossaryentry{B_inlet_pos}{
  type=symbolslist,
  name={\ensuremath{B}},
  description={Inlet horizontal position},
  user1={mm}
}
\newglossaryentry{C_inlet_angle}{
  type=symbolslist,
  name={\ensuremath{C}},
  description={Inlet angle},
  user1={deg}
}
\newglossaryentry{D_vert_offset}{
  type=symbolslist,
  name={\ensuremath{D}},
  description={Inlet height},
  user1={mm}
}
\newglossaryentry{E_outlet_size}{
  type=symbolslist,
  name={\ensuremath{E}},
  description={Outlet size},
  user1={mm}
}
\newglossaryentry{F_height}{
  type=symbolslist,
  name={\ensuremath{F}},
  description={Cavity height},
  user1={mm}
}
\newglossaryentry{G_width}{
  type=symbolslist,
  name={\ensuremath{G}},
  description={Cavity width},
  user1={mm}
}
\newglossaryentry{radius}{
  type=symbolslist,
  name={\ensuremath{r_d}},
  description={Ball radius},
  user1={mm}
}
\newglossaryentry{ball}{
  type=symbolslist,
  name={\ensuremath{G_b}},
  description={Generic point (ball $B_r$)},
  user1={mm}
}
\newglossaryentry{ballr}{
  type=symbolslist,
  name={\ensuremath{B_{r_d}}},
  description={Ball of radius $r_d$, center $x_i$},
  user1={mm}
}
\newglossaryentry{G_desc}{
  type=symbolslist,
  name={\ensuremath{\mathcal{G}}},
  description={Geometry descriptor},
  user1={--}
}
\newglossaryentry{domega}{
  type=symbolslist,
  name={\ensuremath{D_\Omega}},
  description={Input mesh},
  user1={--}
}
\newglossaryentry{ss}{
  type=symbolslist,
  name={\ensuremath{s_s}},
  description={Spatial subsampling factor},
  user1={--}
}
\newglossaryentry{st}{
  type=symbolslist,
  name={\ensuremath{s_t}},
  description={Temporal subsampling factor},
  user1={--}
}
\newglossaryentry{sd}{
  type=symbolslist,
  name={\ensuremath{s_d}},
  description={Data availability percentage},
  user1={--}
}

\newglossaryentry{delta_t}{
  type=symbolslist,
  name={\ensuremath{\Delta t}},
  description={Time step size in simulations},
  user1={s}
}

%% file: 1_Abstract.tex
A mold filling problem is a flow in which a fluid advances into a cavity and occupies it under given process and geometric constraints. In metal casting, mold filling is a representative filling problem where the hydrodynamics govern defect formation, microstructure, and the final cast part quality. Evaluating candidate designs that improve these outcomes often requires running many expensive transient computational fluid dynamics simulations, which slows the exploration of large configuration and parameter spaces. With this in mind, we borrow a real-world casting example and pose it as a simplified $2$D operator learning problem. In our proposed method, a graph based encoder aggregates local neighborhood information on an input unstructured mesh and encodes geometry and boundary data. Then, a Fourier spectral core acts on a regular latent grid and captures global interactions across the domain. Finally, a graph based decoder projects the latent fields to a target mesh. Our model simultaneously forecasts velocities, pressure, and volume fraction over a fixed horizon and generalizes across varying ingate locations and process settings. On held out geometries and inlet conditions, it reproduces large scale advection and the fluid-air interface evolution with localized errors only near steep gradients. Mean relative $L_2$ errors are about \SI{5}{\percent} across velocities, pressure and volume fraction fields. The solver runs \numrange{e2}{e3} faster than traditional computational fluid dynamics and provides rapid predictions for design workflows. We additionally perform ablation studies that show monotonic accuracy loss with stronger spatial subsampling of input vertices, and a gentler deterioration with temporal subsampling. Lastly, investigations on the importance of the training dataset size show that our model presents only a small error growth when the training data are reduced by \SI{50}{\percent}. These results establish neural operators as efficient surrogates for $2$D mold filling and filling problems in general, and enable fast design in the loop exploration and optimization of gating systems in metal casting.

%% file: 3_Introduction.tex
\section{Introduction}
\label{sec:INTRO}

The description and analysis of fluid flows and their intrinsic properties constitute a central focus of research due to their significant influence on phenomena across various scientific disciplines and everyday life. Understanding fluid behavior, governed by parameters such as viscosity, flow regime, and forces, is critical for applications that range from biological processes to engineering systems. In many of these fields, the flow of fluid filling cavities has become an area of growing interest. In manufacturing, fluid flows are particularly relevant in metal casting, where the flow of molten metal inside the mold strongly affects metallurgical properties, casting geometry, and the occurrence of defects, which in turn influence the mechanical performance of the final casting. Metal casting is among the oldest manufacturing processes of humanity, with archaeological evidence dating back over $7,000$ years \citep{RADIVOJEVIC20102775}. Over time, numerous casting variants, from simple hand casting in sand to \gls{hpdc} in complex metal molds, have been developed. Despite their differences, each variant involves molten metal being poured into a cavity, representing a specific filling problem. The design of the gating system, the casting geometry, melt properties, and process parameters together govern the filling dynamics. These factors control turbulence, air and oxide entrapment, heat transfer, and solidification pathways, thereby influencing defect formation, microstructure evolution, and ultimately, the mechanical properties of the casting \citep{NourianAvval2020, ma13163539}.

\subsection{CFD simulation and optimization of mold filling in metal casting}
\label{ssec:casting}

Metal casting has evolved from early gravity-based methods such as lost wax and sand casting to modern pressure-assisted processes. Along this trajectory, \gls{hpdc} consolidated controlled filling, thermal management, and automation, enabling thin walls, short cycles, and consistent quality for aluminum and magnesium components \citep{met12111959}. Understanding mold filling is essential for assessing casting quality. The gating system governs the hydrodynamics of melt flow and thereby the formation of defects and the evolution of microstructure. Poor layouts increase turbulence, entrain oxides, and promote bifilm formation, providing nucleation sites for porosity \citep{NourianAvval2020,met14030312}. The design of the gating system and the cross section of the ingates influence pressure losses, flow distribution, and solidification rates. Unsuitable choices lead to shrinkage and microstructural inhomogeneity \citep{ma13163539,NourianAvval2020}. In thin-walled parts, multiple ingates shorten flow paths and homogenize filling, supporting grain refinement \citep{Ramadan}. Flow kinematics also control the dispersion of particles and additives \citep{2013_Gunasegaram}, while even the orientation of the gating system alters head pressure and porosity distribution, affecting strength and ductility \citep{RAZA2021201}. These mechanisms align with observations in aluminum castings, where large or clustered pores reduce ductility and increase property scatter \citep{ma13163539,NourianAvval2020,LORDAN2020139107,KANG2022103673,MAYER2003245}. Optimized gating reduces porosity and enhances strength by producing cleaner, denser microstructures
 \citep{KANG2022103673}. Thus, gating design is a primary control variable for defect incidence, microstructure, and mechanical performance.
 
Mold design is therefore widely assisted by \gls{cfd} simulation to predict melt flow and solidification. Since the late $20$th century, commercial \gls{cfd} tools \citep{magmasoft,procast,flow3d_cast} have enabled virtual exploration of mold filling by solving coupled flow and heat transfer, thereby accelerating design iteration cycles compared to trial-and-error. In practice, simulations have been applied to optimize gating layouts. \citet{kwon2019} used iterative \gls{cae} analysis to refine a \gls{hpdc} gating and overflow system to reduce air entrapment, while \citet{zhao2018} optimized thin-walled AlSi10MnMg \gls{hpdc} beams with Flow-$3$D simulations, demonstrating reduced air defects and improved surface quality. In sand casting, \citet{brna2021} combined simulation with experiments to show how gating design affects melt velocity, turbulence, and bifilm formation.

Recent advances in multiphase \gls{cfd} extend these capabilities by capturing phenomena such as air entrainment, surface turbulence, and thermal gradients. Interface capturing approaches include the \gls{vof} method \citep{kohlstdt2021} and phase-field formulations. For instance,  \citet{fuwa2019} applied a Cahn-Hilliard model to predict lamination defects in zinc alloy die casting. Advanced models also account for turbulence and Fourier type heat conduction \citep{kohlstdt2021}. While such high fidelity simulations improve predictive accuracy, fully coupled three-dimensional, two-phase, non-isothermal models remain computationally demanding \citep{kohlstdt2021}. Each candidate design still requires a numerically intensive transient simulation, which hinders the rapid exploration of large design spaces. As a result, gating design often depends on engineering heuristics, and selected solutions may only be locally optimal.  

To address these challenges, optimization frameworks have been introduced. Adjoint based gradient optimizers such as the method of moving asymptotes \citep{Svanberg1987} and its globally convergent variant \citep{Svanberg2002}, as well as multiobjective evolutionary algorithms such as NSGA-II \citep{Deb2002NSGAII}, are commonly applied to engineering design problems. In casting, surrogate assisted approaches are increasingly explored. \citet{shahane2020} demonstrated neural network surrogates trained on finite volume simulations combined with NSGA-II for multi-objective optimization of die casting solidification, while \citet{papanikolaou2018} coupled \gls{cfd} with NSGA-II in counter gravity casting, illustrating both the promise and computational constraints of such methods.

\subsection{Neural solvers for filling problems}
\label{ssec:ai_filling}

In the early $21$st century, machine learning approaches began to complement traditional scientific methods \citep{msr_fifthparadigm}. The advent of \gls{CNNs}, evolved from multilayer perceptron networks \citep{rosenblatt1958perceptron}, represent an early machine learning success. \gls{CNNs} take advantage of the structured grid, e.g. of images, and the hierarchical nature of the features via convolution
and pooling layers, respectively, to fight the curse of dimensionality and simplify the learning problem. Recent research increasingly focuses on embedding biases into neural networks operating on structured and unstructured grid data \citep{DBLP:journals/corr/BronsteinBLSV16}. \citet{gilmer2017neural} unified several graph models under the message passing neural network framework, where learned messages along edges and permutation invariant readout capture local interactions that compose into accurate global predictions on molecular benchmarks. \citet{velickovic2017graph} introduced attention on graphs so that a node can weight its neighbors adaptively, which improved both transductive and inductive node classification and showed that data driven neighborhood selection strengthens representation quality when graph structure is irregular. Complementing these advances, \citet{kipf2016semi} proposed graph convolutional networks that approximate spectral filters with localized operations, enabling efficient semi supervised learning on citation networks and establishing a simple baseline that supports deeper architectural variants. On spherical signals from omnidirectional vision and climate data, \citet{cohen2018spherical} designed spherical convolutions computed in the spectral domain to ensure rotation equivariance and reported gains on shape recognition and physics regression. In geometry processing of meshes, \citet{masci2015geodesic} constructed intrinsic geodesic patches so the network learns curvature aware features that support segmentation, retrieval, and correspondence on non-Euclidean surfaces. In \gls{CFD}, \citet{li2023geometryinformedneuraloperatorlargescale} developed \gls{PDE} surrogates on irregular shapes by combining graph encodings of geometry with Fourier layers on a latent grid, achieving large computation speedups for three dimensional aerodynamics tasks such as pressure prediction and drag estimation.  

Neural solvers are a set machine learning approaches employing neural network to solve differential equations. Neural solvers approaches can be broadly categorized into \gls{PINNs}, neural operators, and hybrid techniques combining traditional numerical methods with neural networks \citep{faroughi2022physics}. In \gls{PINNs} \citep{raissi2017physics}, the solution field is represented by a neural network trained to penalize \gls{PDE} residuals at interior collocation points while simultaneously enforcing \gls{IC}/\gls{BC} at initial and boundary collocation points, respectively. \gls{PINNs} have been successfully applied across fluids, structures, and dynamical systems \citep{mahmoudabadbozchelou2021rheology,katsikis2022gentle,raissi2017physics,stiasny2021physics}, among others. However, because the loss of the \gls{PINNs} typically targets a specific \gls{PDE} instance, geometry, and parameter set, models must be retrained or substantially fine-tuned when either changes, limiting scalability across the many distinct \gls{PDE} instances encountered in a design optimization problem.

Neural operators learn mappings from function spaces to solution fields, approximating the solution operator itself rather than an individual \gls{PDE} instance \citep{faroughi2022physics, kovachki2023neural}. Because of their broad applicability across geometric settings and parameter regimes, neural operators have attracted growing attention. For instance, \citet{lu2019deeponet} introduced DeepONets, a class of neural operators that combine a branch network for input functions and a trunk network for output coordinates and proved a universal approximation theorem for operators, establishing a practical architecture for supervised operator regression. Building on this foundation, \citet{lin2021operator} used operator learning to predict multiscale bubble growth where a network captured interface evolution across strong scale separation, while \citet{oommen2022learning} coupled neural operators with autoencoders to represent two-phase microstructure evolution and linked latent dynamics to physical order parameters. \citet{li2020fourier} proposed \gls{FNO}, a class of neural operators that lifts fields to Fourier space, learns global integral kernels on selected modes, and projects back to the spatial domain. This design delivers resolution invariance and zero shot super resolution together with strong accuracy on Burgers, Darcy, and Navier Stokes systems and large speedups over spectral solvers. Follow up theory established universal approximation and error bounds that clarify conditions under which Fourier layers recover solution operators and thereby strengthened the mathematical foundation of this approach \citep{li2024physics}. \citet{qin2024toward} further analyzed the spectral behavior of Fourier operators, identified a parametrization bias toward dominant frequencies, and proposed spectral boosting modules that recover non dominant content and reduce error on a range of \gls{PDE} benchmarks. Beyond regular grids, \citet{li2023fourier} introduced a geometry aware variant that learns a deformation to a uniform latent mesh so that Fourier layers act on irregular discretizations while preserving discretization convergence and accelerating inference across diverse geometries. \citet{wen2022u} designed an augmented operator for multiphase flow that injects U-Net style multi-resolution mixing into the spectral core. Taken together, all these developments ground Fourier and other types of neural solvers for filling problems by enabling operator learning surrogates on unstructured meshes that accelerate design exploration while maintaining accuracy across geometries.

\subsection{Aim and contribution}
\label{ssec:aim}
Metal casting represents an important manufacturing method that can be characterized as a filling problem. Specifically, the hydrodynamics of mold filling decisively govern defect formation, microstructure evolution, and thus final cast part quality. \gls{CFD} simulations are the industry standard approach for analyzing mold filling and therefore for guiding mold and gating system designs. Yet, these simulations are typically executed in iterative workflows, where each candidate layout demands a numerically intensive transient simulation. While running a single simulation is generally viable, the outcome design configuration is likely suboptimal. At the same time, the exploration of large design spaces that are often encountered in engineering problems is slow. Consequently, research about surrogate models that can approximate filling problems are becoming a relevant research avenue.

In the recent years, artificial intelligence has begun to reshape scientific computing by complementing first principles modeling with data driven surrogates. In scientific machine learning, operator learning frameworks such as DeepONet \citep{lu2019deeponet} and the \gls{FNO} \citep{li2020fourier} approximate \gls{PDE} solution operators directly in function space, enabling surrogate evaluations that are orders of magnitude faster than conventional numerical solvers while preserving accuracy on established benchmarks. Yet, filling problems in the manufacturing industry remain comparatively underexplored due to the difficulty of representing strongly coupled, geometry-dependent spatio-temporal dynamics under varying \gls{IC}/\gls{BC}, and the large, structured design space encountered in design optimization. To address this gap, this work develops an operator learning surrogate for a filling problem.  Specifically, we study a simplified yet representative two-phase mold filling problem and train a neural solver to approximate the coupled \gls{CHNS} solution operator on unstructured meshes, with generalization across diverse gate geometries and placements. Our main contributions are:

\begin{itemize}[noitemsep,topsep=0pt,parsep=0pt,partopsep=0pt, leftmargin=.5cm]
\item[--] This work formulates $2$D mold filling, a representative case of the broader family of filling problems, as an operator learning problem and propose a Fourier-Graph neural solver that maps problem definitions such as geometry, initial and boundary conditions, and process parameters to transient fields of velocity, pressure, and volume fraction.
\item[--] The resulting neural solver models generalize across parametric gating design spaces, amortizing learning over many \gls{PDE} instances and delivering orders-of-magnitude speedups relative to \gls{CFD} while retaining predictive accuracy for fluid-air interface and flow fields.
\item[--] This study analyses the model susceptibility to spatial and temporal subsampling and long-horizon error growth, and ablate architectural and training choices such as Fourier modes, latent resolution, and temporal rollout to identify stability and accuracy drivers.
% \item[--] We show that aggregating point-wise \gls{PDE} knowledge from multiple designs yields robust surrogates that can be conditioned on new gate configurations to produce fast, reliable flow predictions to accelerate mold filling gate design.
\item[--] We show that aggregating point-wise \gls{PDE} knowledge from multiple designs yields robust surrogates that can be conditioned on new gate configurations to produce fast, reliable flow predictions.
\end{itemize}

%On held-out geometries and inlet conditions, our solver attains a mean relative $L_2$ of $\approx 6$--$7\%$ on $(v_x, v_y, p, \alpha)$ while delivering wall-clock speedups over CFD (see Section \ref{sec:res}). 

\textbf{Paper organization:} Sections \ref{formulation} and \ref{strong_form_eq} formalize the problem setup and governing equations, respectively. Section \ref{data_generation} present our \gls{CFD} dataset generation strategy and Section \ref{method} our proposed Fourier-Graph neural operator learning setup. Section \ref{ablation_studies} summarizes the ablation studies performed. Section \ref{Quantitative_results} broadly investigates the general solver capabilities. Sections \ref{spatial_subsampling}, \ref{temporal_subsampling}, and \ref{training_data} evaluate the performance of the proposed neural solver under spatial and temporal subsampling, as well as data efficiency, respectively. Finally, Section \ref{conclusion} concludes and outlines future research opportunities.

%% file: 4_Methodology.tex
\section{Methodology}
\label{Methodology}

The following methodology is adopted to learn a parametrized two-phase mold filling simulation using neural solvers. It begins by introducing the \gls{CHNS} problem and a parametric mold cavity model, which, together with \gls{IC} and \gls{BC} data, define the mold filling design space. Then, we formalize the governing equations in strong and numerical form. Next, the generation of supervised datasets via \gls{CFD} on the parametric cavity model is detailed and the neural solver theory and training recipe are presented. Finally, the metrics employed in the evaluation protocol and the ablation studies on spatial and temporal subsampling, and the availability of data are outlined.

\subsection{Problem formulation and assumptions}
\label{formulation}
Isothermal two-phase mold filling is modeled by the incompressible Navier-Stokes \citep{Navier1823,Stokes1845} equations coupled with a diffuse-interface Cahn-Hilliard \citep{CahnHilliard1958} formulation. The unknowns are the velocity field $\gls{u_vec}(\gls{x},\gls{time}) \in\mathbb{R}^2$, pressure $\gls{p}(\gls{x},\gls{time}) \in\mathbb{R}$, and the volume fraction $\gls{alpha}(\gls{x},\gls{time}) \in[0,1]$\footnote{Bold symbols denote vectors (e.g., \gls{u_vec}), plain symbols denote scalars (e.g., \gls{p},\gls{phi})} for some time $\gls{time} \in \gls{Thorizon}$ with $\gls{time} \ge0$. Thermal and solidification effects are neglected to isolate the fluid-dynamic contribution to filling patterns.

\vspace{1em}
\begin{figure}[htb] 
    \centering                  
  % Panel 1
\begin{subfigure}[b]{0.33\textwidth}
    \centering
    \small
    \def\svgwidth{135pt}
    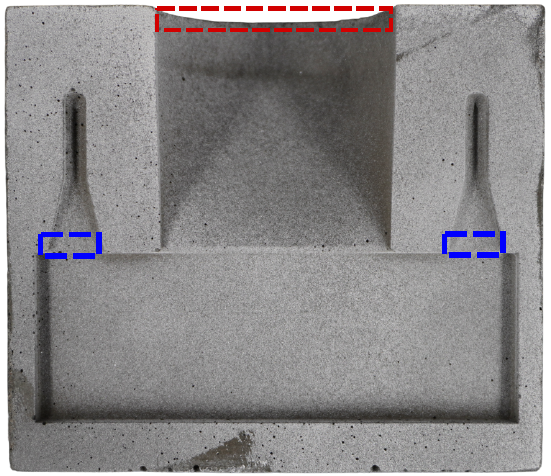
    \caption{Experimental concrete mold}
    \label{mold_exp}
  \end{subfigure}\hfill
  \begin{subfigure}[b]{0.33\textwidth}
    \centering
    \footnotesize
    \def\svgwidth{150pt}
    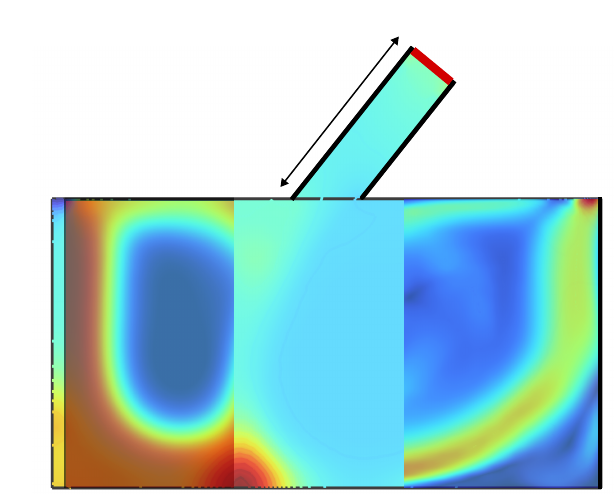
    % Koordinatensystem in der linken unteren Ecke
    \begin{tikzpicture}[overlay, remember picture]
        \coordinate (origin) at (-4.65,0.2);
        \draw[-{Stealth}, thick, white] (origin) -- ++(.55,0) node[right] {x};
        \draw[-{Stealth}, thick, white] (origin) -- ++(0,.55) node[above] {y};
    \end{tikzpicture}
    \caption{Mold cavity model}
    \label{cavity}
  \end{subfigure}\hfill{}
  % Panel 4
  \begin{subfigure}[b]{0.33\textwidth}
    \centering
    \def\svgwidth{135pt}
    \small
    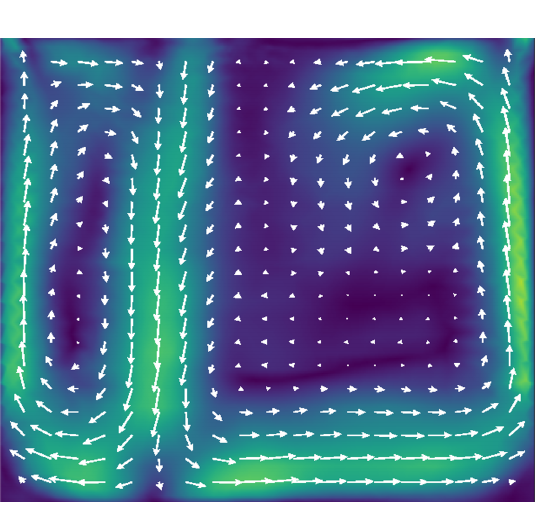
    \caption{Exemplaric field prediction}
    \label{field_prediction}
  \end{subfigure}\hfill
    \caption[Experimental concrete mold, mold cavity model, and velocity field prediction]{\textbf{Casting model}. Figure \ref{mold_exp} shows an experimental concrete mold with sprue, cavity, and two risers. Figure \ref{cavity} depicts the parametric cavity model with filling colors representing numerical simulation outcomes. Figure \ref{field_prediction} illustrates an exemplarily velocity field prediction on a rectangular domain affinely normalized to $[0,1]^2$.}
    \label{fig:model_annotated}
\end{figure}

Let $\gls{domain} \subset\mathbb{R}^2$ denote the mold model composed of the ingate pipe and the rectangular cavity. Its boundary $\partial\gls{domain}$ is partitioned into a single inlet \gls{inletBoundary} at the top wall, two outlets \gls{outletBoundary}, and no-slip walls \gls{wallBoundary}. Figure \ref{fig:model_annotated} outlines this setup: Figure \ref{mold_exp} shows an experimental concrete mold, taken from the study published by \citet{Link2025}, with sprue and two risers that inspired our model. Figure \ref{cavity} highlights the parametric cavity model with inlet size \gls{A_inlet_size}, horizontal position \gls{B_inlet_pos}, angle \gls{C_inlet_angle}, vertical offset \gls{D_vert_offset}, outlet size \gls{E_outlet_size}, height \gls{F_height}, and width \gls{G_width}. Finally, Figure \ref{field_prediction} illustrates a predicted velocity field on the affinely normalized domain $[0,1]^2$.

At $\gls{time}=\SI{0}{\second}$, a steady inflow with a diffuse interface consistent with the \gls{CHNS} formulation is imposed on \gls{inletBoundary}. The inflow has a Dirichlet velocity of magnitude \gls{Vmag} and direction \gls{zeta}. On the walls \gls{wallBoundary}, we impose no-slip condition. The outlets \gls{outletBoundary} are designed to suppress backflow and to compensate hydrostatic pressure. Given a geometry descriptor \gls{G_desc} defining \gls{domain}, process parameters \gls{Process_param}, and inlet data $(\gls{Vmag},\gls{zeta})$, the task is twofold: (i) numerically solve the forward \gls{CHNS} problem to obtain $(\gls{u_vec},\gls{p},\gls{alpha})$ fields on $\gls{domain}\times(0,\gls{Thorizon}]$ and (ii) learn an operator \gls{operator} that approximates the solution map \gls{solution_map}
\[
\mathcal{S} : (\mathcal{G},\theta,V,\zeta) \;\mapsto\; (\mathbf{u},p,\alpha),
\tag{1}
\]
from supervised pairs generated by the numerical solver, yielding fast surrogate predictions $(\gls{u_vec_hat},\gls{p_hat},\gls{alpha_hat})$
\[
(\hat{\mathbf{u}},\hat{p},\hat{\alpha})=\mathcal{N}_{\vartheta}(\mathcal{G},\theta,V,\alpha).
\tag{2}
\]

\subsection{Governing equations}
\label{strong_form_eq}
On $\gls{domain}\times(0,\gls{Thorizon}]$, the governing incompressible \gls{CHNS} \citep{GurtinPolignoneVinals1996} system for the velocity field $\gls{u_vec}=(\gls{u},\gls{v})$, pressure \gls{p}, and volume fraction \gls{alpha} reads

%We solve the incompressible Navier-Stokes/Cahn-Hilliard system on $\Omega \times (0,T]$ for velocity $\mathbf{u}=(u,v)$, pressure $p$, and volume fraction $\alpha \in [0,1]$ using \textsc{COMSOL Multiphysics}\cite{comsol2024}. The continuity equation describing the mass conservation reads
\[
\nabla \cdot \mathbf{u} = 0
\tag{3}
\]
for the continuity equation and
\[
\rho\left(
\frac{\partial u}{\partial t}
+ u \frac{\partial u}{\partial x}
+ v \frac{\partial u}{\partial y}
\right)
= -\frac{\partial p}{\partial x}
+ \mu\left(
\frac{\partial^2 u}{\partial x^2}
+ \frac{\partial^2 u}{\partial y^2}
\right)
+ F_x,
\tag{4}
\]

\[
\rho\left(
\frac{\partial v}{\partial t}
+ u \frac{\partial v}{\partial x}
+ v \frac{\partial v}{\partial y}
\right)
= -\frac{\partial p}{\partial y}
+ \mu\left(
\frac{\partial^2 v}{\partial x^2}
+ \frac{\partial^2 v}{\partial y^2}
\right)
+ F_y
+ \rho g,
\tag{5}
\]
for the momentum conservation, with \gls{g} acting in the $-\gls{ycoord}$ direction, $(\gls{Fx},\gls{Fy})$ being the surface tension force components, and \gls{rho} the fluid density. The volume fractions of the two fluids are modeled with a phase field model based on the Cahn-Hilliard equations
\[
\frac{\partial \phi}{\partial t} 
+ u \frac{\partial \phi}{\partial x} 
+ v \frac{\partial \phi}{\partial y} 
= \frac{\gamma \lambda}{\varepsilon^{2}}
\left( \frac{\partial^{2} \Psi}{\partial x^{2}} 
     + \frac{\partial^{2} \Psi}{\partial y^{2}} \right)
\tag{6}
\]

\[
\Psi = -\varepsilon^{2}
\left( \frac{\partial^{2} \phi}{\partial x^{2}} 
     + \frac{\partial^{2} \phi}{\partial y^{2}} \right)
+ (\phi^{2}-1)\phi
\tag{7}
\]
with phase-field variable \gls{phi}, chemical potential \gls{psi}, mixing energy density \gls{lambda_mix}, interface thickness \gls{epsilon}, and mobility parameter $\gls{gamma}=\gls{chi}\gls{epsilon}^2$ for mobility tuning parameter \gls{chi}. This leads to the volume fractions of both fluids via
\[
\alpha_{2} =
\begin{cases}
0, & \phi < -1, \\[6pt]
\dfrac{1+\phi}{2}, & -1 \leq \phi \leq 1, \\[6pt]
1, & \phi > 1,
\end{cases}
\qquad
\alpha_{1} = 1 - \alpha_{2}.
\tag{8, 9}
\]
The surface tension contributions to the Navier–Stokes equations in each direction are
\[
F_x = \frac{\lambda}{\varepsilon^{2}} \, \Psi \, \frac{\partial \phi}{\partial x},
\qquad
F_y = \frac{\lambda}{\varepsilon^{2}} \, \Psi \, \frac{\partial \phi}{\partial y}.
\tag{10, 11}
\]
and the density and dynamic viscosity are interpolated via
\[
\rho = \rho_1 \alpha_{1} + \rho_2 \alpha_{2},
\qquad
\mu = \mu_1 \alpha_{1} + \mu_2 \alpha_{2}
\tag{12,13}
\]
for the density \gls{rho1} and dynamic viscosity \gls{mu1} of fluid one and the density \gls{rho2} and dynamic viscosity \gls{mu2} of fluid two. On the inlet \gls{inletBoundary}, Figure \ref{cavity}, the boundary conditions read: 
\[
u=V\cos (C), \qquad v=V\sin (C).
\tag{14, 15}
\]
On wetted walls \gls{wallBoundary}, the no-slip $\gls{u}=\gls{v}=0$ conditions read
\begin{equation}
\frac{\gamma \lambda}{\varepsilon^{2}}
\left(
\cos(\beta_n)\frac{\partial \Psi}{\partial x}
+ \sin(\beta_n)\frac{\partial \Psi}{\partial y}
\right) = 0,
\tag{16}
\end{equation}

\begin{equation}
\varepsilon^{2}
\left(
\cos(\beta_n)\frac{\partial \phi}{\partial x}
+ \sin(\beta_n)\frac{\partial \phi}{\partial y}
\right)
= \varepsilon^{2}\cos(\theta_w)
\left(
\left(\frac{\partial \phi}{\partial x}\right)^{2}
+ \left(\frac{\partial \phi}{\partial y}\right)^{2}
\right),
\tag{17}
\end{equation}
for the respective angle \gls{beta_n} of the normal vector of the wall, static contact angle at the wall \gls{theta_w}, and chemical potential \gls{psi}. At the outlets \gls{outletBoundary} the backflow is suppressed and on the whole domain the hydrostatic pressure is compensated. The initial phase field \gls{phi0} for the part of the domain filled with the first fluid is set to
\[
\phi_{0} = -\tanh\!\left(\frac{D_{wi}}{\sqrt{2}\,\varepsilon}\right),
\tag{18}
\]
and for the part of the domain filled with the second fluid to
\[
\phi_{0} = \tanh\!\left(\frac{D_{wi}}{\sqrt{2}\,\varepsilon}\right),
\tag{19}
\]
for the distance to the initial interface \gls{Dwi}. Unless otherwise stated, any variable dependent initialization or constant not explicitly specified adheres to the default COMSOL Multiphysics \citep{comsol2024} settings used in our simulations.

\subsection{Dataset generation}
\label{data_generation}

To probe the capabilities of neural solvers, we generate two synthetic datasets by varying the model parameters of the filling problem presented in Figure \ref{cavity} and numerically solving the \gls{CHNS} equations in COMSOL. These two datasets, DS$_1$ and DS$_2$, were conceptualized to represent increasing levels of learning complexity for neural solvers. In DS$_1$, the inlet horizontal position is fixed and only the inlet velocity magnitude and direction vary, representing a learning problem with localized momentum injection. In DS$_2$, the inlet horizontal position also varies, producing more complex flow regimes. 

The parametric design space of the mold filling problem is summarized in the Table~\ref{tab:params}. The geometric parameter ranges and step counts follow Figure  \ref{fig:model_annotated} and Table~\ref{tab:params}: $\gls{A_inlet_size}\in\qtylist[list-units=bracket,list-open-bracket={[},list-close-bracket={]}]{10;25}{\mm}$, $\gls{B_inlet_pos}\in\qtylist[list-units=bracket,list-open-bracket={[},list-close-bracket={]}]{10;90}{\mm}$ (fixed in DS$_1$, variable in DS$_2$), $\gls{C_inlet_angle}\in\qtylist[list-units=bracket,list-open-bracket={[},list-close-bracket={]}]{10;90}{\degree}$,  $\gls{D_vert_offset}= \SI{30}{\milli\meter},$ $\gls{E_outlet_size}=\SI{10}{\milli\meter}$, 
$\gls{F_height}=\SI{50}{\milli\meter}$, 
$\gls{G_width}=\SI{100}{\milli\meter}$, and $\gls{Vmag}\in\qtylist[list-units=bracket,list-open-bracket={[},list-close-bracket={]}]{0.1;0.9}{\percent}$.

% DS$_1$ varies ingate size, direction, and velocity, while the inlet horizontal position is held fixed. DS$_2$ extends DS$_1$ by sweeping the inlet horizontal position $B$, thereby inducing qualitatively distinct flow topologies as the momentum injection point moves across the top wall. 

%This construction yields a controlled progression in difficulty: DS$_1$ isolates changes in magnitude and orientation at a fixed location. DS$_2$ adds spatial relocation of the ingate, stressing non-local couplings between boundary geometry and interior fields. 

In all simulation cases, a steady inflow on the inlet \gls{inletBoundary} with Dirichlet velocity of magnitude \gls{Vmag} and direction \gls{zeta} is imposed at $\gls{time}=\SI{0}{\second}$, the walls \gls{wallBoundary} are considered no-slip, and the outlets \gls{outletBoundary} suppress backflow and compensate hydrostatic effects. As initial condition, we consider the inlet pipe filled with water at $\SI{373}{\kelvin}$, whereas the rectangular mold is filled with air at $\SI{373}{\kelvin}$. As an outcome, each simulation produces time-resolved velocity $\gls{u_vec_hat}=(\gls{u_hat},\gls{v_hat})$, pressure \gls{p_hat}, and volume fraction \gls{alpha_hat} fields on an unstructured mesh, yielding sequences suitable for mesh-aware operator learning. The number of mesh elements stays below $3000$ elements and depends on the \gls{BC} and geometric characteristics. The total temporal integration spans \gls{time} $\in \qtylist[list-units=bracket,list-open-bracket={[},list-close-bracket={]}]{0;5}{\second}$ with a fixed time step of $\gls{delta_t}=\SI{0.01}{\second}$. The resulting datasets comprise $939$ simulations for the simplified DS$_1$ setting and $6320$ for the complete DS$_2$ setting. Each simulation runs for about \SI{2}{\minute} on a \gls{HPC} cluster while using two \textit{Intel Xeon Gold $\mathit{6326}$ Ice Lake CPUs} with $2\times16$ cores at $\SI{2.9}{\giga\hertz}$.

\begin{table}[htbp] 
    \newcolumntype{C}{>{\centering\arraybackslash}m{1.5cm}} 
    \newcolumntype{F}{>{\centering\arraybackslash}m{2.5cm}}
    \small
    \centering 
    \caption{\textbf{Mold filling design space}. Parametric ranges employed for the generation of the DS$_1$ and DS$_2$ mold filling datasets. DS$_2$ extends DS$_1$ by additionally varying the inlet horizontal position, thereby allowing momentum injection at different locations of the upper rectangle wall.}
    \vspace{.5em}
    \begin{tabularx}{.95\textwidth}{XCFFF}
    \toprule
       \textbf{Parameter} & \textbf{Unit} & \textbf{Value range} & \textbf{Steps DS$_1$} & \textbf{Steps DS$_2$}\\
    \midrule
        Inlet size (A) & \SI{}{\milli\meter} & \numrange{10}{25} & $10$ & $10$\\
        Inlet horizontal position (B) & \SI{}{\milli\meter} & \numrange{10}{90} & fixed & $10$\\
        Inlet angle (C) & \SI{}{\degree} & \numrange{10}{90} & $10$ & $10$\\
        Inlet vertical offset from top wall (D) & \SI{}{\milli\meter} & 30 & fixed & fixed\\
        %Outlet size (E) & \SI{}{\milli\meter} & $10$ & fixed & fixed\\
        %Cavity height (F) & \SI{}{\milli\meter} & $50$ & fixed & fixed\\
        %Cavity width (G) & \SI{}{\milli\meter} & $100$ & fixed & fixed\\
        %Cavity width (G) & \SI{}{\milli\meter} & $100$ & fixed & fixed\\
        Inlet velocity (V) & \SI{}{\percent} & \numrange{0}{100}\footnotemark & $10$ & $10$\\
    \bottomrule
    \end{tabularx}
    \label{tab:params}
\end{table}

\footnotetext{Velocity fractions that enforce Reynolds number $< 2300$ at the ingate pipe.}

\subsection{Proposed method}
\label{method}
The main goal of the current study is to understand to what extent Fourier neural solvers can approximate parametric transient filling problems with varying \gls{IC} and \gls{BC} conditions. Therefore, we borrow and simplify a casting problem from an existing study, see Section \ref{formulation}, and seek to map parametric mold geometries, \gls{IC} and \gls{BC} conditions as well as process parameters to time-resolved velocity $(\gls{u_hat},\gls{v_hat})$, pressure \gls{p_hat}, and volume fraction \gls{alpha_hat} fields. Towards this goal, we introduce a Fourier-Graph neural solver that is based on and extends the \gls{GINO} \citep{li2023geometryinformedneuraloperatorlargescale}. 
In \gls{GINO}, a \gls{GNO} encoder first aggregates local features from an unstructured input mesh and maps them to a regular latent grid. On this grid, a \gls{FNO} stack captures long-range couplings while modulating Fourier features according to the inlet velocity via \gls{AdaIN} \citep{huang2017arbitrarystyletransferrealtime}. Finally, a \gls{GNO} decoder projects the learned latent predictions back onto an unstructured mesh that is not necessarily the same as the input. In summary, the \gls{GINO} design combines the ability of \gls{GNO} in handling unstructured meshes and learning local features with the efficiency of \gls{FNO} for modeling global interactions. However, GINO presents the drawback of only modeling single field, steady-state problems, whereas the regarded casting problem is a transient, multi-field problem. 

To address these limitations, we introduce a Fourier-Graph method that extends \gls{GINO} in four different ways, 
see Figure \ref{fig:model}. First, we modify the output of the \gls{GNO} encoder to broadcast the $2$D latent grid to $3$D, thereby augmenting the latent space to handle transient features. Second, given the parametric nature of our mold filling problem, we set \gls{AdaIN} to modulate the learned Fourier features based on the inlet setup \gls{inlet_setup} composed of inlet velocity \gls{Vmag}, size \gls{A_inlet_size}, horizontal position \gls{B_inlet_pos}, and angle \gls{C_inlet_angle}. Third, we condition the input unstructured meshes on the initial field conditions and inlet mask to assist model learning. Finally, we employ an adapted causal rollout loss, inspired by \citet{wang2024respecting}, which reweighs temporal errors to prioritize earlier time steps during training, mirroring the forward-marching behavior of standard time-stepping solvers. With these extensions, our proposed Fourier-Graph neural operator is capable of predicting $3$D multi-field transient mold filling simulations. In the following, Section~\ref{FNO_} presents the Fourier neural solver theory, Section~\ref{objective_optmization} formalizes our Fourier-Graph learning objective and optimization, and Section~\ref{sec:metrics} the metrics we used for training and evaluation.

\subsubsection{Fourier Neural Solver}
\label{FNO_}
\begin{figure}[htbp] 
    \centering                  
    \def\svgwidth{\textwidth}
    \scriptsize
  {%
    %\fontsize{9}{10}\selectfont
    %\adjustbox{trim=225pt 170pt 55pt 158pt, clip}{%
    %\adjustbox{trim=250pt 185pt 55pt 172pt, clip}{%
    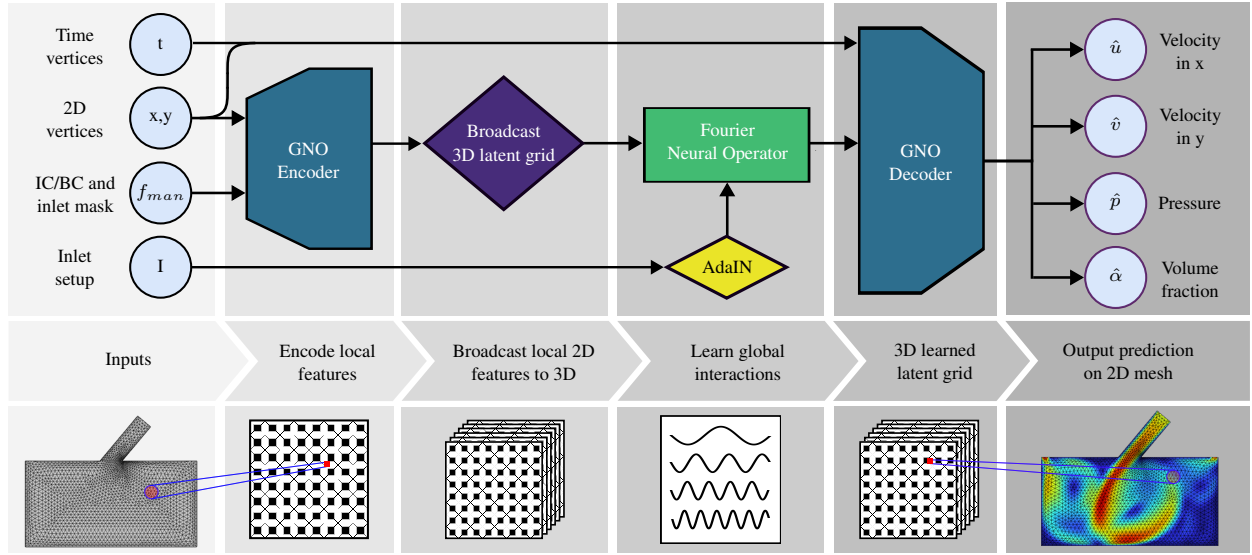
    %}%
  }
    \caption{\textbf{Fourier–Graph neural solver.} The model receives as input the unstructured mold geometry mesh, the inlet mask, and the initial fields at $\gls{time}_0=0$. A geometry-aware \gls{GNO} encoder lifts the inputs to a regular $2$D latent grid that is broadcasted to $3$D. The Fourier spectral core \gls{FNO} captures long-range couplings while modulating Fourier features through \gls{AdaIN} based on the inlet setup \gls{inlet_setup} composed of inlet velocity \gls{Vmag}, size \gls{A_inlet_size}, horizontal position \gls{B_inlet_pos}, and angle \gls{C_inlet_angle}. Finally, for each time step \gls{time} $ \in $ \gls{Thorizon}, a \gls{GNO} decoder maps the $3$D latent grid to $(\gls{u_vec_hat},\gls{p_hat},\gls{alpha_hat})$ fields at arbitrary $(\gls{xcoord},\gls{ycoord})$ spatial locations within the mold domain.}
    \label{fig:model}
\end{figure}

The proposed Fourier-Graph approach is composed of a encoder–spectral–decoder architecture, as depicted in Figure \ref{fig:model}. The \gls{GNO} encoder follows the idea of \gls{GNNs} \citep{1555942}, where the neighborhood information of a vertice is aggregated. Let \gls{xi} denote a query node in the unstructured input mesh \gls{domega} connected to all mesh points within a ball $\gls{ball}\in\gls{ballr}(\gls{x})\subset\gls{domega}$ with radius $\gls{radius}\in\mathbb{R}_{+}$ centered at \gls{xi}. This construction realizes the integral operator
\[
K(f_{man})(\mathbf{x}) \;=\; \int_{B_{r_d}(\mathbf{x})} \kappa_{\hat{\theta}}(\mathbf{x},b)\,f_{man}(\mathbf{x})\,db
\;,\;\;  \mathbf{x_i} \in D_\Omega.
\tag{20}
\]
for an input function \gls{fman} describing the geometry, inlet mask, and initial conditions, and a learnable neural network kernel \gls{kappa} with parameters \gls{theta_hat}. The \gls{GNO} encoder produces mesh-aware features on a regular $2$D latent grid, which are broadcasted to form a $3$D latent representation that serves as input to the spectral core.

On the $3$D latent grid, each Fourier layer \gls{flayer} of the spectral core transforms the latent signal into the frequency space, applies a linear mapping to the retained low-frequency modes, and brings it back to the spatial domain via an inverse Fourier transform. In parallel, the $3$D latent grid is processed through a pointwise linear transformation. Finally, the two data streams are summed and passed through a nonlinear activation. In compact form, for a layer $\ell$, we therefore have
\[
U^{\ell+1} = \sigma\!\left(W^\ell U^\ell + \mathcal{F}^{-1}\!\left(R^\ell \odot \mathcal{F}(U^\ell)\right)\right),
\tag{21}
\label{eq:fno_layer}
\]
where \gls{latent} denotes the latent signal (e.g. at the first layer $\gls{latent}=\gls{integral})$, \gls{FFF} is the \gls{FFT} on the $3$D latent grid, \gls{rweights} are learnable spectral weights supported on a truncated set of modes, and \gls{plinearmap} is a point-wise linear map. This truncated spectral parametrization is modulated through \gls{AdaIN} based on the inlet setup \gls{inlet_setup} composed of inlet velocity \gls{Vmag}, size \gls{A_inlet_size}, horizontal position \gls{B_inlet_pos}, and angle \gls{C_inlet_angle}, and captures the dominant global couplings of the incompressible flow and interface transport, while the linear maps preserve local information. At the end, the stack of \gls{FNO} layers produces a fixed-horizon sequence on a regular latent grid, which is used by the \gls{GNO} decoder to predict velocities $(\gls{u_hat},\gls{v_hat})$, pressure \gls{p_hat}, and volume fraction \gls{alpha_hat} at each time step \gls{time} $ \in $ \gls{Thorizon}. Note that the output mesh does not necessarily have the same $(\gls{xcoord},\gls{ycoord})$ coordinates as the input mesh.

\vspace{1em}

% \begin{figure}[htbp] 
%     \centering                  
%    % \def\svgwidth{1000pt}
%     \def\svgwidth{920pt}
%   {%
%     \fontsize{9}{10}\selectfont
%     \adjustbox{trim=225pt 170pt 55pt 158pt, clip}{%
%     %\adjustbox{trim=250pt 185pt 55pt 172pt, clip}{%
%     \input{Plots/Methodology/Zeichnung2_workflow9.pdf_tex}
%     }%
%   }
%     \caption{\textbf{Fourier–Graph neural solver.} The model receives as input the unstructured mold geometry mesh, the inlet mask, and the initial fields at $\gls{time}_0=0$. A geometry-aware \gls{GNO} encoder lifts the inputs to a regular $2$D latent grid that is broadcasted to $3$D. The Fourier spectral core \gls{FNO} captures long-range couplings while modulating Fourier features through \gls{AdaIN} based on the inlet setup \gls{inlet_setup} composed of inlet velocity \gls{Vmag}, size \gls{A_inlet_size}, horizontal position \gls{B_inlet_pos}, and angle \gls{C_inlet_angle}. Finally, a \gls{GNO} decoder maps the $3$D latent fields to mesh vertices to predict $(\gls{u_vec_hat},\gls{p_hat},\gls{alpha_hat})$ at each time step to arbitrary $(\gls{xcoord},\gls{ycoord})$ locations within the mold domain.}
%     \label{fig:model}
% \end{figure}

\subsubsection{Learning objective and optimization}
\label{objective_optmization}

Let the input bundle \gls{input_a} $=$ (\text{geometry mask}, ${\gls{IC}/\gls{BC}}, \gls{A_inlet_size},\dots,\gls{Vmag})$ be defined on \gls{domain}.
We learn a discretization–invariant neural operator as a supervised, fixed–horizon operator regressor over the four $(\gls{u},\gls{v},\gls{p},\gls{alpha})$ simulation field targets. For prediction steps $\gls{steps}=1,\ldots,\gls{stepsT}$, target fields $\gls{fields}\in({\gls{u},\gls{v},\gls{p},\gls{alpha}})$, and prediction fields $\gls{fieldsp}=(\gls{u_hat},\gls{v_hat},\gls{p_hat},\gls{alpha_hat})$, the per–step, per–field relative \(L_2\) error over all \gls{verticesN} vertices is given by
\begin{equation}
\ell_q^{(k)}
=
\frac{\frac{1}{N}\sum_{i=1}^{N}\big(\hat q_i^{(k)}-q_i^{(k)}\big)^2}
     {\frac{1}{N}\sum_{i=1}^{N}\big(q_i^{(k)}\big)^2}
\tag{22}
\label{eq:opt}
\end{equation}
To stabilize multi–step forecasts during training we apply a causal rollout weighting \citep{wang2024respecting} that increases the weight on a time step only after the residuals at earlier steps are reduced. In order to do that, we form the cumulative past loss for each field $\gls{closs}_q^{(k)}=\sum_{j<k}\gls{lloss}_q^{(j)}$ and assign a single temporal weight $\gls{tweight}^{(k)}$ to step \gls{steps} by combining all the distinct field contributions
\begin{equation}
\gamma_t^{(k)}=\min_{q}\exp\!\big(-\tau\,c_q^{(k)}\big)
            =\exp\!\Big(-\tau\,\max_{q}\sum_{j<k}\ell_q^{(j)}\Big),\qquad \tau>0 .
\tag{23}
\label{eq:tau}
\end{equation}
where \gls{tau} denotes the causality parameter that controls the steepness of the weights. Notice that we adopt a conservative temporal weight approach by selecting the minimum weight among all fields instead of a per-field minimum. Combining Equations (\ref{eq:opt}) and (\ref{eq:tau}) finally leads to our optimization loss \gls{trainingloss}

\begin{equation}
L_{\text{opt}}
= \frac{1}{4H}\sum_{q\in\{u,v,p,\alpha\}}\sum_{k=1}^{H} \gamma^{(k)}\,\ell_q^{(k)}
\tag{24}
\end{equation}
which was employed in all experiments we performed. 

During training, all models were optimized with the same training recipe. Specifically, we employed the AdamW \citep{loshchilov2019adamw} optimizer, a reduce-on-plateau learning rate schedule, and early stopping on a held-out validation split. Furthermore, we considered single-simulation mini-batches on DS$_1$ and DS$_2$ with a fixed, non-autoregressive forecast horizon of $\gls{stepsT}=10$ steps. For training, a single \textit{NVIDIA A$\mathit{100}$ GPU} with $\SI{40}{\giga\byte}$ VRAM was employed, and training convergence was typically achieved within \SIrange{48}{72}{\hour} per model. Under the available computational budget, the best performing configuration, used as the default unless stated otherwise, employs a Fourier-Graph operator with \qtyproduct{48x48x48}{} modes in the Fourier core, a graph radius of \SI{24}{\mm} for message passing, a latent-grid resolution of $80$, an initial learning rate of $1\times10^{-4}$, and weight decay of $10^{-5}$.

\subsubsection{Evaluation metrics}
\label{sec:metrics}
During evaluation, we report the time-averaged relative $L_2$ percentage error \emph{per field} and a \emph{mean} across all fields. The relative $L_2$ percentage  error \gls{rerror} at step \gls{steps} for a target field \gls{fields} is given by
\begin{equation}
r_q^{(k)} \;=\;\,\frac{\lVert \hat{q}^{(k)} - q^{(k)}\rVert_2}{\lVert q^{(k)}\rVert_2}\,
\tag{25}
\end{equation}
The time-averaged relative error \gls{rerror}$_q$ for field \gls{fields} is
\begin{equation}
r_q \;=\; \frac{1}{H}\sum_{k=1}^H r^{(k)}_q \,
\tag{26}
\end{equation}
When the mean is reported, we consider the mean time relative average error across all fields:
\begin{equation}
r \;=\; \tfrac{1}{4}\sum_{q\in\{u,v,p,\alpha\}} r_q\,
\tag{27}
\end{equation}

\subsection{Ablation studies}
\label{ablation_studies}
Neural operator models learn \gls{PDE} solution operators that are mesh-agnostic, generalizing across distinct discretization strategies and resolutions \citep{li2020fourier,kovba}. However, this characteristic has not been evaluated extensively for filling problems with extensive parametric design spaces. To characterize how the discretization and amount of data impact the neural solver performance, we therefore conducted controlled ablations over spatial and temporal subsampling, as well as training set size. Table~\ref{tab:exp} summarizes all the ablation studies we performed.

First, we investigate the impact of different spatial subsampling factors $\gls{ss}\in \{1, \ldots, 5\}$ on DS$_1$ and DS$_2$ by uniformly sampling vertices during training. In this scenario, our motivation is to evaluate how coarsening the input training meshes impact the modeling of high frequency field data by reporting the model performance at full mesh resolutions. Second, we evaluate the model robustness against temporal subsampling factors $\gls{st}\in \{1, \ldots, 6\}$ on DS$_1$ and D$S_2$ by varying the time step size \gls{delta_t} between steps $\gls{steps}=1,\ldots,\gls{stepsT}$ used for training. In this case, we seek to evaluate whether changing the simulation time step impacts the model learning ability or not, i.e. evaluate the model ability to learn more abrupt field changes between each transient step prediction. Finally, we investigate the role of data availability \gls{sd} on DS$_2$ by training with \qtylist[list-units=brackets,list-open-bracket = \{,list-close-bracket =\}]{100;90;80;70;60;50}{\percent} of the available data corpus. In this scenario, we progressively shrink the size of the training dataset by sampling from it and report the mean and standard deviation. Our motivation is to estimate the amount of data necessary to achieve certain performance levels with neural solvers. Together, these ablation studies map the Fourier-Graph vulnerability to temporal and spatial samplings and training data volume. 

\begin{table}[htbp]
    \newcolumntype{E}{>{\centering\arraybackslash}m{2.5cm}} 
    \centering
    \caption{\textbf{Summary of experiments}. We investigate the spatial subsampling of mesh vertices and the temporal subsampling of trajectories on both datasets (DS$_1$, DS$_2$). Data-efficiency is evaluated on DS$_2$ by training with \SIrange{100}{50}{\percent} of the available data.}
    \small
    \vspace{.5em}
    \begin{tabularx}{.95\linewidth}{XlEccccccE}
    \toprule
    \textbf{No} & \textbf{Experimental variable} & \textbf{Unit} & \multicolumn{6}{c}{\bfseries Variable range} & \textbf{Dataset} \\
    \midrule
    $1$ & Spatial subsampling factor $s_{s}$ & - & $1$ & $2$ & $3$ & $4$ & $5$ & - & DS$_1$, DS$_2$ \\
    $2$ & Temporal subsampling factor $s_{t}$ & - & $1$ & $2$ & $3$ & $4$ & $5$ & $6$ & DS$_1$, DS$_2$ \\
    $3$ & Amount of training data $s_{d}$ & \SI{}{\percent} & $100$ & $90$ & $80$ & $70$ & $60$ & $50$ & DS$_2$ \\
    \bottomrule
    \end{tabularx}
    \label{tab:exp}
\end{table}

\begin{comment}
\begin{table}[htbp] 
    \centering
    \caption{\textbf{Summary of experiments}. We investigate the spatial subsampling of mesh vertices and the temporal subsampling of trajectories on both datasets (DS$_1$, DS$_2$). Data-efficiency is evaluated on DS$_2$ by training with 100–50\% of the available data.}
    \small
    \begin{tabular}{l|c|c|cccccc|c|c}
    \toprule
    % the optional [-0.5ex] shifts the stack up by 0.5ex
    \textbf{No} &\textbf{Experiment variable} & \textbf{Symbol} & \multicolumn{6}{c|}{\bfseries Variable range} & \textbf{Unit} & \textbf{Dataset} \\
    \midrule 
    1 & Spatial Subsampling Factor & $s_{s}$ & 1 & 2 & 3 & 4 & 5 & - & - & DS$_1$, DS$_2$ \\
    2 & Temporal Subsampling Factor & $s_{t}$ & 1 & 2 & 3 & 4 & 5 & 6 & - & DS$_1$, DS$_2$\\
    3 & Amount of Training Data & $s_{d}$ & 100 & 90 & 80 & 70 & 60 & 50 & $\%$ & DS$_2$\\

    \bottomrule
    \end{tabular}
    \label{tab:exp}
\end{table}
\end{comment}

%% file: Plots/Methodology/drawing5_a.pdf_tex
%% Creator: Inkscape 1.3.2 (091e20e, 2023-11-25, custom), www.inkscape.org
%% Accompanies image file 'Plots/Methodology/drawing5_a.pdf' (pdf, eps, ps)
\begingroup%
  \makeatletter%
  \providecommand\color[2][]{%
    \errmessage{(Inkscape) Color is used for the text in Inkscape, but the package 'color.sty' is not loaded}%
    \renewcommand\color[2][]{}%
  }%
  \providecommand\transparent[1]{%
    \errmessage{(Inkscape) Transparency is used (non-zero) for the text in Inkscape, but the package 'transparent.sty' is not loaded}%
    \renewcommand\transparent[1]{}%
  }%
  \providecommand\rotatebox[2]{#2}%
  \newcommand*\fsize{\dimexpr\f@size pt\relax}%
  \newcommand*\lineheight[1]{\fontsize{\fsize}{#1\fsize}\selectfont}%
  \ifx\svgwidth\undefined%
    \setlength{\unitlength}{264.22581951bp}%
    \ifx\svgscale\undefined%
      \relax%
    \else%
      \setlength{\unitlength}{\unitlength * \real{\svgscale}}%
    \fi%
  \else%
    \setlength{\unitlength}{\svgwidth}%
  \fi%
  \global\let\svgwidth\undefined%
  \global\let\svgscale\undefined%
  \makeatother%
  \begin{picture}(1,0.86305706)%
    \lineheight{1}%
    \setlength\tabcolsep{0pt}%
    \put(0,0){\includegraphics[width=\unitlength,page=1]{Plots/Methodology/drawing5_a.pdf}}%
    \put(0.32121999,0.23426067){\color[rgb]{1,1,1}\transparent{0.95846599}\makebox(0,0)[lt]{\lineheight{1.25}\smash{\begin{tabular}[t]{l}Mold Cavity\end{tabular}}}}%
    \put(0.40981909,0.73766248){\color[rgb]{1,1,1}\transparent{0.95846599}\makebox(0,0)[lt]{\lineheight{1.25}\smash{\begin{tabular}[t]{l}Sprue\end{tabular}}}}%
    \put(0.78293189,0.72627899){\color[rgb]{1,1,1}\transparent{0.95846599}\makebox(0,0)[lt]{\lineheight{1.25}\smash{\begin{tabular}[t]{l}Riser\end{tabular}}}}%
    \put(0.04700238,0.72627899){\color[rgb]{1,1,1}\transparent{0.95846599}\makebox(0,0)[lt]{\lineheight{1.25}\smash{\begin{tabular}[t]{l}Riser\end{tabular}}}}%
  \end{picture}%
\endgroup%

%% file: Plots/Methodology/drawing5_b.pdf_tex
%% Creator: Inkscape 1.3.2 (091e20e, 2023-11-25, custom), www.inkscape.org
%% Accompanies image file 'Plots/Methodology/drawing5_b.pdf' (pdf, eps, ps)
\begingroup%
  \makeatletter%
  \providecommand\color[2][]{%
    \errmessage{(Inkscape) Color is used for the text in Inkscape, but the package 'color.sty' is not loaded}%
    \renewcommand\color[2][]{}%
  }%
  \providecommand\transparent[1]{%
    \errmessage{(Inkscape) Transparency is used (non-zero) for the text in Inkscape, but the package 'transparent.sty' is not loaded}%
    \renewcommand\transparent[1]{}%
  }%
  \providecommand\rotatebox[2]{#2}%
  \newcommand*\fsize{\dimexpr\f@size pt\relax}%
  \newcommand*\lineheight[1]{\fontsize{\fsize}{#1\fsize}\selectfont}%
  \ifx\svgwidth\undefined%
    \setlength{\unitlength}{294.50455072bp}%
    \ifx\svgscale\undefined%
      \relax%
    \else%
      \setlength{\unitlength}{\unitlength * \real{\svgscale}}%
    \fi%
  \else%
    \setlength{\unitlength}{\svgwidth}%
  \fi%
  \global\let\svgwidth\undefined%
  \global\let\svgscale\undefined%
  \makeatother%
  \begin{picture}(1,0.80392954)%
    \lineheight{1}%
    \setlength\tabcolsep{0pt}%
    \put(0,0){\includegraphics[width=\unitlength,page=1]{Plots/Methodology/drawing5_b.pdf}}%
    \put(0.896891,0.53518342){\color[rgb]{0,0,0}\makebox(0,0)[lt]{\lineheight{1.25}\smash{\begin{tabular}[t]{l}E\end{tabular}}}}%
    \put(0.14815301,0.53518342){\color[rgb]{0,0,0}\makebox(0,0)[lt]{\lineheight{1.25}\smash{\begin{tabular}[t]{l}E\end{tabular}}}}%
    \put(0.71940961,0.7455845991){\color[rgb]{0,0,0}\rotatebox{-39.321168}{\makebox(0,0)[lt]{\lineheight{1.25}\smash{\begin{tabular}[t]{l}A\end{tabular}}}}}%
    \put(0,0){\includegraphics[width=\unitlength,page=2]{Plots/Methodology/drawing5_b.pdf}}%
    \put(0.7093739,0.5258714){\color[rgb]{0,0,0}\makebox(0,0)[lt]{\lineheight{1.25}\smash{\begin{tabular}[t]{l}C\end{tabular}}}}%
    \put(0.52068678,0.4178576){\color[rgb]{0,0,0}\makebox(0,0)[lt]{\lineheight{1.25}\smash{\begin{tabular}[t]{l}B\end{tabular}}}}%
    \put(0.5252969761,0.6150009931){\color[rgb]{0,0,0}\rotatebox{51.872077}{\makebox(0,0)[lt]{\lineheight{1.25}\smash{\begin{tabular}[t]{l}D\end{tabular}}}}}%
    \put(0.52019294,0.814359126){\color[rgb]{0,0,0}\makebox(0,0)[lt]{\lineheight{1.25}\smash{\begin{tabular}[t]{l}G\end{tabular}}}}%
    \put(0,0){\includegraphics[width=\unitlength,page=3]{Plots/Methodology/drawing5_b.pdf}}%
    \put(0.02527952,0.22679974){\color[rgb]{0,0,0}\makebox(0,0)[lt]{\lineheight{1.25}\smash{\begin{tabular}[t]{l}F\end{tabular}}}}%
    \put(0,0){\includegraphics[width=\unitlength,page=4]{Plots/Methodology/drawing5_b.pdf}}%
    \put(0.52953043,0.467235051){\color[rgb]{0.89803922,0.04313725,0.04313725}\makebox(0,0)[lt]{\lineheight{1.25}\smash{\begin{tabular}[t]{l}\textbf{x}\end{tabular}}}}%
  \end{picture}%
\endgroup%

%% file: Plots/Methodology/drawing5_c.pdf_tex
%% Creator: Inkscape 1.3.2 (091e20e, 2023-11-25, custom), www.inkscape.org
%% Accompanies image file 'Plots/Methodology/drawing5_c.pdf' (pdf, eps, ps)
\begingroup%
  \makeatletter%
  \providecommand\color[2][]{%
    \errmessage{(Inkscape) Color is used for the text in Inkscape, but the package 'color.sty' is not loaded}%
    \renewcommand\color[2][]{}%
  }%
  \providecommand\transparent[1]{%
    \errmessage{(Inkscape) Transparency is used (non-zero) for the text in Inkscape, but the package 'transparent.sty' is not loaded}%
    \renewcommand\transparent[1]{}%
  }%
  \providecommand\rotatebox[2]{#2}%
  \newcommand*\fsize{\dimexpr\f@size pt\relax}%
  \newcommand*\lineheight[1]{\fontsize{\fsize}{#1\fsize}\selectfont}%
  \ifx\svgwidth\undefined%
    \setlength{\unitlength}{258.81347416bp}%
    \ifx\svgscale\undefined%
      \relax%
    \else%
      \setlength{\unitlength}{\unitlength * \real{\svgscale}}%
    \fi%
  \else%
    \setlength{\unitlength}{\svgwidth}%
  \fi%
  \global\let\svgwidth\undefined%
  \global\let\svgscale\undefined%
  \makeatother%
  \begin{picture}(1,0.93782927)%
    \lineheight{1}%
    \setlength\tabcolsep{0pt}%
    \put(0,0){\includegraphics[width=\unitlength,page=1]{Plots/Methodology/drawing5_c.pdf}}%
    \put(0.2624857,0.89070922){\color[rgb]{0,0,0}\makebox(0,0)[lt]{\lineheight{1.25}\smash{\begin{tabular}[t]{l}Inlet\end{tabular}}}}%
    \put(0.77720529,0.89086013){\color[rgb]{0,0,0}\makebox(0,0)[lt]{\lineheight{1.25}\smash{\begin{tabular}[t]{l}Outlet\end{tabular}}}}%
    \put(0.02112401,0.89086013){\color[rgb]{0,0,0}\makebox(0,0)[lt]{\lineheight{1.25}\smash{\begin{tabular}[t]{l}Outlet\end{tabular}}}}%
    \put(0,0){\includegraphics[width=\unitlength,page=2]{Plots/Methodology/drawing5_c.pdf}}%
  \end{picture}%
\endgroup%

%% file: Plots/Methodology/flowchart.pdf_tex
%% Creator: Inkscape 1.3.2 (091e20e, 2023-11-25, custom), www.inkscape.org
%% Accompanies image file 'Plots/Methodology/flowchart.pdf' (pdf, eps, ps)
%%   \input{<filename>.pdf}
%%   \input{<filename>.pdf}
%%   \import{<path to file>}{<filename>.pdf}
\begingroup%
  \makeatletter%
  \providecommand\color[2][]{%
    \errmessage{(Inkscape) Color is used for the text in Inkscape, but the package 'color.sty' is not loaded}%
    \renewcommand\color[2][]{}%
  }%
  \providecommand\transparent[1]{%
    \errmessage{(Inkscape) Transparency is used (non-zero) for the text in Inkscape, but the package 'transparent.sty' is not loaded}%
    \renewcommand\transparent[1]{}%
  }%
  \providecommand\rotatebox[2]{#2}%
  \newcommand*\fsize{\dimexpr\f@size pt\relax}%
  \newcommand*\lineheight[1]{\fontsize{\fsize}{#1\fsize}\selectfont}%
  \ifx\svgwidth\undefined%
    \setlength{\unitlength}{675.90658834bp}%
    \ifx\svgscale\undefined%
      \relax%
    \else%
      \setlength{\unitlength}{\unitlength * \real{\svgscale}}%
    \fi%
  \else%
    \setlength{\unitlength}{\svgwidth}%
  \fi%
  \global\let\svgwidth\undefined%
  \global\let\svgscale\undefined%
  \makeatother%
  \begin{picture}(1,0.44297665)%
    \lineheight{1}%
    \setlength\tabcolsep{0pt}%
    \put(0,0){\includegraphics[width=\unitlength,page=1]{Plots/Methodology/flowchart.pdf}}%
    \put(0.24188798,0.32038758){\color[rgb]{1,1,1}\makebox(0,0)[t]{\lineheight{1.25}\smash{\begin{tabular}[t]{c}GNO \\Encoder\end{tabular}}}}%
    \put(0,0){\includegraphics[width=\unitlength,page=2]{Plots/Methodology/flowchart.pdf}}%
    \put(0.124685,0.40615218){\color[rgb]{0,0,0}\makebox(0,0)[rt]{\lineheight{1.25}\smash{\begin{tabular}[t]{r}t\end{tabular}}}}%
    \put(0,0){\includegraphics[width=\unitlength,page=3]{Plots/Methodology/flowchart.pdf}}%
    \put(0.39918696,0.33451983){\color[rgb]{1,1,1}\makebox(0,0)[t]{\lineheight{1.25}\smash{\begin{tabular}[t]{c}Broadcast \\3D latent grid\end{tabular}}}}%
    \put(0,0){\includegraphics[width=\unitlength,page=4]{Plots/Methodology/flowchart.pdf}}%
    \put(0.1229081,0.22570397){\color[rgb]{0,0,0}\makebox(0,0)[t]{\lineheight{1.25}\smash{\begin{tabular}[t]{c}I\end{tabular}}}}%
    \put(0.05373004,0.23393049){\color[rgb]{0,0,0}\makebox(0,0)[t]{\lineheight{1.25}\smash{\begin{tabular}[t]{c}Inlet\\setup\end{tabular}}}}%
    \put(0.05380084,0.35313921){\color[rgb]{0,0,0}\makebox(0,0)[t]{\lineheight{1.25}\smash{\begin{tabular}[t]{c}2D \\vertices\end{tabular}}}}%
    \put(0,0){\includegraphics[width=\unitlength,page=5]{Plots/Methodology/flowchart.pdf}}%
    \put(0.12321677,0.28615886){\color[rgb]{0,0,0}\makebox(0,0)[t]{\lineheight{1.25}\smash{\begin{tabular}[t]{c}$f_{man}$\end{tabular}}}}%
    \put(0.05427658,0.29267494){\color[rgb]{0,0,0}\makebox(0,0)[t]{\lineheight{1.25}\smash{\begin{tabular}[t]{c}IC/BC and \\inlet mask\end{tabular}}}}%
    \put(0.05380084,0.41247179){\color[rgb]{0,0,0}\makebox(0,0)[t]{\lineheight{1.25}\smash{\begin{tabular}[t]{c}Time\\vertices\end{tabular}}}}%
    \put(0.25729331,0.15738244){\color[rgb]{0,0,0}\makebox(0,0)[t]{\lineheight{1.25}\smash{\begin{tabular}[t]{c}Encode local \\features\end{tabular}}}}%
    \put(0.41530816,0.15738244){\color[rgb]{0,0,0}\makebox(0,0)[t]{\lineheight{1.25}\smash{\begin{tabular}[t]{c}Broadcast local 2D \\features to 3D\end{tabular}}}}%
    \put(0.58772666,0.15738244){\color[rgb]{0,0,0}\makebox(0,0)[t]{\lineheight{1.25}\smash{\begin{tabular}[t]{c}Learn global\\interactions\end{tabular}}}}%
    \put(0.74633627,0.15849253){\color[rgb]{0,0,0}\makebox(0,0)[t]{\lineheight{1.25}\smash{\begin{tabular}[t]{c}3D learned\\latent grid\end{tabular}}}}%
    \put(0,0){\includegraphics[width=\unitlength,page=6]{Plots/Methodology/flowchart.pdf}}%
    \put(0.57916309,0.225602){\color[rgb]{0,0,0}\makebox(0,0)[t]{\lineheight{1.25}\smash{\begin{tabular}[t]{c}AdaIN\end{tabular}}}}%
    \put(0,0){\includegraphics[width=\unitlength,page=7]{Plots/Methodology/flowchart.pdf}}%
    \put(0.57951628,0.33543799){\color[rgb]{1,1,1}\makebox(0,0)[t]{\lineheight{1.25}\smash{\begin{tabular}[t]{c}Fourier\\Neural Operator\end{tabular}}}}%
    \put(0,0){\includegraphics[width=\unitlength,page=8]{Plots/Methodology/flowchart.pdf}}%
    \put(0.89190892,0.34031078){\color[rgb]{0,0,0}\makebox(0,0)[t]{\lineheight{1.25}\smash{\begin{tabular}[t]{c}$\hat{v}$\end{tabular}}}}%
    \put(0,0){\includegraphics[width=\unitlength,page=9]{Plots/Methodology/flowchart.pdf}}%
    \put(0.89184589,0.40224868){\color[rgb]{0,0,0}\makebox(0,0)[t]{\lineheight{1.25}\smash{\begin{tabular}[t]{c}$\hat{u}$\end{tabular}}}}%
    \put(0,0){\includegraphics[width=\unitlength,page=10]{Plots/Methodology/flowchart.pdf}}%
    \put(0.89169083,0.27909892){\color[rgb]{0,0,0}\makebox(0,0)[t]{\lineheight{1.25}\smash{\begin{tabular}[t]{c}$\hat{p}$\end{tabular}}}}%
    \put(0,0){\includegraphics[width=\unitlength,page=11]{Plots/Methodology/flowchart.pdf}}%
    \put(0.89221195,0.21840686){\color[rgb]{0,0,0}\makebox(0,0)[t]{\lineheight{1.25}\smash{\begin{tabular}[t]{c}$\hat{\alpha}$\end{tabular}}}}%
    \put(0.95196116,0.40816723){\color[rgb]{0,0,0}\makebox(0,0)[t]{\lineheight{1.25}\smash{\begin{tabular}[t]{c}Velocity\\in x\end{tabular}}}}%
    \put(0.95213055,0.34748384){\color[rgb]{0,0,0}\makebox(0,0)[t]{\lineheight{1.25}\smash{\begin{tabular}[t]{c}Velocity\\in y\end{tabular}}}}%
    \put(0.95156676,0.27705433){\color[rgb]{0,0,0}\makebox(0,0)[t]{\lineheight{1.25}\smash{\begin{tabular}[t]{c}Pressure\end{tabular}}}}%
    \put(0.95182175,0.22457746){\color[rgb]{0,0,0}\makebox(0,0)[t]{\lineheight{1.25}\smash{\begin{tabular}[t]{c}Volume\\fraction\end{tabular}}}}%
    \put(0,0){\includegraphics[width=\unitlength,page=12]{Plots/Methodology/flowchart.pdf}}%
    \put(0.90222855,0.15846136){\color[rgb]{0,0,0}\makebox(0,0)[t]{\lineheight{1.25}\smash{\begin{tabular}[t]{c}Output prediction \\on 2D mesh\end{tabular}}}}%
    \put(0,0){\includegraphics[width=\unitlength,page=13]{Plots/Methodology/flowchart.pdf}}%
    \put(0.73484146,0.31888943){\color[rgb]{1,1,1}\makebox(0,0)[t]{\lineheight{1.25}\smash{\begin{tabular}[t]{c}GNO \\Decoder\end{tabular}}}}%
    \put(0,0){\includegraphics[width=\unitlength,page=14]{Plots/Methodology/flowchart.pdf}}%
    \put(0.09697521,0.15140156){\color[rgb]{0,0,0}\makebox(0,0)[t]{\lineheight{1.25}\smash{\begin{tabular}[t]{c}Inputs\end{tabular}}}}%
    \put(0,0){\includegraphics[width=\unitlength,page=15]{Plots/Methodology/flowchart.pdf}}%
    \put(0.12291093,0.34812119){\color[rgb]{0,0,0}\makebox(0,0)[t]{\lineheight{1.25}\smash{\begin{tabular}[t]{c}x,y\end{tabular}}}}%
    \put(0,0){\includegraphics[width=\unitlength,page=16]{Plots/Methodology/flowchart.pdf}}%
  \end{picture}%
\endgroup%

%% file: 5_Results_and_Discussion.tex
\section{Results and Discussion}
\label{sec:res}
In this section, we examine the performance and  robustness of the proposed Fourier-Graph neural solver across different scenarios, see Table \ref{tab:exp} and Section \ref{Quantitative_results}. Unless stated otherwise, all models were trained with the loss and training procedures highlighted in Section \ref{objective_optmization}, and evaluated with the metrics outlined in Section \ref{sec:metrics}. For clarity of presentation, the predicted fields are resampled and visualized on a uniform $1000\times1000$ grid.

\subsection{General Exploration}
\label{Quantitative_results}
\begin{figure}[bth] 
  \centering
  % Row 1: Prediction
  \begin{subfigure}[b]{1\linewidth}
  \centering
    % left label (vertically centered)
    \begin{minipage}[c]{0.05\linewidth}
      \centering
      \rotatebox{90}{Prediction $\mathbf{\hat{u}}$}
    \end{minipage}%
    % four panels
    \begin{minipage}[c]{0.95\linewidth}
      \noindent
      \begin{subfigure}[c]{\field\linewidth}
        \includegraphics[width=\linewidth,trim={0 50 0 50},clip]{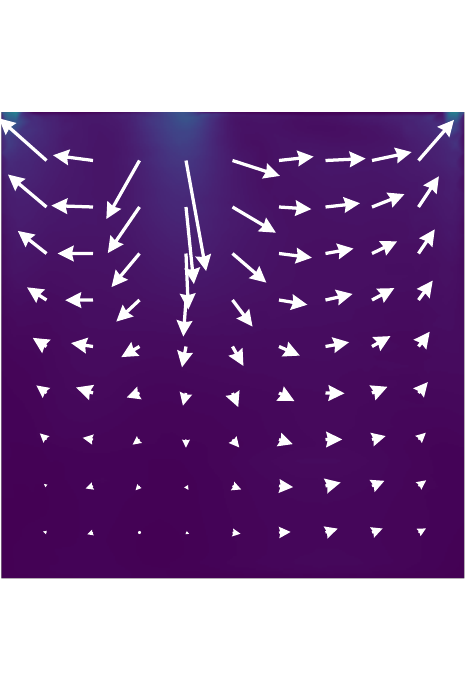}
      \end{subfigure}
      \begin{subfigure}[c]{\field\linewidth}
        \includegraphics[width=\linewidth,trim={0 50 0 50},clip]{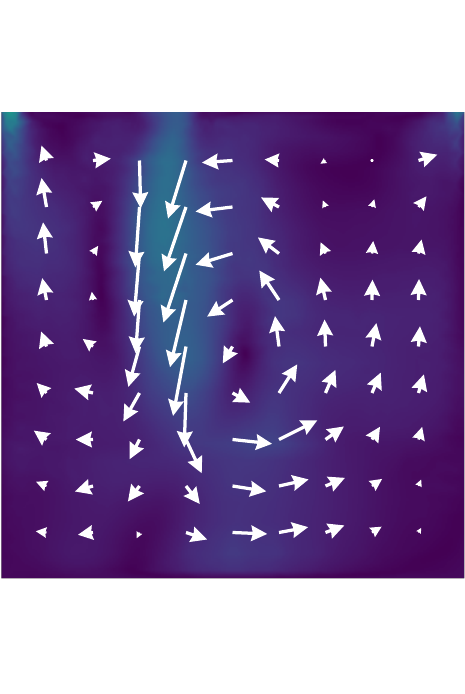}
      \end{subfigure}
      \begin{subfigure}[c]{\field\linewidth}
        \includegraphics[width=\linewidth,trim={0 50 0 50},clip]{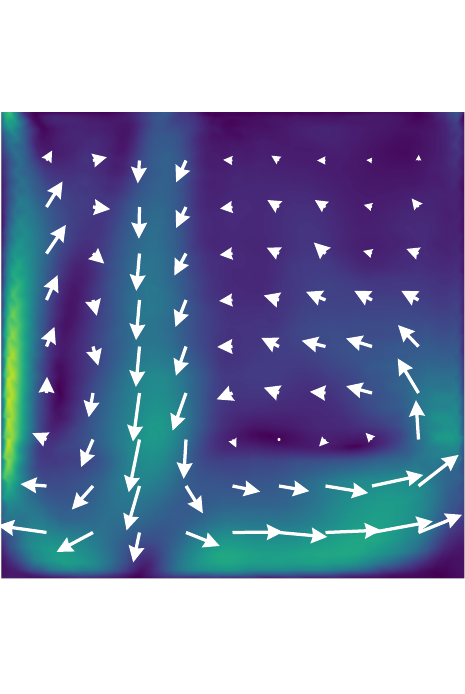}
      \end{subfigure}
      \begin{subfigure}[c]{\field\linewidth}
        \includegraphics[width=\linewidth,trim={0 50 0 50},clip]{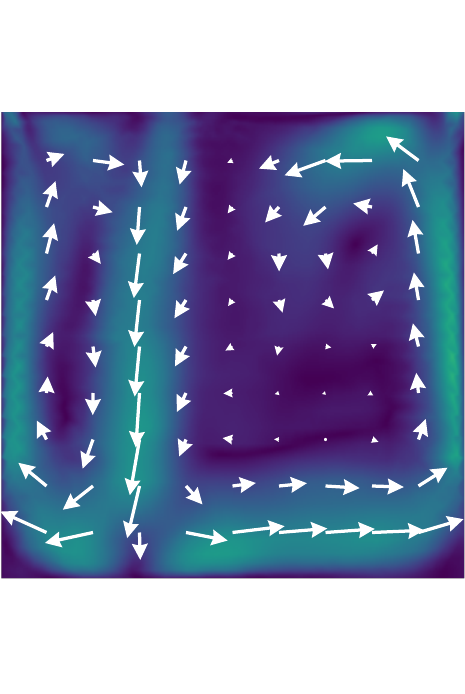}
      \end{subfigure}
      \begin{subfigure}[c]{\field\linewidth}
        \includegraphics[width=\linewidth,trim={0 50 0 50},clip]{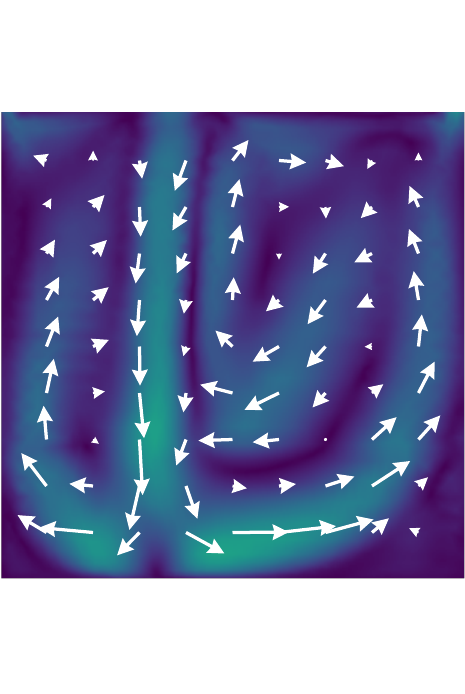}
      \end{subfigure}
      \begin{subfigure}[c]{\field\linewidth}
        \includegraphics[width=\linewidth,trim={0 50 0 50},clip]{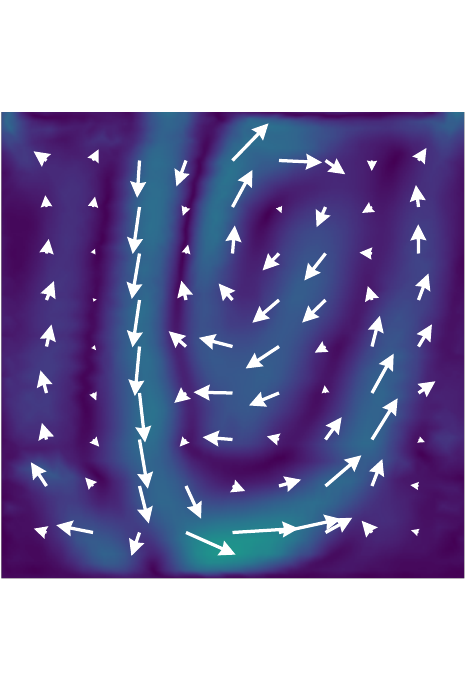}
      \end{subfigure}
    \end{minipage}
  \end{subfigure}\par\medskip

  % Row 2: Target
  \begin{subfigure}[b]{1\linewidth}
    \begin{minipage}[c]{0.05\linewidth}
      \centering
      \rotatebox{90}{Target $\mathbf{u}$}
    \end{minipage}%
    \begin{minipage}[c]{0.95\linewidth}
      \noindent
      \begin{subfigure}[c]{\field\linewidth}
        \includegraphics[width=\linewidth,trim={0 50 0 50},clip]{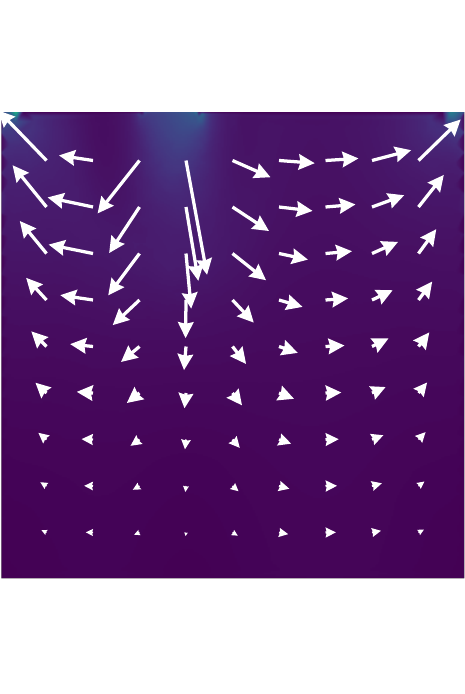}
      \end{subfigure}
      \begin{subfigure}[c]{\field\linewidth}
        \includegraphics[width=\linewidth,trim={0 50 0 50},clip]{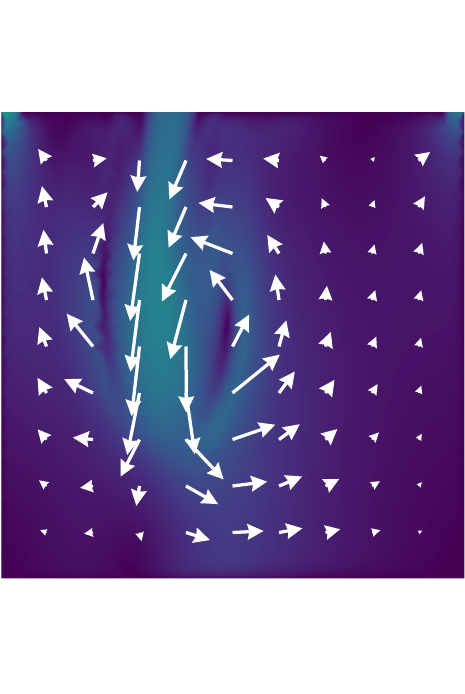}
      \end{subfigure}
      \begin{subfigure}[c]{\field\linewidth}
        \includegraphics[width=\linewidth,trim={0 50 0 50},clip]{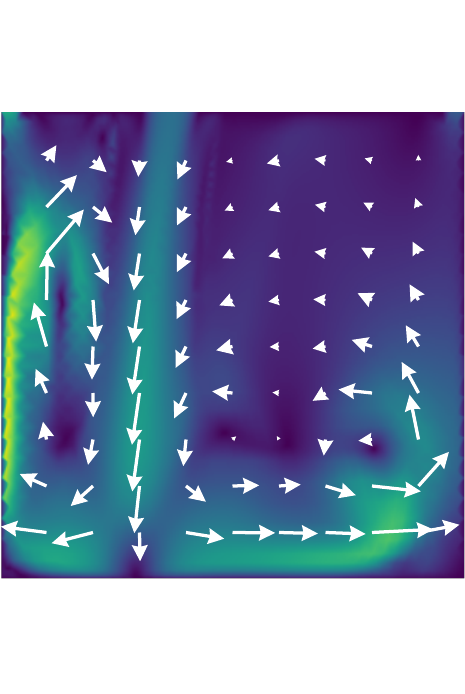}
      \end{subfigure}
      \begin{subfigure}[c]{\field\linewidth}
        \includegraphics[width=\linewidth,trim={0 50 0 50},clip]{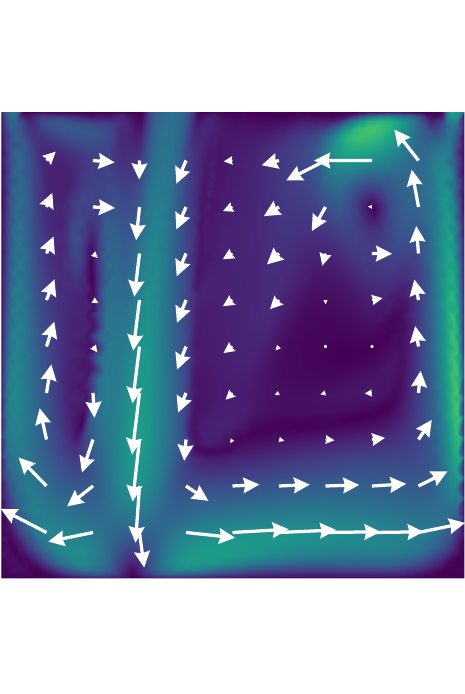}
      \end{subfigure}
      \begin{subfigure}[c]{\field\linewidth}
        \includegraphics[width=\linewidth,trim={0 50 0 50},clip]{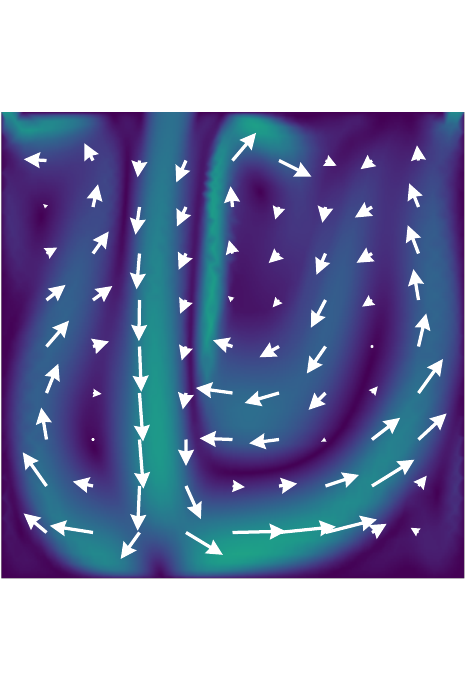}
      \end{subfigure}
      \begin{subfigure}[c]{\field\linewidth}
        \includegraphics[width=\linewidth,trim={0 50 0 50},clip]{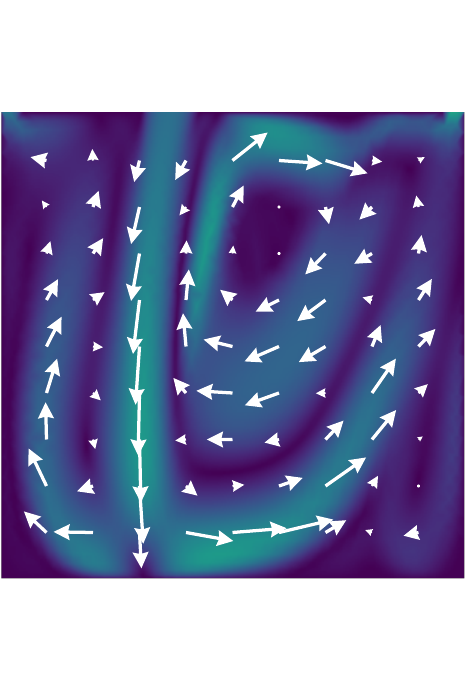}
      \end{subfigure}
    \begin{subfigure}[c]{0.01\linewidth}
    % Don't let this object contribute to the row height/depth
    \hspace*{0.0cm}
    \raisebox{-0.50\height}[0pt][0pt]{%
    \begin{adjustbox}{height=235pt, trim=78pt 0pt 0pt 20pt,clip}
     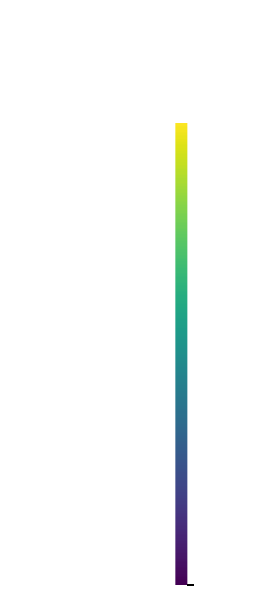%
    \end{adjustbox}}%
    \end{subfigure}
    \end{minipage}
  \end{subfigure}\par\medskip

% Row 3: Error (with properly centered row‐label and aligned captions)
  \begin{subfigure}[b]{1\linewidth}
  \centering
    % vertical row label, centered alongside the 4 panels
    \begin{minipage}[c]{0.05\linewidth}
      \centering
       \vspace*{-0.8cm}% <- adjust this length to taste
      \rotatebox{90}{$\lVert \mathbf{\hat{u}} - \mathbf{u} \rVert_2$}
    \end{minipage}%
    % the four error panels, each with a bottom caption
    \begin{minipage}[c]{0.95\linewidth}
    \noindent
    \begin{subfigure}[c]{\field\linewidth}
          \centering
            \begingroup
        \def\svgwidth{\linewidth}
    \adjustbox{trim=0 18 0 15,clip,width=\linewidth}{%
    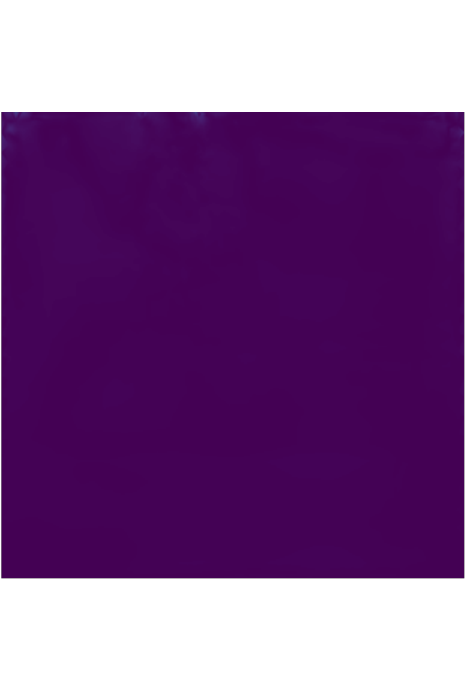}
      \endgroup
    \caption*{$r_{\mathbf{u}}^{(0)} = \SI{18.24}{\percent}$}
    \end{subfigure}
        \begin{subfigure}[c]{\field\linewidth}
          \centering
            \begingroup
        \def\svgwidth{\linewidth}
    \adjustbox{trim=0 18 0 15,clip,width=\linewidth}{%
    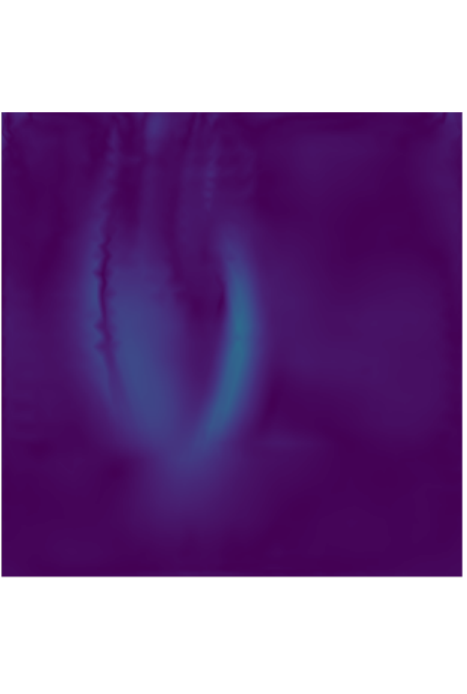}
      \endgroup
    \caption*{$r_{\mathbf{u}}^{(1)} = \SI{44.73}{\percent}$}
    \end{subfigure}
            \begin{subfigure}[c]{\field\linewidth}
          \centering
            \begingroup
        \def\svgwidth{\linewidth}
    \adjustbox{trim=0 18 0 15,clip,width=\linewidth}{%
    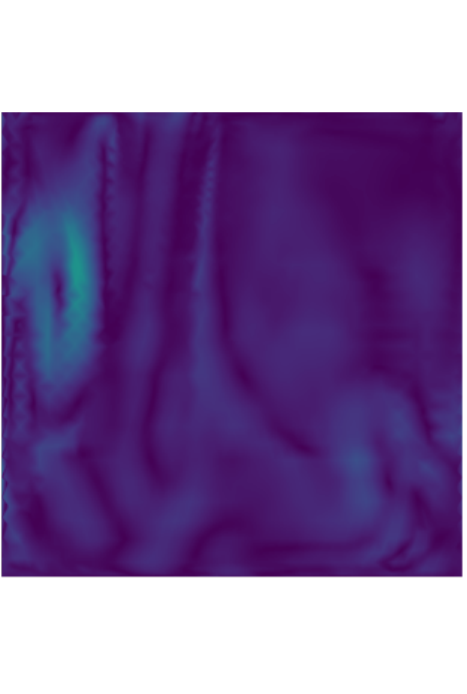}
      \endgroup
    \caption*{$r_{\mathbf{u}}^{(2)} = \SI{34.16}{\percent}$}
    \end{subfigure}
    \begin{subfigure}[c]{\field\linewidth}
        \centering
        \begingroup
        \def\svgwidth{\linewidth}
        \adjustbox{trim=0 18 0 15,clip,width=\linewidth}{%
  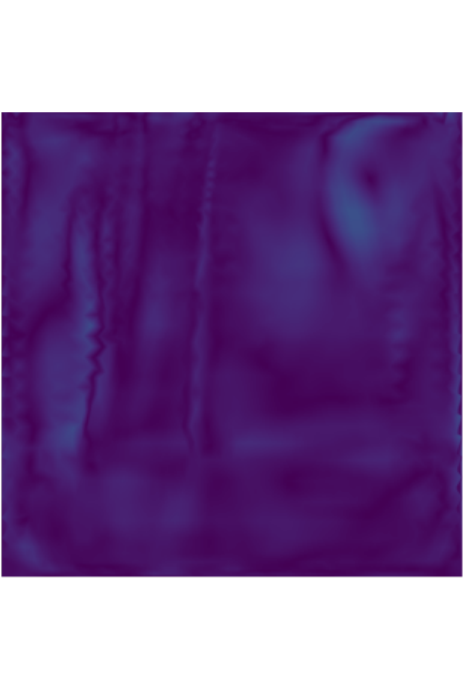}
        \endgroup
        \caption*{$r_{\mathbf{u}}^{(3)} = \SI{24.43}{\percent}$}
      \end{subfigure}
      \begin{subfigure}[c]{\field\linewidth}
        \centering
        \begingroup
        \def\svgwidth{\linewidth}
        \adjustbox{trim=0 18 0 12,clip,width=\linewidth}{%
  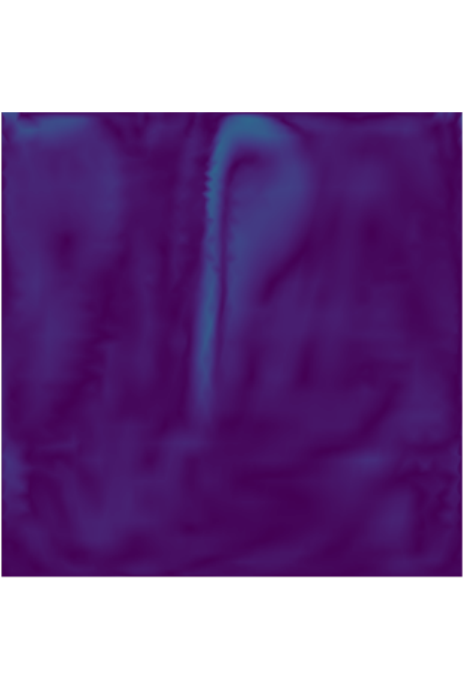}
        \endgroup
        \caption*{$r_{\mathbf{u}}^{(6)} = \SI{28.25}{\percent}$}
      \end{subfigure}
      \begin{subfigure}[c]{\field\linewidth}
        \centering
        \def\svgwidth{\linewidth}
        \begingroup
        \adjustbox{trim=0 18 0 12,clip,width=\linewidth}{%
        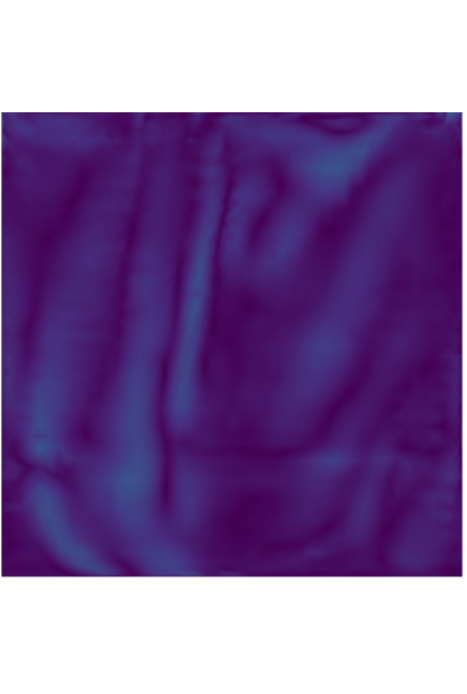}  
        \endgroup
        \caption*{$r_{\mathbf{u}}^{(9)} = \SI{38.36}{\percent}$}
      \end{subfigure}    
    \end{minipage}
  \end{subfigure}
  % Check once again the formulas.
  \caption{\textbf{Velocity field prediction}. Velocity field prediction $\gls{u_vec_hat}$ (row $1$), target \gls{CFD} velocity field \gls{u_vec} (row $2$), pointwise Euclidean error $\lVert\gls{u_vec_hat}-\gls{u_vec}\rVert_2$ (row $3$), and relative $L_2$ error $\gls{rerror}_{\mathbf{u}}^{(k)}$(row $4$) at simulation steps $\gls{steps}\in\{0,1,2,3,6,9\}$ with $\gls{steps} \times\SI{0.06}{\second}$ of a mold filling simulation. Quiver overlays indicate flow direction (array orientation) and velocity magnitude (arrow length).}
  \label{fig:velocity_fields}
\end{figure}

We first evaluate the ability of the neural solver to predict transient flow fields in a representative mold filling scenario. Figure \ref{fig:velocity_fields} presents the predicted velocity field $\gls{u_vec_hat}= (\gls{u_hat},\gls{v_hat})$ for a selected DS$_2$ sample, alongside the target \gls{CFD} velocity field \gls{u_vec}, and the associated pointwise Euclidean error $\lVert\gls{u_vec_hat}-\gls{u_vec}\rVert_2$. The velocity fields were produced with a model trained with temporal subsampling factor $\gls{st}=6$ that forecasts \SI{0.6}{\second} into the future. To assess the temporal consistency of the model, we display snapshots at prediction steps $\gls{steps}\in\{0,1,2,3,6,9\}$ with each snapshot corresponding to a prediction time $\gls{time}=\gls{steps}\times\SI{0.06}{\second}$. Qualitatively, the figure shows that the proposed Fourier-Graph neural solver reproduces the target flow patterns with fidelity, capturing the motion and spreading of the fluid jet within the cavity and the overall advection dynamics. More generally, the model generalizes across different ingate locations and process conditions present in the dataset, indicating that it has learned an abstract representation of the flow physics of different mold geometries rather than overfitting to a single sample instance. This ability to handle varying geometries and inputs is a hallmark of operator learning approaches \citep{li2023fourier}. Combined with a graph-based encoder, this yields mesh-invariant predictions consistent with recent advances in Fourier operators on arbitrary domains \citep{li2023geometryinformedneuraloperatorlargescale}.

The largest discrepancies between prediction and simulation appear in regions of sharp gradients and complex small-scale structures, specially near the fluid-air interface, in shear layers, and in recirculation zones. In these localized areas, the neural solver tends to smooth out fine details, as seen by attenuated velocity peaks and blurred fluid-air interfaces in the error maps. We hypothesize two main contributing factors to that: First, the spectral truncation in the Fourier layers acts as a low-pass filter, limiting the highest Fourier modes, and therefore diminishing the model capacity to represent sharp, high-frequency features. This spectral bias of \gls{FNO}-based models has been widely discussed in the literature and is considered a fundamental trade-off for efficiency in Fourier-based models \citep{DBLP:journals/corr/abs-2010-08895, 10.1063/5.0254681}. Second, because we train by minimizing a mean squared error loss, the learned solution in regions with inherent ambiguity, such as turbulent eddies or diffuse interfaces, approximates an average of possible outcomes. Such averaging leads to over-smoothed predictions that underestimate extreme values or rapid fluctuations. These effects are clearly visible in the velocity field, where the predicted flow \gls{u_vec_hat} lacks some of the sharpest velocity spikes present in the target velocity field \gls{u_vec}, and very thin splashes or filamentary structures are slightly smeared out. We note however that this behavior is not unique to our model, and that even enhanced \gls{FNO} architectures require special measures, e.g. attention mechanisms or learned deformations, to resolve boundary layers and other steep features \citep{10.1063/5.0254681, li2023fourier}. Despite these localized errors, the global flow organization is preserved across all reported time steps, with the predicted velocity vectors (quiver arrows in Figure \ref{fig:velocity_fields}) maintaining the correct directionality and relative magnitude, therefore indicating that the proposed neural solver captures the bulk transport of momentum through the mold filling domain. 

The per step relative $L_2$ errors for the velocity field show the temporal accuracy trajectory of the model. In this particular sample, the $\gls{rerror}_{\mathbf{u}}^{k}$ error rise sharply from roughly \SI{18.24}{\percent} at step $\gls{steps}= 0$ to about \SI{45}{\percent} at $\gls{steps}=1$, then fall to about \SI{24}{\percent} at $\gls{steps}=3$, before increasing again toward about \SI{38}{\percent} by step $\gls{steps}=9$. This pattern aligns well with the velocity field predictions \gls{u_vec_hat} of the Figure \ref{fig:velocity_fields}. Namely, the early loss peak follows the initial jet formation and acceleration, when sharp gradients appear. Furthermore, the mid horizon predictions reflects a period when the bulk advection is well organized and easier to predict, while the later rise coincides with the increasing number of regions with interface and in recirculation zones, which increases local discrepancies. The non-monotonic trajectory of loss values confirms that the model corrects after the first transient yet gradually loses fine scale detail over a longer horizon due to the flow complexity, which is consistent with the smoothing effect previously mentioned.

\begin{figure}[tbh] 
  \centering
  \begin{subfigure}[b]{0.33\textwidth}
  \centering
  \small
  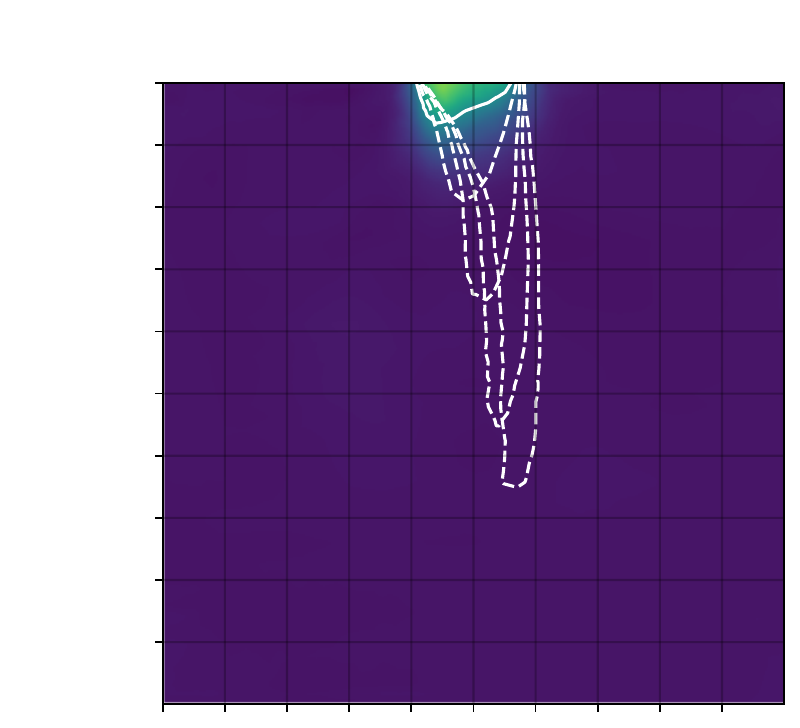
  \caption{Prediction}\label{fig:pred}
  \end{subfigure}
  \hfill
  \begin{subfigure}[b]{0.272\textwidth}
  \centering
  \small
  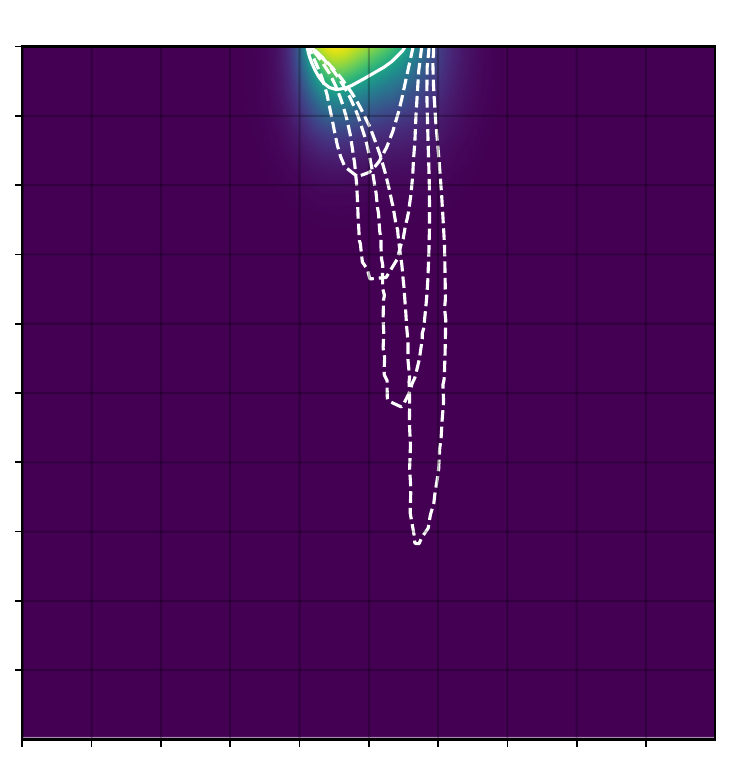
  \caption{Simulation Target}\label{fig:targ}
  \end{subfigure}
  \hfill
  \begin{subfigure}[b]{0.385\textwidth}
  \centering
  \small
  \def\svgwidth{175pt}
  \adjustbox{trim=5 22
  0 15, clip}{%
  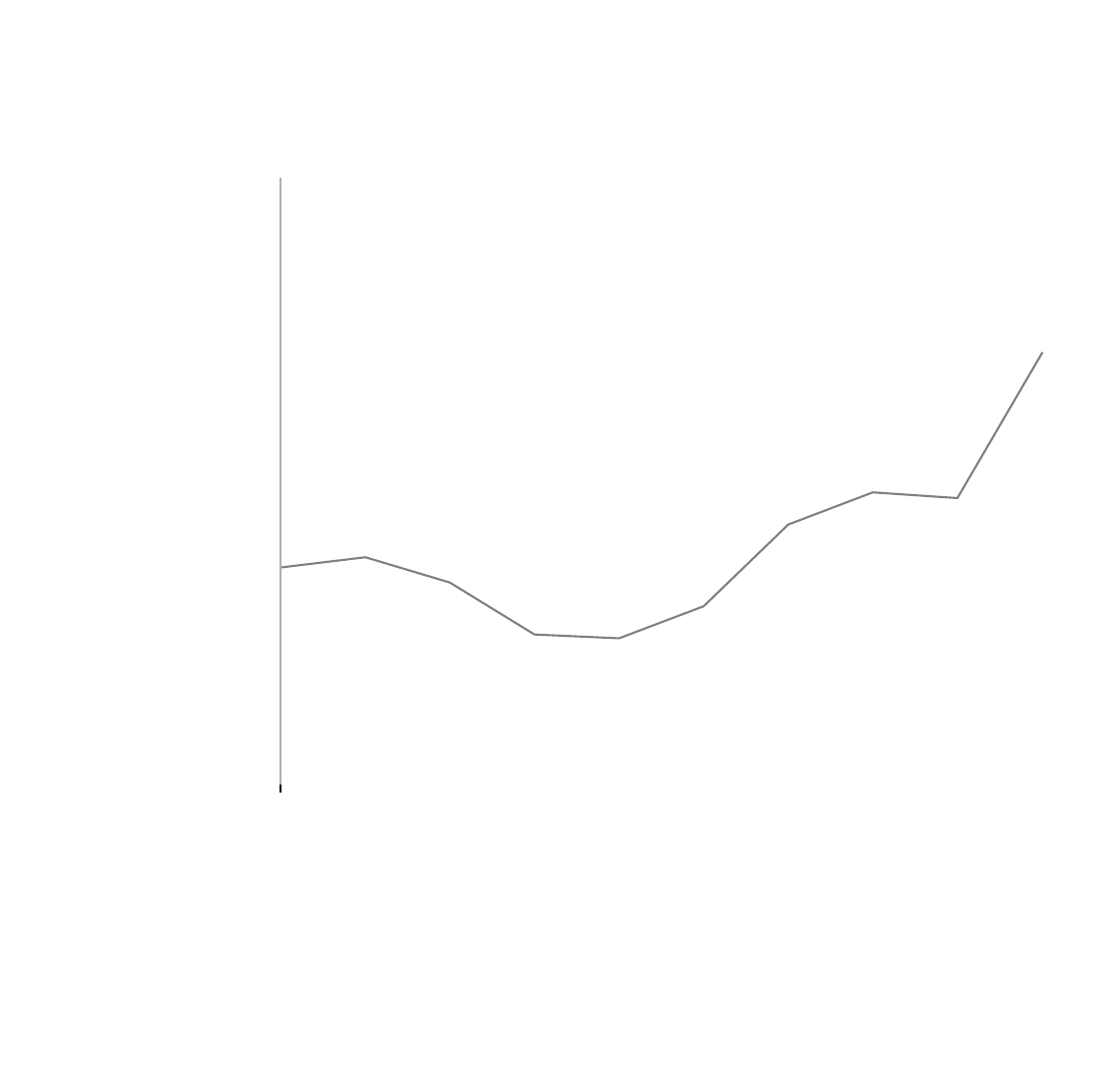}
  \caption{Relative $L_2$ $r_{\alpha}^{(k)}$ Error}\label{fig:RMSE}
  \end{subfigure}
\caption{\textbf{Fluid-air interface tracking with iso-contour $\gls{alpha}=0.5$ as a proxy}. Figs. \ref{fig:pred} and \ref{fig:targ} show the predicted \gls{alpha_hat} and simulation target volume fraction \gls{alpha} fields, respectively. The background colormap and the continuous fluid-air interface overlay curves corresponds to step $\gls{steps}=0$, while dashed curves denote the fluid-air interfaces at steps $\gls{steps}\in \{3,5,7,9\}$ with $\SI{0.1}{\second}\times\gls{steps}$ increments. Figure \ref{fig:RMSE} shows the relative $L_2$ error $\gls{rerror}_{\alpha}^{(k)}$ at each step \gls{steps}, with filled circle markers corresponding to the errors between fluid-air interfaces depicted in Figs. \ref{fig:pred} and \ref{fig:targ}.}
\label{fig:pred_vs_target_error}
\end{figure}

Predicting the fluid-air interface through the volume fraction field \gls{alpha} is critical for mold filling applications. With this in mind, Figure \ref{fig:pred_vs_target_error} illustrates the accuracy of the proposed Fourier-Graph model in predicting the fluid-air interface for a model with $H=$ \SI{0.1}{\second} total prediction horizon. Although our simulation data were generated with a diffuse interface method (Cahn-Hilliard phase-field), we extract an approximate interface location by plotting the $\gls{alpha}=0.5$ iso-contour of the volume fraction field. Figure \ref{fig:pred} shows the predicted volume fraction field \gls{alpha_hat} at several time steps, while Figure \ref{fig:targ} shows the corresponding target volume fraction field \gls{alpha}. In order to allow for direct comparison, both prediction and target plots present an overlay of the fluid-air interface. At early times, e.g. at $\gls{time}=\SI{0.01}{\second}$ ($\gls{steps}=1$), the predicted and true interfaces coincide almost exactly, indicating that the neural solver initially advects the fluid-air interface front at the correct speed. Over longer rollouts, a gradual divergence becomes evident. At $\gls{time}=\SI{0.09}{\second}$ ($\gls{steps}=9$), the interface front slight shifts between the predicted and actual interface, particularly in regions of high curvature. Nonetheless, the overall shape and extent of the filling front remain well captured. We quantify the interface prediction error in Figure \ref{fig:RMSE}, which plots the relative $L_2$ error $\gls{rerror}_{\alpha}^{(k)}$ of the volume fraction field \gls{alpha} as a function of the prediction step \gls{steps}. The error grows roughly monotonically with time horizon, reaching about \SI{18}{\percent} relative discrepancy at the final step. Notably, the increase is gradual and there is no explosive error accumulation, suggesting the rollout remains stable over the tested \SI{0.1}{\second} horizon. Together, the velocity field predictions and fluid-air interface results indicate that the Fourier-Graph network aligns well with the parabolic nature of momentum transport and the hyperbolic advection of the phase front. In contrast, the pressure field, governed by an elliptic constraint, poses a greater challenge, which we analyze next. 

\begin{figure}[thb] 
% Row 0: Legend
\begin{subfigure}[b]{1.0\linewidth}
  \centering
  \begin{adjustbox}{trim=0pt 63pt 70pt 85pt,clip}
    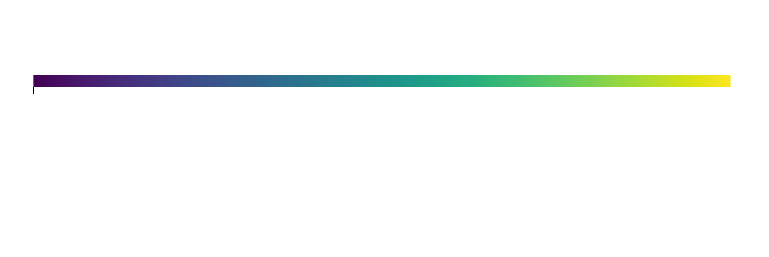
  \end{adjustbox}
\end{subfigure}\par\medskip

  \centering
  % Row 1: Prediction
  \begin{subfigure}[b]{1\linewidth}
    \centering
    % left label (vertically centered)
    \begin{minipage}[c]{0.05\linewidth}
      \centering
      \rotatebox{90}{Prediction $\hat{p}$}
    \end{minipage}%
    % four panels
    \begin{minipage}[c]{0.95\linewidth}
      \noindent
      \begin{subfigure}[c]{\field\linewidth}
        \includegraphics[width=\linewidth,trim={0 0 0 0},clip]{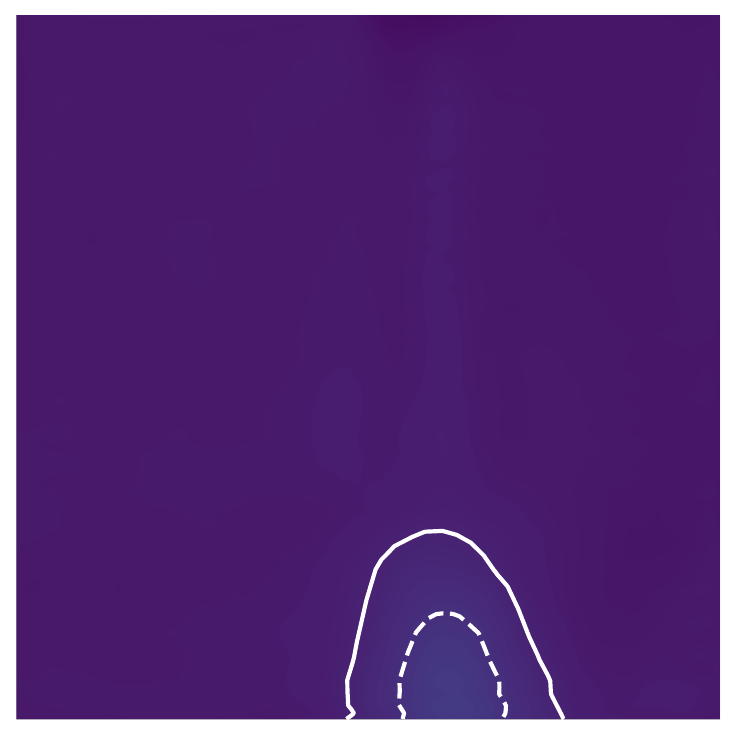}
      \end{subfigure}
      \begin{subfigure}[c]{\field\linewidth}
        \includegraphics[width=\linewidth,trim={0 0 0 0},clip]{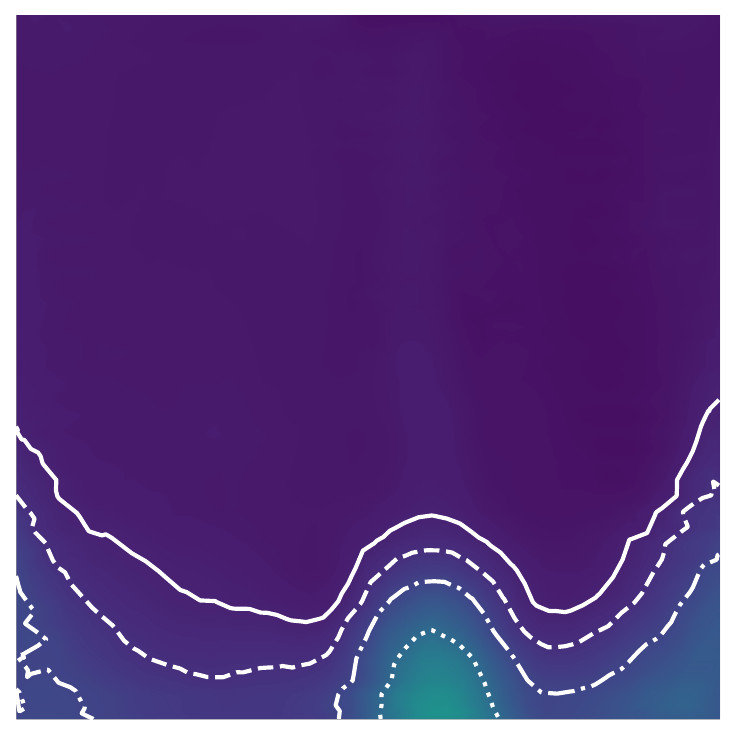}
      \end{subfigure}
      \begin{subfigure}[c]{\field\linewidth}
        \includegraphics[width=\linewidth,trim={0 0 0 0},clip]{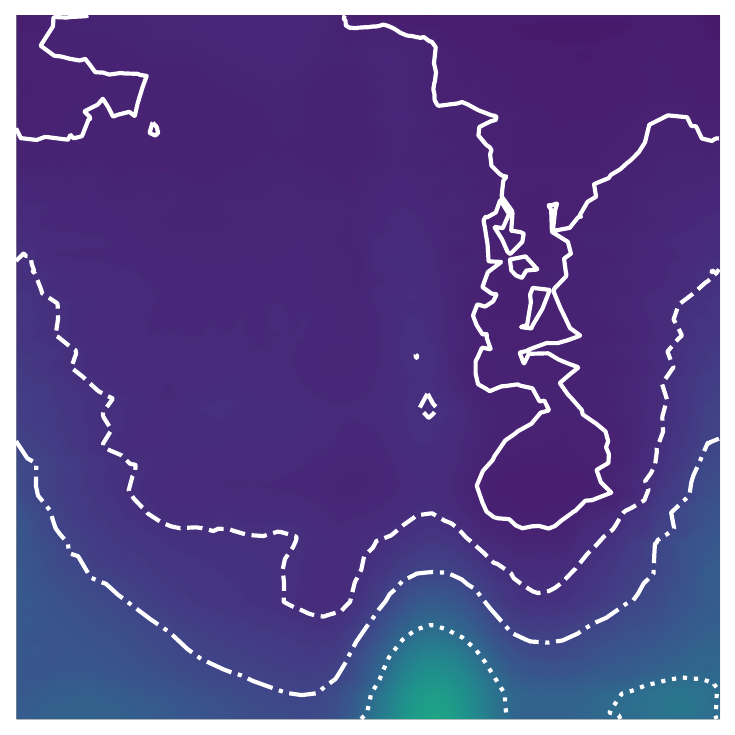}
      \end{subfigure}
      \begin{subfigure}[c]{\field\linewidth}
        \includegraphics[width=\linewidth,trim={0 0 0 0},clip]{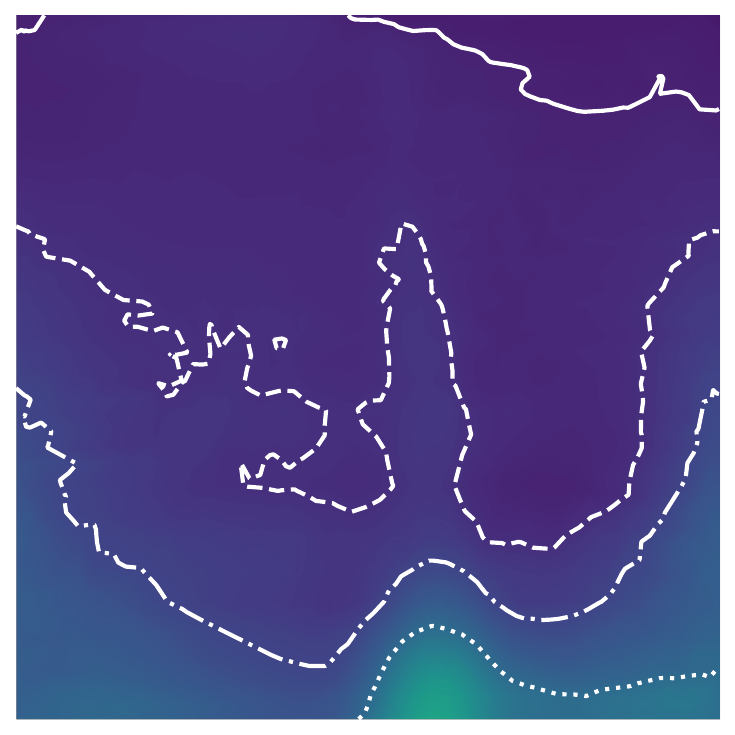}
      \end{subfigure}
      \begin{subfigure}[c]{\field\linewidth}
        \includegraphics[width=\linewidth,trim={0 0 0 0},clip]{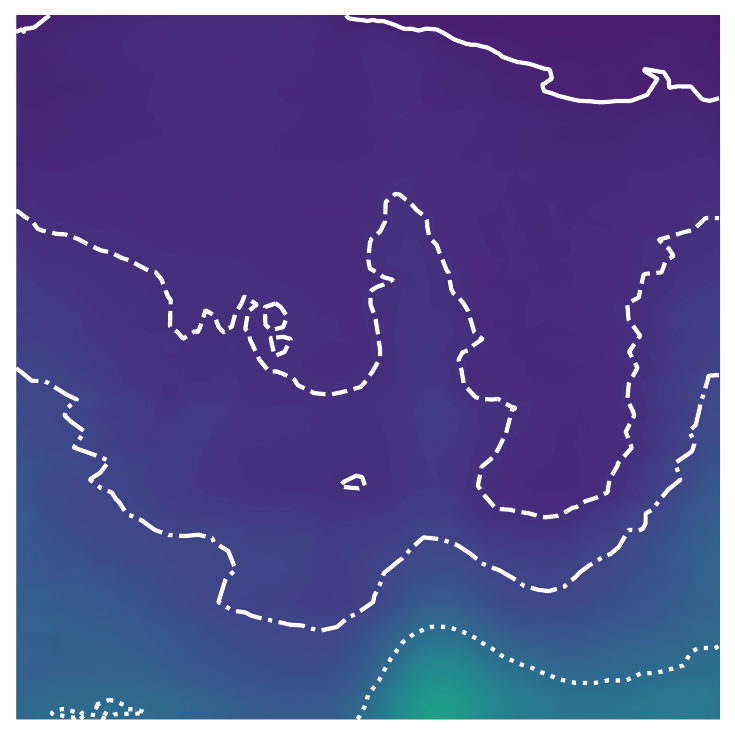}
      \end{subfigure}
      \begin{subfigure}[c]{\field\linewidth}
        \includegraphics[width=\linewidth,trim={0 0 0 0},clip]{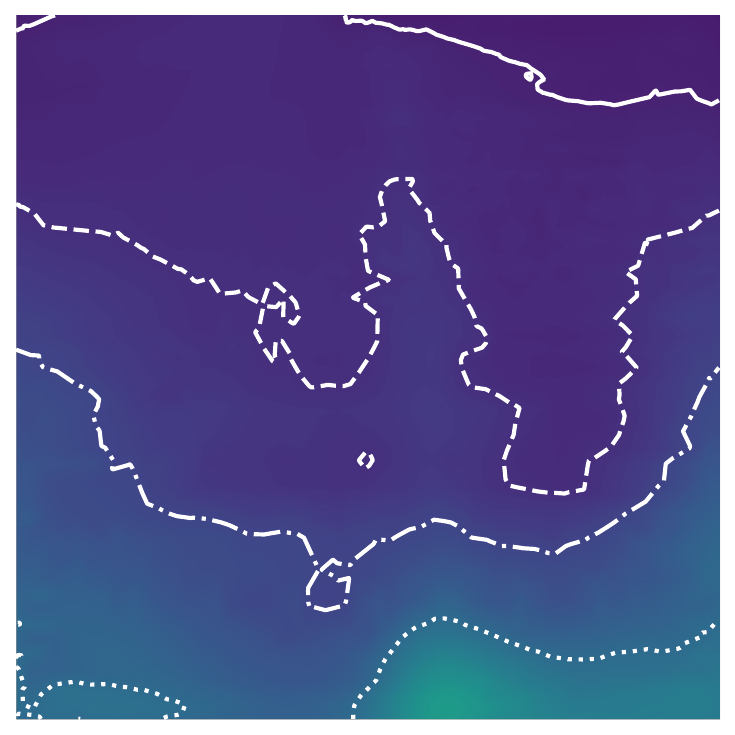}
      \end{subfigure}
    \end{minipage}
  \end{subfigure}

  % Row 2: Target
  \begin{subfigure}[b]{1\linewidth}
    \begin{minipage}[c]{0.05\linewidth}
      \centering
      \rotatebox{90}{Target $p$}
    \end{minipage}%
    \begin{minipage}[c]{0.95\linewidth}
      \noindent
      \begin{subfigure}[c]{\field\linewidth}
        \includegraphics[width=\linewidth,trim={0 0 0 0},clip]{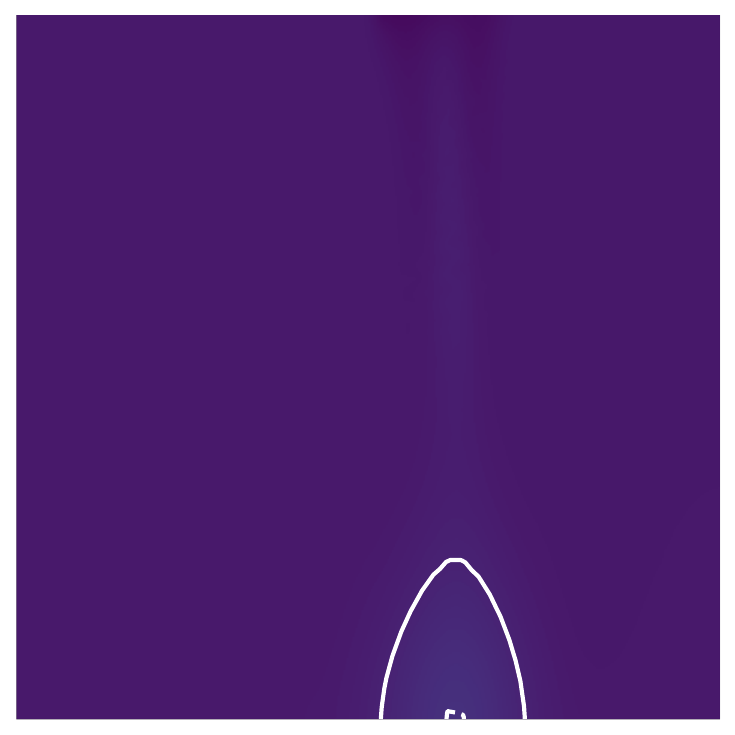}
      \end{subfigure}
      \begin{subfigure}[c]{\field\linewidth}
        \includegraphics[width=\linewidth,trim={0 0 0 0},clip]{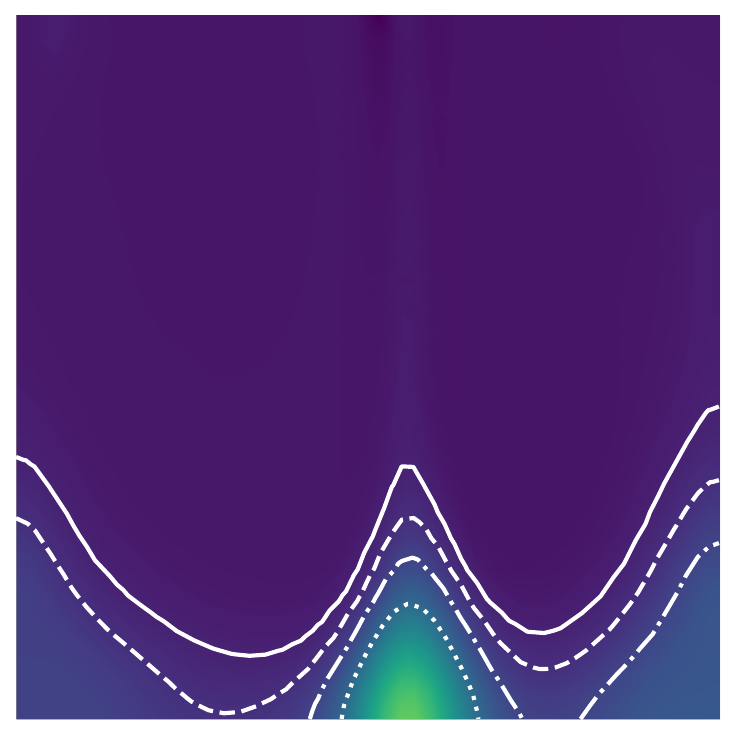}
      \end{subfigure}
      \begin{subfigure}[c]{\field\linewidth}
        \includegraphics[width=\linewidth,trim={0 0 0 0},clip]{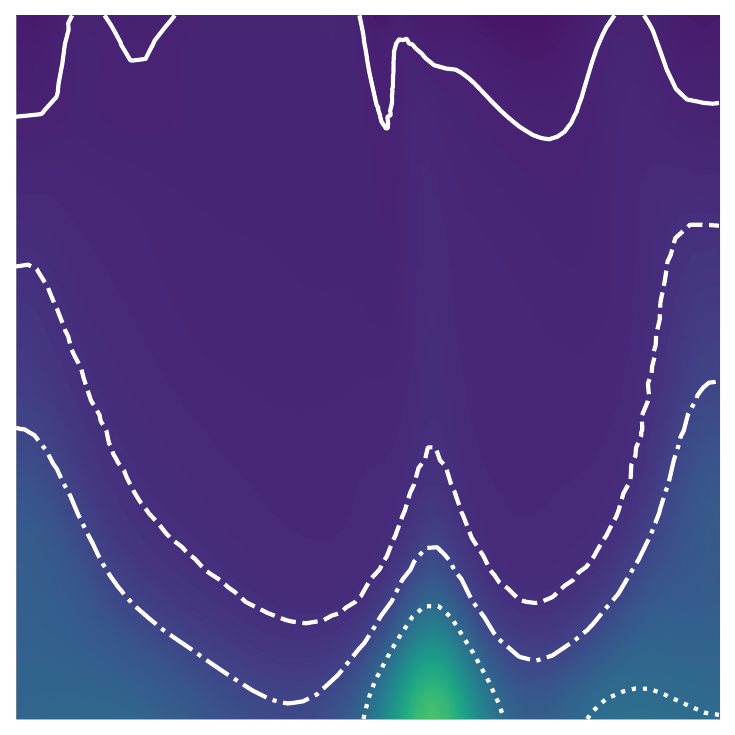}
      \end{subfigure}
      \begin{subfigure}[c]{\field\linewidth}
        \includegraphics[width=\linewidth,trim={0 0 0 0},clip]{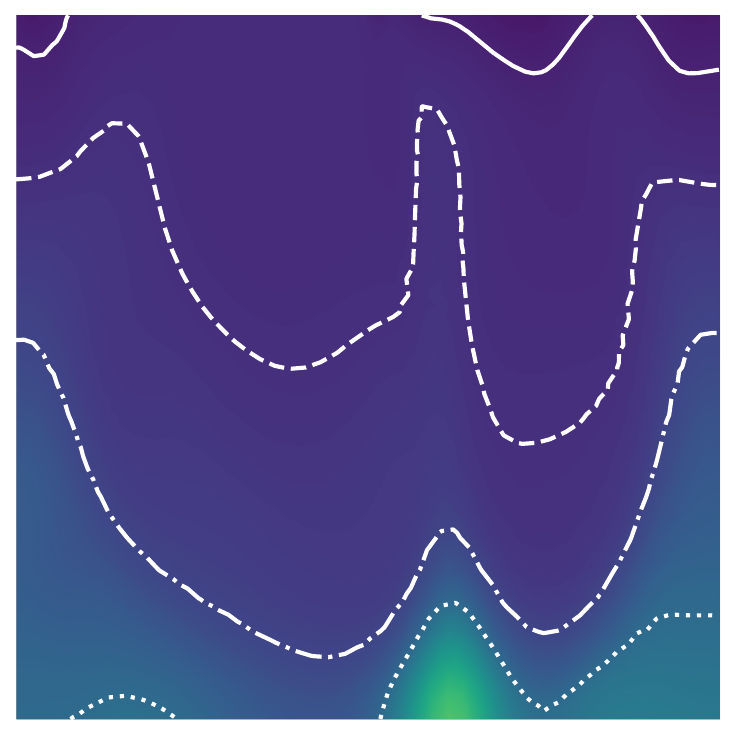}
      \end{subfigure}
      \begin{subfigure}[c]{\field\linewidth}
        \includegraphics[width=\linewidth,trim={0 0 0 0},clip]{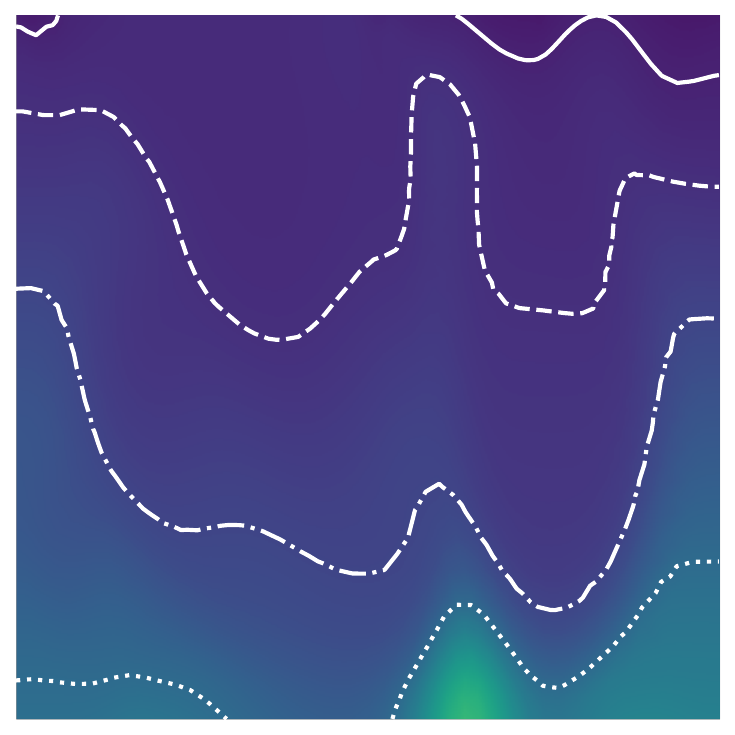}
      \end{subfigure}
      \begin{subfigure}[c]{\field\linewidth}
        \includegraphics[width=\linewidth,trim={0 0 0 0},clip]{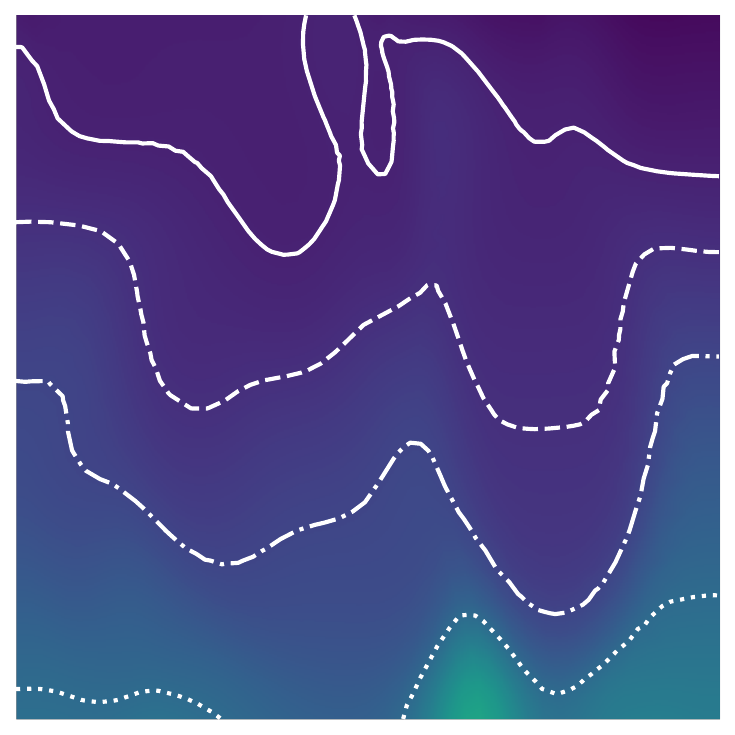}
      \end{subfigure}
    \begin{subfigure}[c]{0.01\linewidth}
    % Don't let this object contribute to the row height/depth
    \raisebox{-0.719\height}[0pt][0pt]{%
    \begin{adjustbox}{height=335pt, trim=135pt 100pt 60pt 29pt,clip}
     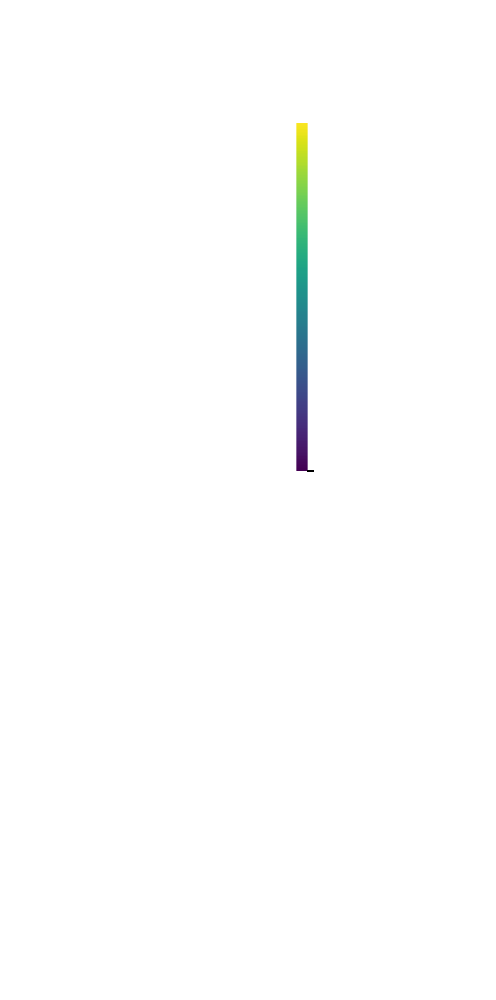%
    \end{adjustbox}}%
    \end{subfigure}
    \end{minipage}
  \end{subfigure}

% Row 3: Error (with properly centered row‐label and aligned captions)
  \begin{subfigure}[c]{1\linewidth}
    % vertical row label, centered alongside the 4 panels
    \begin{minipage}[c]{0.05\linewidth}
      \centering
       \vspace*{-0.8cm}% <- adjust this length to taste
      \rotatebox{90}{\textbf{$\lVert \hat{p} - p\rVert_2 $}}
    \end{minipage}%
    % the four error panels, each with a bottom caption
    \begin{minipage}[c]{0.95\linewidth}
    \noindent
    \begin{subfigure}[b]{\field\linewidth}
    \def\svgwidth{\linewidth}% 
    \adjustbox{trim=0 0 0 0,clip}{%
    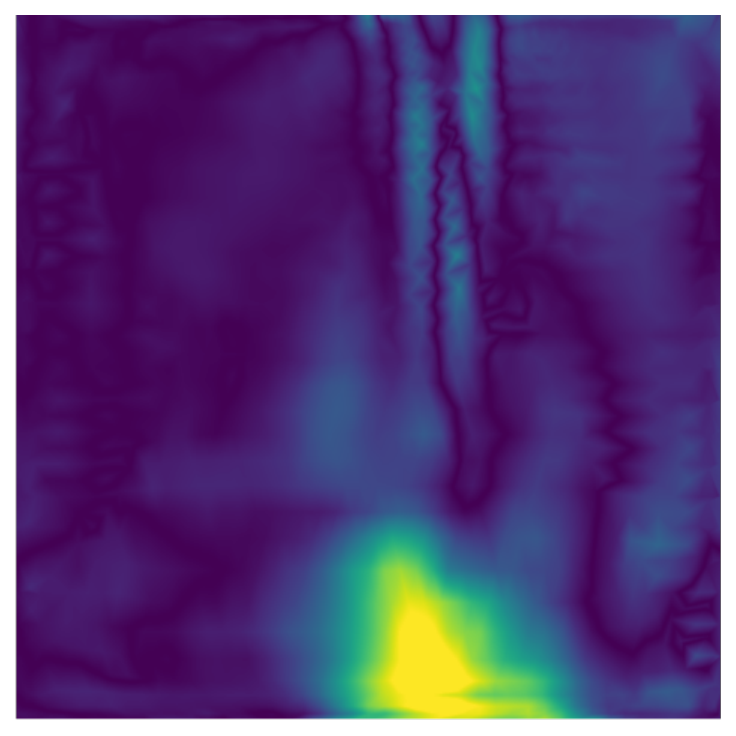}
    \caption*{$r_{p}^{(2)} = \SI{72.93}{\percent}$}
    \end{subfigure}
    \begin{subfigure}[b]{\field\linewidth}
    \def\svgwidth{\linewidth}% 
    \adjustbox{trim=0 0 0 0,clip}{%
    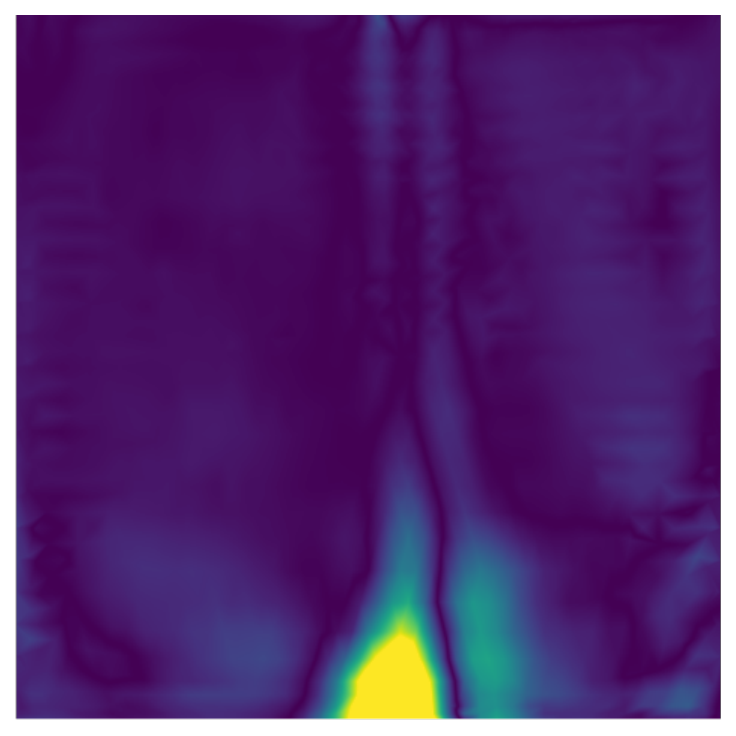}
    \caption*{$r_p^{(4)} = \SI{36.10}{\percent}$}
    \end{subfigure}
    \begin{subfigure}[b]{\field\linewidth}
    \def\svgwidth{\linewidth}% 
    \adjustbox{trim=0 0 0 0,clip}{%
    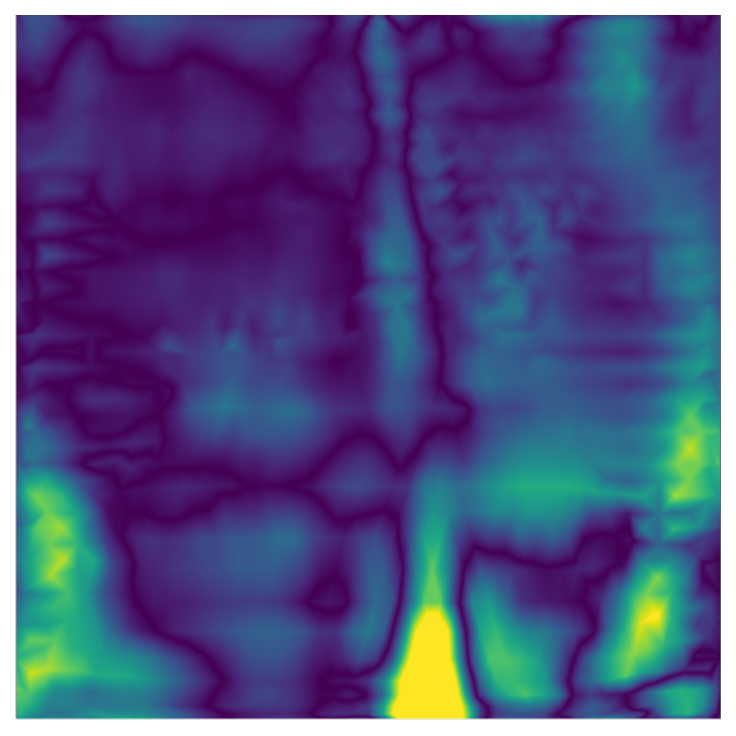}    \caption*{$r_p^{(6)} = \SI{15.96}{\percent}$}
    \end{subfigure}
    \begin{subfigure}[b]{\field\linewidth}
    \def\svgwidth{\linewidth}% 
    \adjustbox{trim=0 0 0 0,clip}{%
    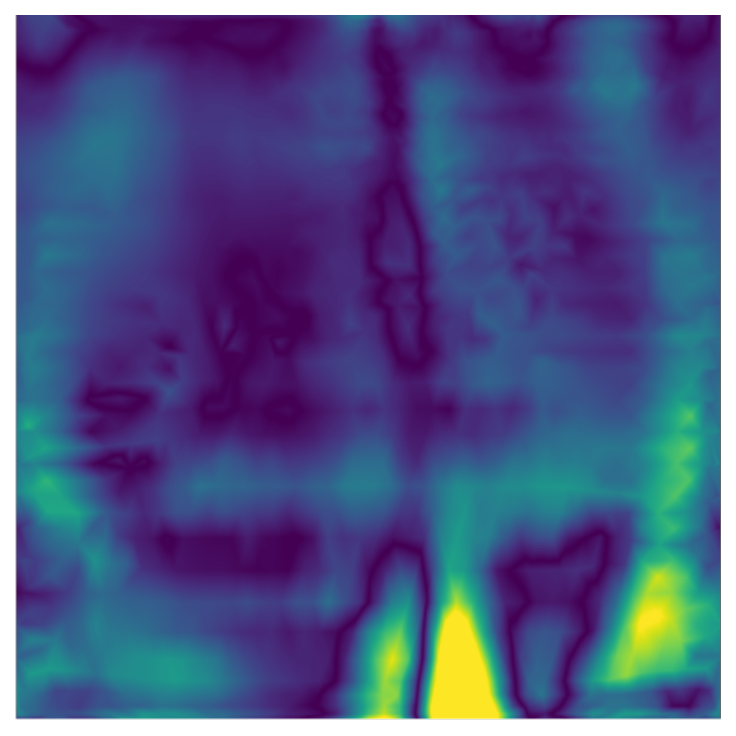}    \caption*{$r_p^{(7)} = \SI{18.20}{\percent}$}
    \end{subfigure}
    \begin{subfigure}[b]{\field\linewidth}
    \def\svgwidth{\linewidth}% 
    \adjustbox{trim=2.5 0 0 0,clip}{
    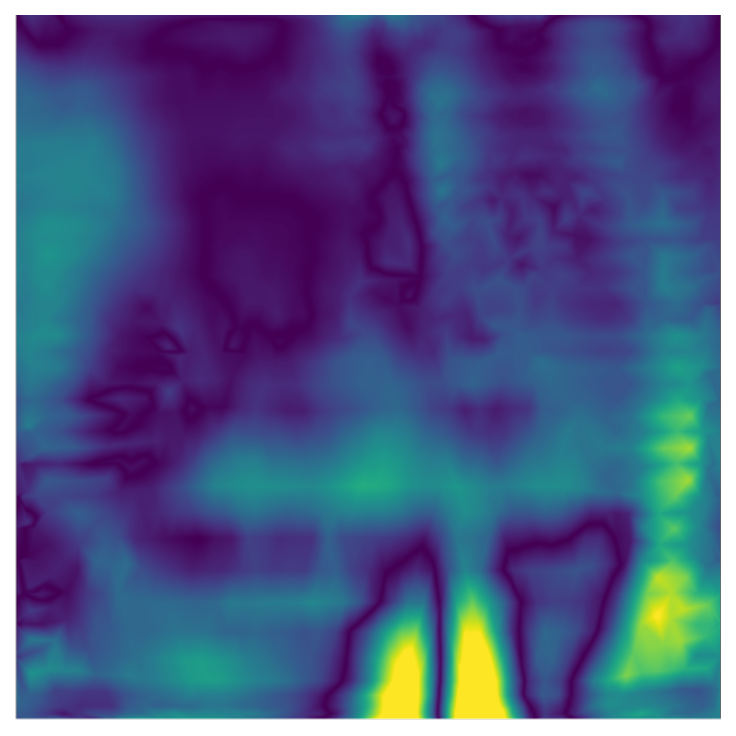}     \caption*{$r_p^{(8)} = \SI{19.02}{\percent}$}
    \end{subfigure}
    \begin{subfigure}[b]{\field\linewidth}
    \def\svgwidth{\linewidth}% 
    \adjustbox{trim=2.5 0 0 0,clip}{
    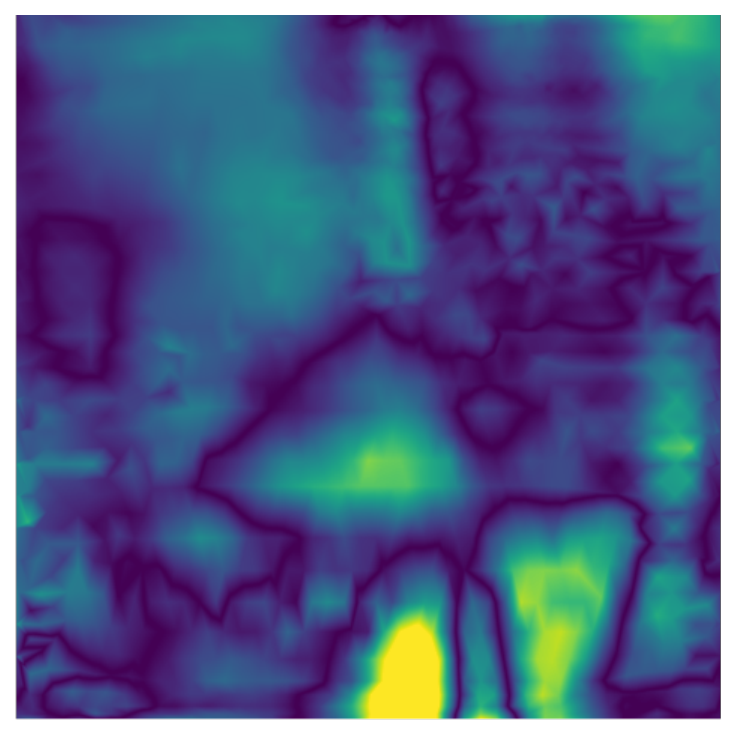}     \caption*{$r_p^{(9)} = \SI{18.44}{\percent}$}
    \end{subfigure}
    \end{minipage}
  \end{subfigure}

\caption{\textbf{Pressure field prediction}. Pressure field prediction \gls{p_hat} (row $1$), target \gls{CFD} pressure field \gls{p} (row $2$), pointwise Euclidean error $\lVert\gls{p_hat}-\gls{p}\rVert_2$ (row $3$), and relative L$_2$ error $\gls{rerror}_{\alpha}^{(k)}$ at simulation steps $\gls{steps}\in\{2,4,6,7,8,9\}$ with $\gls{steps}\times\SI{0.06}{\second}$ of a mold filling simulation. Quiver overlays indicate flow direction (array orientation) and velocity magnitude (arrow length). Isocontour lines show distinct pressure levels.}
  \label{fig:pressure_fields}
\end{figure}

Figure \ref{fig:pressure_fields} highlights a sequence of pressure field predictions \gls{p_hat} from our proposed Fourier-Graph neural solver. We show snapshots of the predicted pressure field \gls{p_hat} alongside the target pressure field \gls{p}, as well as the pointwise Euclidean $\lVert\gls{p_hat}-\gls{p}\rVert_2$ and relative $L_2$ percentage errors for steps \gls{steps} $\in(2,4,6,7,8,9)$ with horizon $\SI{0.06}{\second}\times\gls{steps}$. By overlaying pressure iso-contours, we observe that the neural solver can reproduce the broad pressure distribution in the cavity. However, the error maps in Figure \ref{fig:pressure_fields} reveals the presence of concentrated deficiencies at locations of steep pressure gradients. For instance, near the advancing fluid front and more prominently along dynamic impact zones where the flow impinges on the walls of the mold cavity. The largest pointwise pressure errors occur right after the liquid impingement, where the true solution exhibits sharp spikes that the model cannot fully capture. Quantitatively, the pointwise Euclidean error in pressure is in general higher than that of velocity or volume fraction at comparable times, confirming that the pressure field is the most challenging variable to be modeled. This finding is consistent with the reasoning that pressure solutions involve global elliptic constraints, and that small discrepancies can manifest if the global coupling of the model is not reasonably accurate. In the proposed Fourier-Graph architecture, the graph-based decoder does introduce some non-local message passing, which helps mitigate this issue by propagating information across the domain, but a residual gap still remains. Similar observations have been made in other operator-learning studies \citep{wen2022u}, where U-nets backbones helped to improve the predictions Fourier-based neural solvers.

The last row of in Figure \ref{fig:pressure_fields} shows the relative $L_2$ error $\gls{rerror}_p^{(k)}$ for different steps \gls{steps}. At step $\gls{steps}=2$, the model shows a high error of \SI{73}{\percent}, which is followed by a rapid decay to values ranging \SIrange[]{16}{19}{\percent} at later steps. The initial high value matches the first impingement phase when steep pressure gradients appear and the model wrongly approximates fast transients. The subsequent plateau indicates that pressure errors stabilize rather than escalate over the tested $10$-step horizon, which agrees with the absence of spurious oscillations in the maps. Together, these results reinforce that pressure is most vulnerable during the first transient steps and, once the flow organizes, the operator remains stable even though fine gradients remain less accurate.

Overall, the general exploration results establish that our Fourier-Graph neural solver can mimic the key behavior of a two-phase filling simulation by correctly advecting the fluid-air interface and reproducing the velocity and pressure fields. We also observed a systematic tendency to smooth sharp features, yet the results still justify their use as surrogates for design optimization, where consistency along the temporal prediction trajectory is critical. Moreover, the neural solver achieves these results at a fraction of the computational cost of transient \gls{CFD}. In comparable settings, neural operators such as \gls{FNO} have been reported to deliver \numrange{e2}{e3} speedups over traditional numerical solvers \citep{li2020fourier,pfaff2021learningmeshbasedsimulationgraph}, enabling rapid, design-in-the-loop evaluations that were previously impractical.

\subsection{Spatial Subsampling}
\label{spatial_subsampling}
We next investigate the trade-off between simulation resolution and model accuracy by subsampling the training data spatially. Spatial subsampling here means we thin out the mesh vertices used during training, effectively training the network 38on a coarser version of the field data. This experiment reveals how sensitive the neural solver is to missing fine-scale spatial information, and by extension, how it might perform if deployed on lower-resolution simulations for efficiency. We train separate model instances on progressively coarser samplings of the mesh: factors $\gls{ss}=1$ (no subsampling, using all mesh vertices), $2$ (using $1$ out of $2$ vertices), $3$, $4$, and $5$ (using only $1$ in $5$ vertices, i.e. \SI{20}{\percent} of vertices). After training, we evaluate each model on the full-resolution meshes of the DS$_1$ and  DS$_2$ datasets to understand how well the proposed neural solvers recover the high-resolution fields. Moreover, we also consider a DS$_2^{60}$ dataset whose training data are temporally subsampled with factor $\gls{st}=6$ to examine whether a longer predicted time window interacts with spatial resolution changes.

Table \ref{tab:combined_sp_subsampling} reports the relative $L_2$ percentage error of the model on each field (velocity components $\gls{u}, \gls{v}$, pressure \gls{p}, and volume fraction \gls{alpha}) as a function of the spatial sampling factor \gls{ss}. The overall trend is clear and monotonic, with even a modest reduction in spatial resolution during training  significantly impacting accuracy and performance degradation growing steadily with larger \gls{ss}. For instance, on DS$_2$ the mean relative $L_2$ percentage error rises from about \SI{4.5}{\percent} at full resolution ($\gls{ss}=1$) to \SI{6.6}{\percent} at $\gls{ss}=5$. The simplified DS$_1$ shows a similar behavior (from $\sim$ \SIrange[range-phrase={ to }]{3.97}{7.46}{\percent} mean error) when going from no subsampling to $\gls{ss}=5$. The DS$_2^{60}$ case with fewer temporal frames but a longer horizon likewise sees its mean error increase from \SIrange[range-phrase={ to }]{6.28}{7.53}{\percent} at $\gls{ss}=5$. These numbers confirm the intuitive result that training the neural operator on coarser spatial data biases it towards low-frequency content and impairs its ability to reconstruct fine details. Among the individual field variables, pressure is consistently the most sensitive to spatial coarsening. For example, on DS$_1$ the pressure error roughly doubles (\SIrange[range-phrase={ to }]{4.58}{9.07}{\percent}) when \gls{ss} increases from $1$ to $5$. By contrast, the volume fraction \gls{alpha} is less affected by subsampling (error increasing from \SIrange[range-phrase={ to }]{3.40}{6.66}{\percent} on DS$_1$), reflecting that the diffuse interface can still be roughly located even with fewer points. Lastly, the velocity components show only intermediate sensitivity to spatial subsampling. Overall, these field-wise trends mirror the qualitative behavior seen in Section \ref{Quantitative_results}, with features that require high spatial resolution such as pressure spikes and thin fluid-air interfaces being the first to suffer when the training data are under-resolved.

\begin{table}[bth]
\newcolumntype{B}{>{\centering\arraybackslash}p{0.195cm}}
\centering
\caption{\textbf{Spatial subsampling performance for datasets DS$_1$, DS$_2$, and DS$_2^{60}$}. Entries report the per-field and mean relative $L_2$ error (\SI{}{\percent}) evaluated at full resolution meshes for models trained with uniform spatial sampling factors $\gls{ss}\in\{1,\dots,5\}$. DS$_1$ and DS$_2$ use no temporal subsampling. DS$_2^{60}$ uses a temporal subsampling factor of $\gls{st}=6$.}
\vspace{.5em}
\small
\setlength{\tabcolsep}{2.5pt}
\renewcommand{\arraystretch}{.8}

% Layout:
% 1 = s_s
% 2 = spacer
% groups of (3 values + spacer), repeated 5 times = u,v,p,alpha,mean
\begin{tabularx}{.95\textwidth}{
  c B
  ccc B
  ccc B
  ccc B
  ccc B
  ccc
}
\toprule
  \multirow{2.5}{*}{\makecell[c]{\bfseries Spatial \\ \bfseries subsampling \\ \bfseries factor $\mathbf{s_{s}}$}}
  & \multicolumn{20}{c}{\bfseries Relative $L_2$ Error [\SI{}{\percent}]} \\
\cmidrule{3-21}
  &  % spacer
  & \multicolumn{3}{c}{$\boldsymbol{u}$}
  &  % spacer
  & \multicolumn{3}{c}{$\boldsymbol{v}$}
  &  % spacer
  & \multicolumn{3}{c}{$\boldsymbol{p}$}
  &  % spacer
  & \multicolumn{3}{c}{$\boldsymbol{\alpha}$}
  &  % spacer
  & \multicolumn{3}{c}{$\textbf{Mean}$} \\
\cmidrule(lr){3-5}\cmidrule(lr){7-9}\cmidrule(lr){11-13}\cmidrule(lr){15-17}\cmidrule(lr){19-21}
  &  % spacer
  & \textbf{DS$_1$} & \textbf{DS$_2$} & \textbf{DS$_2^{60}$}
  &  % spacer
  & \textbf{DS$_1$} & \textbf{DS$_2$} & \textbf{DS$_2^{60}$}
  &  % spacer
  & \textbf{DS$_1$} & \textbf{DS$_2$} & \textbf{DS$_2^{60}$}
  &  % spacer
  & \textbf{DS$_1$} & \textbf{DS$_2$} & \textbf{DS$_2^{60}$}
  &  % spacer
  & \textbf{DS$_1$} & \textbf{DS$_2$} & \textbf{DS$_2^{60}$} \\
\midrule
$1$ & & $3.96$ & $4.49$ & $6.48$ & & $3.92$ & $4.52$ & $6.74$ & & $4.58$ & $5.15$ & $6.28$ & & $3.40$ & $3.90$ & $5.60$ & & $\mathbf{3.97}$ & $\mathbf{4.52}$ & $\mathbf{6.28}$ \\
$2$ & & $5.96$ & $5.62$ & $7.40$ & & $6.11$ & $5.75$ & $7.84$ & & $7.10$ & $6.80$ & $7.92$ & & $5.89$ & $5.37$ & $7.15$ & & $\mathbf{6.27}$ & $\mathbf{5.89}$ & $\mathbf{7.58}$ \\
$3$ & & $6.56$ & $6.37$ & $7.43$ & & $6.54$ & $6.54$ & $7.83$ & & $8.45$ & $7.51$ & $7.77$ & & $6.30$ & $6.19$ & $7.09$ & & $\mathbf{6.96}$ & $\mathbf{6.61}$ & $\mathbf{7.53}$ \\
$4$ & & $6.51$ & $6.13$ & $7.41$ & & $6.67$ & $6.23$ & $7.78$ & & $8.00$ & $8.19$ & $7.80$ & & $6.41$ & $5.97$ & $7.09$ & & $\mathbf{6.90}$ & $\mathbf{6.63}$ & $\mathbf{7.52}$ \\
$5$ & & $7.08$ & $6.26$ & $7.47$ & & $7.02$ & $6.42$ & $7.86$ & & $9.07$ & $7.52$ & $7.67$ & & $6.66$ & $6.10$ & $7.12$ & & $\mathbf{7.46}$ & $\mathbf{6.58}$ & $\mathbf{7.53}$ \\
\bottomrule
\end{tabularx}
\label{tab:combined_sp_subsampling}
\end{table}

\vspace{1em}
\begin{figure}[tbh] 
  %========================
  % Row 0: vertical label + predictions at spatial steps 1, 3, 5, then target
  %========================
  \begin{subfigure}[c]{1\linewidth}
    % Left vertical text column (centered vertically)
    \begin{subfigure}[c]{0.03\linewidth}
      \centering
      \rotatebox{90}{Prediction $\mathbf{\hat{u}}$}
    \end{subfigure}
    % (0,1) Prediction s=1
    \begin{subfigure}[c]{\fields\linewidth}
      \includegraphics[width=\linewidth,trim={0 55 0 55},clip]{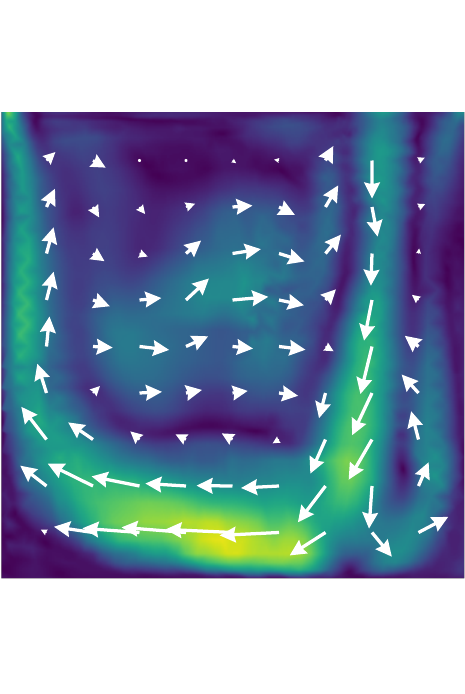}
    \end{subfigure}
    % (0,2) Prediction s=2
    \begin{subfigure}[c]{\fields\linewidth}
      \includegraphics[width=\linewidth,trim={0 55 0 55},clip]{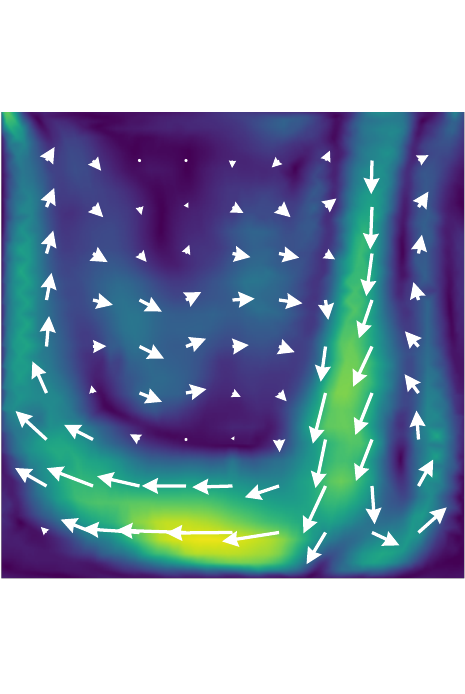}
    \end{subfigure}
    % (0,3) Prediction s=3
    \begin{subfigure}[c]{\fields\linewidth}
      \includegraphics[width=\linewidth,trim={0 55 0 55},clip]{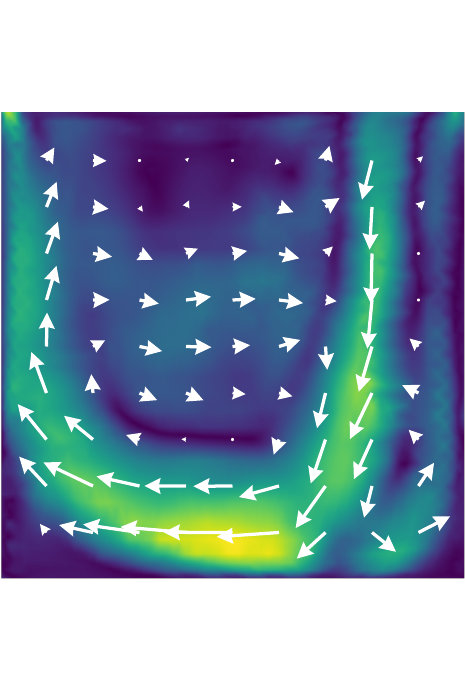}
    \end{subfigure}
    % (0,4) Prediction s=4
    \begin{subfigure}[c]{\fields\linewidth}
      \includegraphics[width=\linewidth,trim={0 55 0 55},clip]{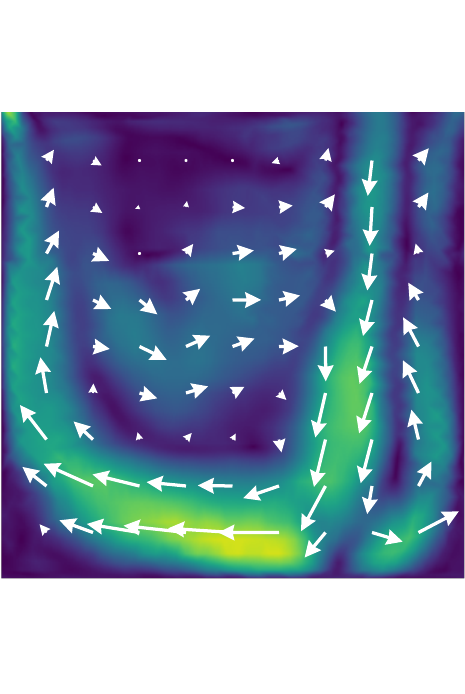}
    \end{subfigure}
    % (0,5) Prediction s=5
    \begin{subfigure}[c]{\fields\linewidth}
      \includegraphics[width=\linewidth,trim={0 55 0 55},clip]{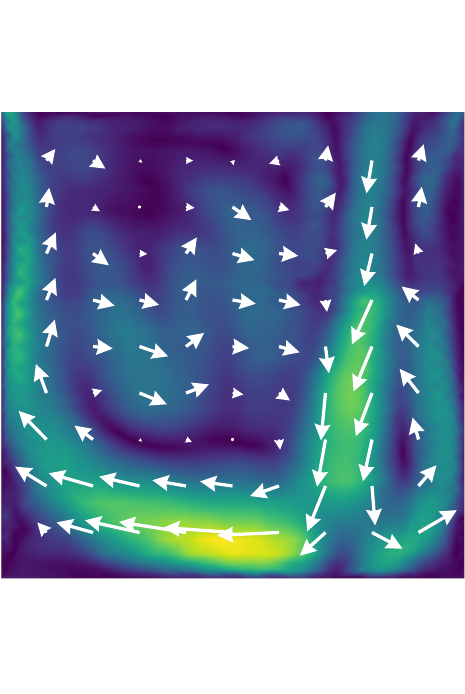}
    \end{subfigure}
    %\hspace{0.5cm}\vrule width 0.5pt\hspace{0.5cm}
    % (0,6) Target
    \begin{subfigure}[c]{\fields\linewidth}
      \includegraphics[width=\linewidth,trim={0 55 0 55},clip]{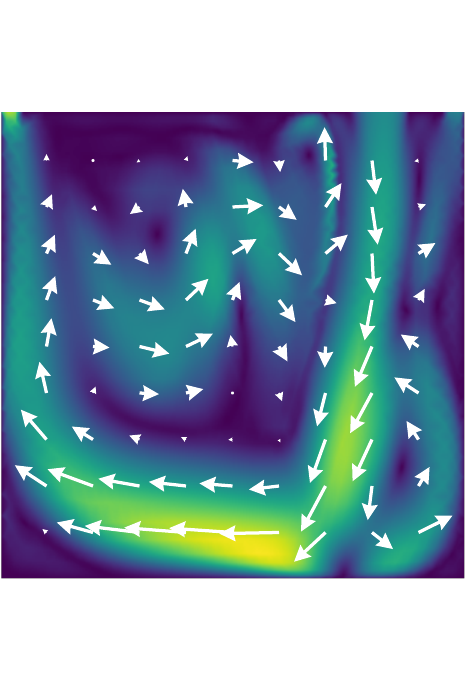}%
    \end{subfigure}
  \end{subfigure}
      \begin{subfigure}[c]{0.05\linewidth}
    % Don't let this object contribute to the row height/depth
    %\hspace*{6.5cm}
    \raisebox{-0.2\height}[0pt][0pt]{%
    \begin{adjustbox}{trim=47pt 0pt 0pt 0pt,clip}
    \def\svgwidth{78pt}%      \phantom{\input{Plots/Results/colorbar_v0_0.98.pdf_tex}}%
    \end{adjustbox}}%
    \end{subfigure}

  %========================
  % Row 1: vertical label + differences at spatial steps 1, 3, 5, then empty
  %========================
  \begin{subfigure}[c]{1\linewidth}
    % Left vertical text column (centered vertically)
    \begin{subfigure}[c]{0.03\linewidth}
    \centering
    \rotatebox{90}{\hspace{1.5em}\textbf{$\lVert\mathbf{\hat{u}} - \mathbf{u}\rVert_2$}}
    \end{subfigure}
    % (1,1) |pred - targ| s=1
    \begin{subfigure}[c]{\fields\linewidth}
    \adjustbox{trim=0 15 0 15,clip}{%
    \def\svgwidth{\linewidth}% 
    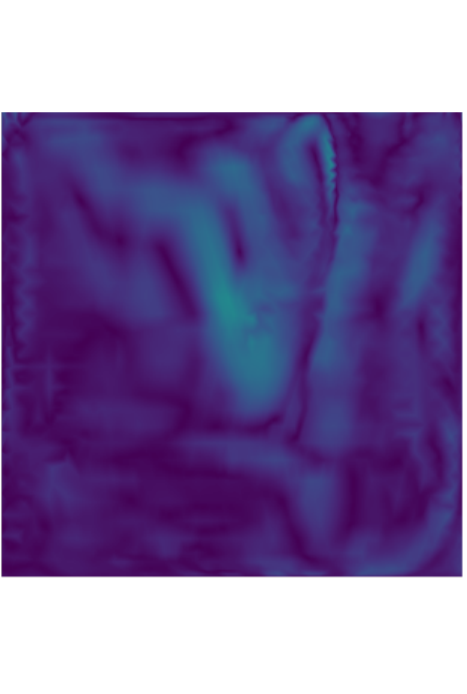}
    \caption*{$r_{\mathbf{u}}^5 = \SI{34.84}{\percent}$}
    \end{subfigure}
    % (1,2) |pred - targ| s=2
    \begin{subfigure}[c]{\fields\linewidth}
    \adjustbox{trim=0 15 0 15,clip}{%
    \def\svgwidth{\linewidth}% 
    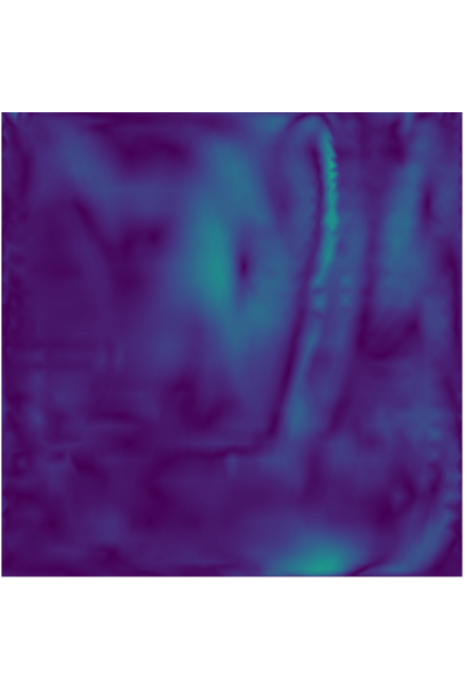}
    \caption*{$r_{\mathbf{u}}^5 = \SI{37.93}{\percent}$}
    \end{subfigure}
    % (1,3) |pred - targ| s=3
    \begin{subfigure}[c]{\fields\linewidth}
    \adjustbox{trim=0 15 0 15,clip}{%
    \def\svgwidth{\linewidth}% 
    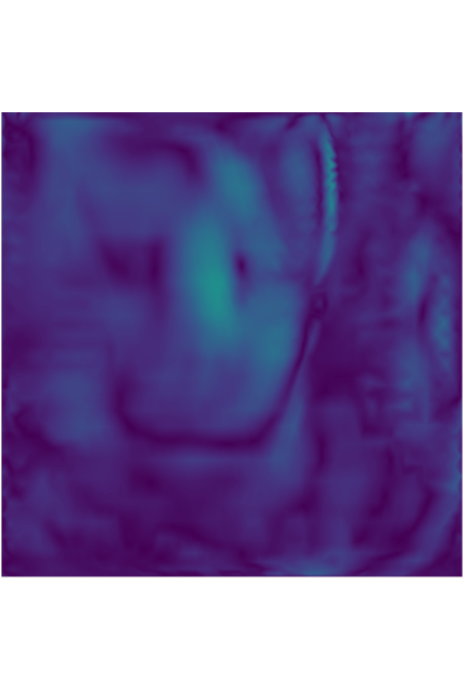}
    \caption*{$r_{\mathbf{u}}^5 = \SI{37.27}{\percent}$}
    \end{subfigure}
    % (1,4) |pred - targ| s=4
    \begin{subfigure}[c]{\fields\linewidth}
    \adjustbox{trim=0 15 0 15,clip}{%
    \def\svgwidth{\linewidth}% 
    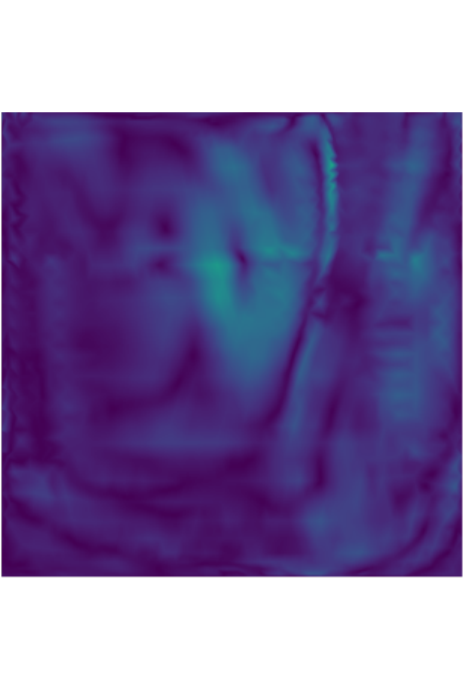}
    \caption*{$r_{\mathbf{u}}^5 = \SI{38.17}{\percent}$}
    \end{subfigure}
    % (1,3) |pred - targ| s=5
    \begin{subfigure}[c]{\fields\linewidth}
    \adjustbox{trim=0 15 0 15,clip}{%
    \def\svgwidth{\linewidth}% 
    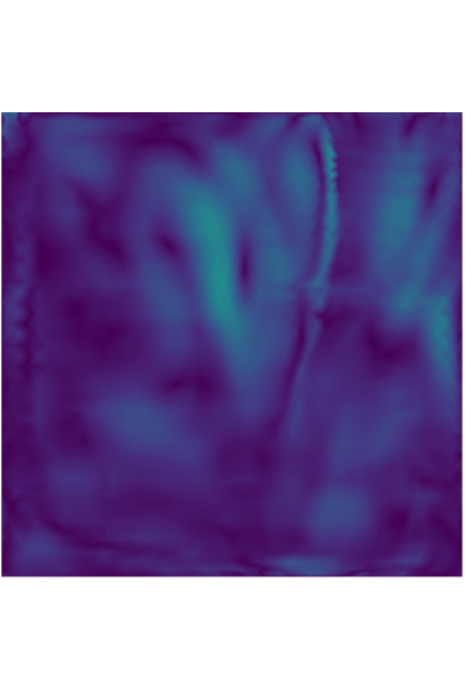}
    \caption*{$r_{\mathbf{u}}^5 = \SI{38.46}{\percent}$}
    \end{subfigure}
    \begin{subfigure}[c]{\fields\linewidth}
    \adjustbox{trim=0 15 0 15,clip}{%
    \def\svgwidth{\linewidth}% 
    \phantom{\input{Plots/Results/ind742_t5_error_sp5.pdf_tex}}}
    \caption*{Target}
    \end{subfigure}
    \begin{subfigure}[c]{0.05\linewidth}
    % Don't let this object contribute to the row height/depth
    %\hspace*{6.5cm}
    \raisebox{-0.32\height}[12pt][30pt]{%
    \begin{adjustbox}{trim=35pt 0pt -45pt 0pt,clip}
    \def\svgwidth{65pt}% 
     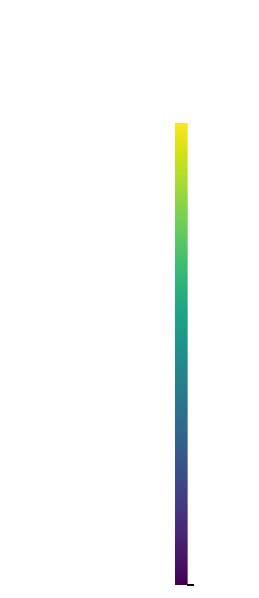%
    \end{adjustbox}}%
    \end{subfigure}
  \end{subfigure}
  \caption{\textbf{Impact of spatial subsampling on the velocity field prediction}. Spatial subsampling predictions at factors $\gls{ss}$ ${\in{1,2,3,4,5}}$ (first five columns) versus the simulation target (last column). The bottom row shows the pointwise Euclidean error $\lVert\gls{u_hat}-\gls{u}\rVert_2$ for each \gls{ss}.}
  \label{fig:velocity_fields_sp}
\end{figure}

\begin{figure}[bth] 
  \centering
  % Left vertical text column (centered vertically)
  \begin{subfigure}[b]{0.025\linewidth}
    \centering
    \rotatebox{90}{\hspace{2.5em}Volume Fraction $[-]$}
    \end{subfigure}
  \begin{subfigure}[b]{0.9\linewidth}
    % (0,1) Target
    \begin{subfigure}[b]{0.2925\linewidth} 
      \adjustbox{trim=0 1 2 12,clip,width=\linewidth}{%
    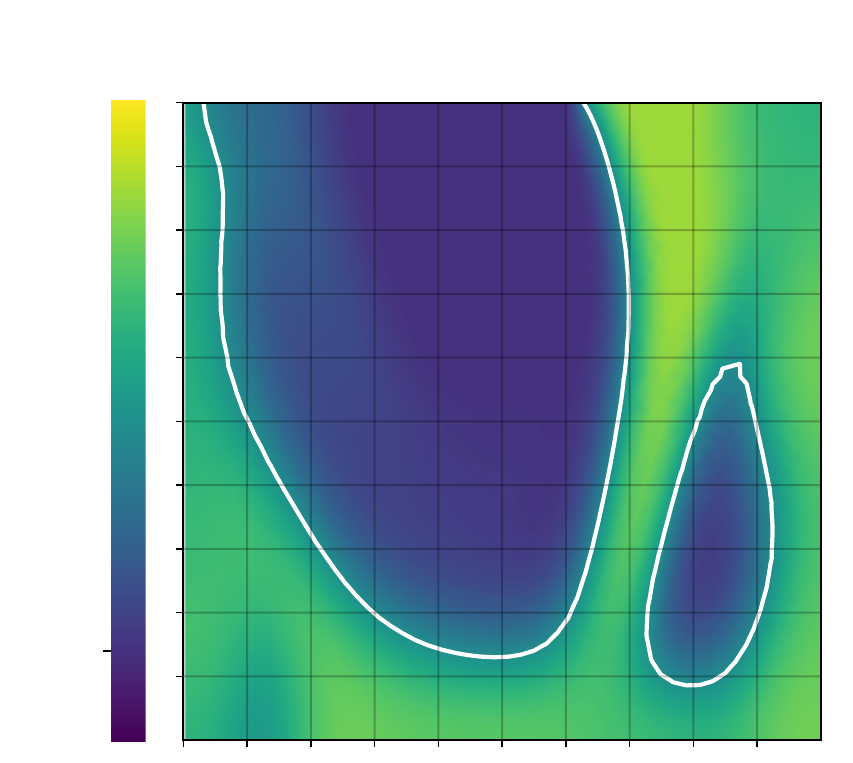}
      \caption{Simulation target $(\gls{ss}=1$)}\label{vf_targ}
      \label{sim_target}
    \end{subfigure}\hfill
    % (0,2) Prediction s=1
    \begin{subfigure}[b]{0.233\linewidth}
      \includegraphics[width=\linewidth,trim={0 0 0 0},clip]{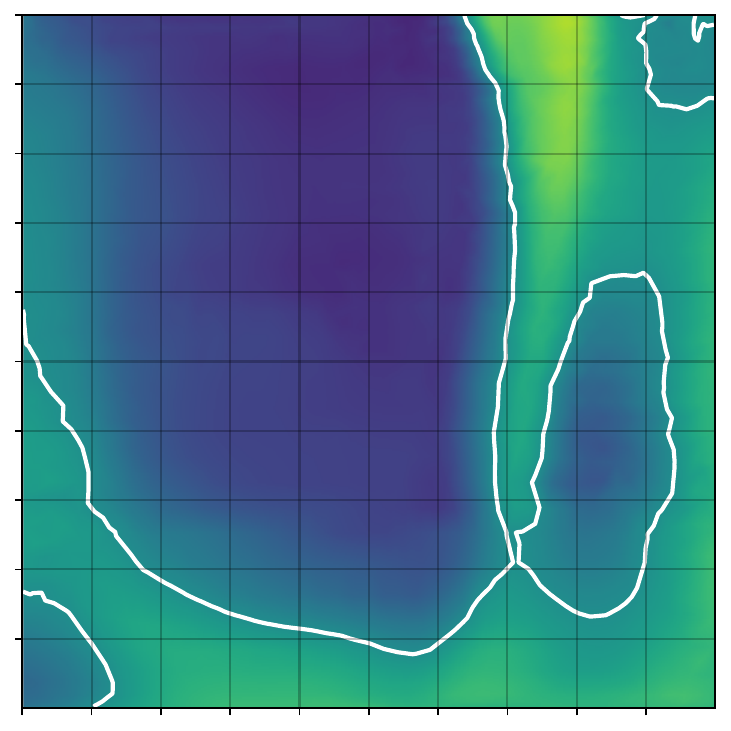}
      \caption{$\gls{ss}=1$}\label{fig:a_no_sp}
    \end{subfigure}\hfill
    % (0,3) Prediction s=3
    \begin{subfigure}[b]{0.233\linewidth}
      \includegraphics[width=\linewidth,trim={0 0 0 0},clip]{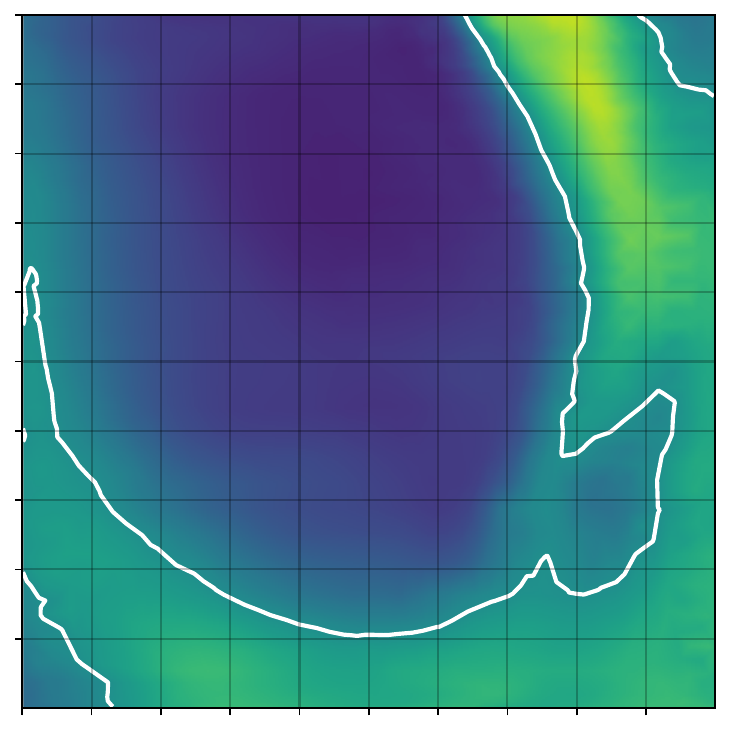}
      \caption{$\gls{ss}=3$}\label{fig:a_sp_3}
    \end{subfigure}\hfill
    % (0,4) Prediction s=5
    \begin{subfigure}[b]{0.233\linewidth}
      \includegraphics[width=\linewidth,trim={0 0 0 0},clip]{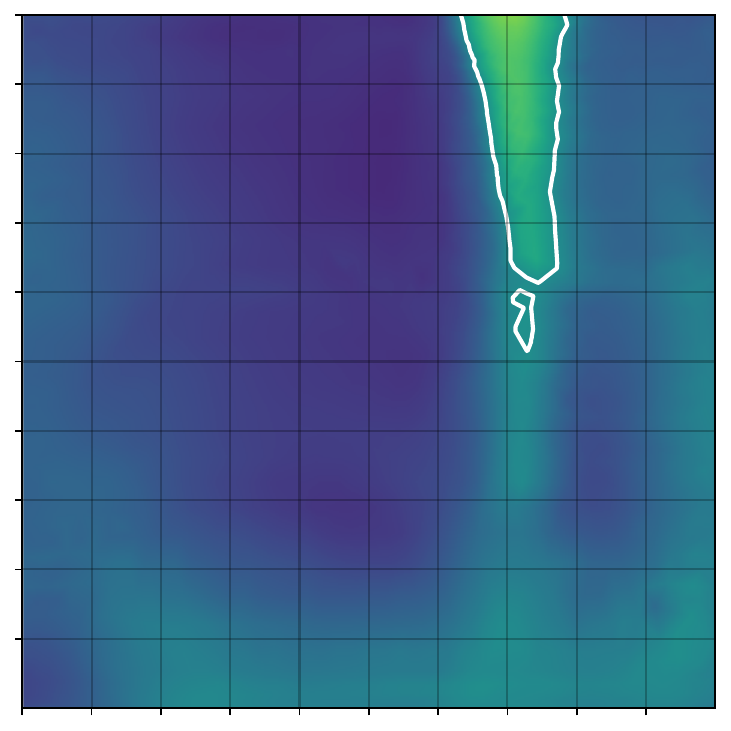}
      \caption{$\gls{ss}=5$}\label{fig:a_sp_5}
    \end{subfigure}
  \end{subfigure}%\par\medskip
  \caption{\textbf{Effect of spatial subsampling \gls{ss} on the fluid-air interface fidelity}. Predicted volume fraction fields \gls{alpha_hat} with interface overlays for sampling factors $\gls{ss}\in\{1,3,5\}$ are compared with the simulation target in Figure \ref{sim_target}. As \gls{ss} increases, the fluid-air interface (white contour, $\gls{alpha}=\SI{50}{\percent}$) prediction accuracy degrades with thickening and drifting, while the large-scale flow pattern remains similar. All predictions are shown at $\gls{time}=\SI{0.42}{\second}$.}
  \label{fig:volume_fraction_fields_sp}
\end{figure}

Figure \ref{fig:velocity_fields_sp} and Figure \ref{fig:volume_fraction_fields_sp} visually illustrate the impact of spatial subsampling on the prediction of the model. In Figure \ref{fig:velocity_fields_sp}, we compare the predicted velocity field against the target \gls{CFD} velocity field at $\gls{time}=\SI{0.42}{\second}$ for models trained with spatial resolution factor $\gls{ss}=1$ (full resolution), $3$, and $5$. The top row shows the velocity field and the bottom row shows the corresponding pointwise Euclidean error $\lVert\gls{u_hat}-\gls{u}\rVert_2$. We see that the large-scale flow structures, such as the general direction of the fluid jet and the circulation pattern in the cavity, remain largely intact even for the heavily subsampled model ($\gls{ss}=5$). This indicates that the network still captures the bulk momentum transport correctly. However, as \gls{ss} increases, fine details begin to blur or disappear. The error maps confirm a systematic growth of localized errors with higher subsampling, with errors concentrated along the interface and in regions of steep velocity gradients. This visual evidence supports the notion that coarse training effectively acts as a low-pass filter, removing access to high-wavenumber information. As a result, the neural solver cannot reconstruct information absent from training, yielding larger errors precisely where high-frequency details matter most. 

The velocity pointwise Euclidean error $\gls{rerror}_{\mathbf{u}}^5$ at step $k = 5$ increases with stronger spatial subsampling. At full resolution, the error is about \SI{35}{\percent} and rises to roughly \SI{38.5}{\percent} at the highest subsampling factor. This trend mirrors the mean errors in Table \ref{tab:combined_sp_subsampling} and confirms that coarser training emphasizes low frequency content at the cost of local fidelity.

In Figure \ref{fig:volume_fraction_fields_sp}, we focus on the fluid-air interface predictions for $\gls{ss}\in {1,3,5}$. Plotted are the volume fraction fields at $\gls{time}=\SI{0.42}{\second}$ with the $\gls{alpha}=0.5$ iso-contour delineating the fluid-air interface. At full resolution ($\gls{ss}= 1$, Figure \ref{fig:a_no_sp}), the predicted interface aligns closely with the true interface, Figure \ref{vf_targ}. At $\gls{ss}=3$, the interface begins to thicken and drift slightly, and the neural solver starts to miss thin portions of the fluid-air interface. By $\gls{ss}= 5$, the deterioration is clear, with the predicted fluid-air interface being completely degenerate and the model unable to maintain a sharp front prediction. Nevertheless, even in the $\gls{ss}=5$ case, the large-scale shape of the filling region is correct. All these observations reinforce that aliasing from spatial subsampling preferentially removes the high-frequency content needed for sharp gradients, e.g. pressure spikes and thin interfaces, while leaving the coarse flow pattern relatively untouched.

\vspace{1em}

Interestingly, when comparing DS$_2$ and DS$_2^{60}$ in Table \ref{tab:combined_sp_subsampling}, we find that extending the temporal prediction horizon does not dramatically worsen the spatial subsampling effect. The DS$_2^{60}$ errors are uniformly higher than DS$_2$ at each \gls{ss}, but the increment in error from $\gls{ss}=1$ to $5$ is similar for DS$_2$ and DS$_2^{60}$. This suggests that within the examined range, spatial resolution is the limiting factor, and training on a coarser mesh degrades performance, regardless of whether we are predicting \SI{0.1}{\second} or \SI{0.6}{\second} ahead. In practice, this means our neural solver can tolerate reasonably long rollout horizons without catastrophic error growth, as long as the spatial information content is sufficient. It also implies that spatial resolution of training data is critical for capturing fine physics and one cannot simply compensate a coarse mesh with a shorter time horizon or vice versa. For future applications, this points to the value of adaptive meshing or smart sampling, where dedicating more spatial degrees of freedom in regions of interest (e.g. near the interface or ingates) has the potential to improve model accuracy without the cost of training with fine meshes. Indeed, other authors have proposed hybrid strategies that alternate high-fidelity simulation with a neural operator to leap in time \citep{oommen2022learning}, and architectures inspired by multi-grid refinement to better handle multi-scale features \citep{LE2021100176}. Such approaches could potentially alleviate the aliasing issues observed in this work.

\subsection{Temporal Subsampling}
\label{temporal_subsampling}
We now examine the impact of temporal resolution on the accuracy of the neural solver. Similarly to the spatial subsampling strategy, we train models on datasets where intermediate time frames are omitted. A temporal subsampling factor \gls{st} means the network sees only every \gls{st}-th simulation frame during training while still forecasting a $\gls{steps}=10$ step horizon. We experimented with $\gls{st}=1$ (no temporal thinning, i.e. $\gls{delta_t}=\SI{0.01}{\second}$ per step), up to $\gls{st}=6$ (using a $\gls{delta_t}= \SI{0.06}{\second}$ step, which for a $10$-step rollout gives the $\SI{0.6}{\second}$ horizon of DS$_2^{60}$). This tests the  tolerance of the model to missing temporal information, where larger \gls{st} forces it to bridge bigger gaps between input frames, which may challenge its ability to capture fast transient phenomena like pressure spikes or small splashes. On the other hand, training with a coarser time step can act as a form of regularization and reduce the total number of frames needed, which is advantageous for computational efficiency.

Table \ref{tab:combined_temporal_subsampling} summarizes the relative $L_2$ errors for velocity $(\gls{u},\gls{v})$, pressure (\gls{p}), and volume fraction (\gls{alpha}) fields on the test set, for models trained with various \gls{st}, using the datasets DS$_1$ and DS$_2$. We see a small, approximately linear degradation in accuracy as the temporal subsampling factor increases. On DS$_1$, the mean error grows from \SI{3.97}{\percent} at $\gls{st}=1$ to \SI{6.07}{\percent} at $\gls{st}=6$. On DS$_2$, the mean error goes from \SIrange[range-phrase={ to }]{4.52}{6.41}{\percent} over the same range. This is a less expressive decline compared to the spatial subsampling case. In fact, the network seems relatively robust up to about $\gls{st}=3$. Beyond that, errors increase but remain within a narrow band for $\gls{st}=\{4,5,6\}$. One interesting feature is a noticeable peak in pressure error at $\gls{st}=2$. A similar bump is seen for DS$_2$. This suggests a non-monotonic behavior where skipping every other frame ($\gls{st}=2$) misaligns some dynamics and causes a worst-case phase error, whereas skipping more frames leads the model to effectively learn on a consistently coarser timeline, to which it can adapt. In other words, there may be a resonance or aliasing effect at $\gls{st}=2$ where the temporal frequency of certain oscillations, e.g. pressure waves or vortex shedding, is poorly sampled, causing a more pronounced error. Once \gls{st} is larger, those high-frequency modes are entirely filtered out, and the model learns the slower effective dynamics, resulting in slightly better error than the $\gls{st}=2$ case. This phenomenon underscores that the relationship between temporal resolution and accuracy is not strictly linear and that it depends on the interplay between the characteristic timescales of the simulation and the sampling interval.

\begin{table}[bth] 
\newcolumntype{J}{>{\centering\arraybackslash}p{.71cm}} 
\centering
\caption{\textbf{Temporal subsampling performance for datasets DS$_1$ and DS$_2$}. Relative $L_2$ errors (\SI{}{\percent}) are reported for each temporal sampling factor and predicted variable. A temporal subsampling factor of \gls{st} $= 1$ means that we are considering the full temporal resolution available in the dataset, whereas a factor of \gls{st} $= 6$ means considering only one every sixth frame.}
\vspace{.5em}
\small
\setlength{\tabcolsep}{4pt}
\renewcommand{\arraystretch}{.8}

% Layout:
% 1 = s_t
% 2 = spacer
% groups of (2 values + spacer), repeated 5 times = u, v, p, alpha, mean
\begin{tabularx}{.95\textwidth}{
  c J
  cc J
  cc J
  cc J
  cc J
  cc
}
\toprule
\multirow{3}{*}{\makecell{\bfseries Temporal\\\bfseries subsampling\\\bfseries factor $\boldsymbol{s_t}$}}
  & \multicolumn{15}{c}{\bfseries Relative $L_2$ Error [\SI{}{\percent}]} \\
\cmidrule{3-16}

  &  % spacer
  & \multicolumn{2}{c}{$\boldsymbol{u}$}
  &  % spacer
  & \multicolumn{2}{c}{$\boldsymbol{v}$}
  &  % spacer
  & \multicolumn{2}{c}{$\boldsymbol{p}$}
  &  % spacer
  & \multicolumn{2}{c}{$\boldsymbol{\alpha}$}
  &  % spacer
  & \multicolumn{2}{c}{\textbf{Mean}} \\
\cmidrule(lr){3-4}\cmidrule(lr){6-7}\cmidrule(lr){9-10}\cmidrule(lr){12-13}\cmidrule(lr){15-16}
  &  % spacer
  &  \textbf{DS$_1$} & \textbf{DS$_2$}
  &  % spacer
  & \textbf{DS$_1$} & \textbf{DS$_2$}
  &  % spacer
  &  \textbf{DS$_1$} & \textbf{DS$_2$}
  &  % spacer
  &  \textbf{DS$_1$} & \textbf{DS$_2$}
  &  % spacer
  &  \textbf{DS$_1$} & \textbf{DS$_2$} \\
\midrule
$1$ & & $3.96$ & $4.49$ & & $3.92$ & $4.52$ & & $4.58$ & $5.15$ & & $3.40$ & $3.90$ & & $\mathbf{3.97}$ & $\mathbf{4.52}$ \\
$2$ & & $5.56$ & $5.79$ & & $5.45$ & $5.81$ & & $8.72$ & $8.22$ & & $4.97$ & $5.32$ & & $\mathbf{6.18}$ & $\mathbf{6.29}$ \\
$3$ & & $5.91$ & $6.07$ & & $5.94$ & $6.14$ & & $6.87$ & $6.94$ & & $5.18$ & $5.59$ & & $\mathbf{5.98}$ & $\mathbf{6.19}$ \\
$4$ & & $6.15$ & $6.32$ & & $6.15$ & $6.49$ & & $6.54$ & $6.74$ & & $5.21$ & $5.72$ & & $\mathbf{6.01}$ & $\mathbf{6.32}$ \\
$5$ & & $6.18$ & $6.47$ & & $6.31$ & $6.63$ & & $6.16$ & $6.54$ & & $5.42$ & $5.76$ & & $\mathbf{6.02}$ & $\mathbf{6.35}$ \\
$6$ & & $6.29$ & $6.60$ & & $6.58$ & $6.88$ & & $6.07$ & $6.46$ & & $5.34$ & $5.71$ & & $\mathbf{6.07}$ & $\mathbf{6.41}$ \\
\bottomrule
\end{tabularx}
\label{tab:combined_temporal_subsampling}
\end{table}

Overall, the temporal subsampling experiments demonstrate that our neural solver maintains a reasonable accuracy even when trained on data that is temporally sparse. The velocity and interface variables are especially resilient with errors increasing only gradually with \gls{st}. This indicates that the network can internally infer a continuous trajectory through time despite only seeing widely spaced simulation time steps. The pressure field, as expected, is more sensitive to temporal coarsening and displays errors rising more noticeably since rapid pressure transients may be missed at large \gls{delta_t}. Yet, even for pressure, the errors plateau for $\gls{st}\ge3$, suggesting the model does not completely fail but rather converges to solving a slightly smoothed version of the physics. From a practical standpoint, this result is promising and implies one can trade some temporal resolution for speed or data volume reduction, without catastrophic loss of accuracy. Indeed, training with $\gls{st}>1$ effectively means needing fewer time steps from expensive CFD simulations, which can save considerable computational resources. In applications like casting where certain integrated outcomes, e.g. final fill time, overall flow pattern, are more important, a slightly time-thinned training regime could be an acceptable compromise. We caution, however, that if one needs to capture very fast transients, high temporal resolution remains important. In summary, within the \SI{0.6}{\second} prediction horizon studied, our neural solver exhibits robustness to missing frames.

%Our findings also resonate with observations in other studies that coarser temporal training can act as a stabilizer and encourage the neural network to focus on the slower manifold of the system while ignoring small high-frequency oscillations that might destabilize long-term predictions \citep{oommen2022learning}

\subsection{Training Data}
\label{training_data}
Finally, we analyze the  data efficiency of the proposed Fourier-Graph neural solver by varying the size of the training dataset. Gathering extensive CFD simulation data can be costly, so it is important to know how performance degrades if we train with less data, and whether there are diminishing returns to adding more training samples. Starting from the full DS$_2$ dataset, we create reduced training sets containing  \qtylist[list-final-separator={ and }]{90;80;70;60;50}{\percent} of the total number of samples. Each subset is chosen by uniform random sampling without replacement of the DS$_2$ simulations, and we train five separate model instances for each subset size using different random draws to account for any sample selection bias. The network architecture and training hyperparameters remain the same as in Section \ref{objective_optmization} for all these experiments. We evaluate each model on the same held-out test set and report the mean and standard deviation of the relative $L_2$ mean percentage errors over the five runs. Table \ref{tab:training_subsampling} presents the results of this study. 
\begin{table}[bht] 
\newcolumntype{K}{>{\centering\arraybackslash}p{0.68cm}} % adjust as desired\textbf{}
    \centering
    \caption{\textbf{Data efficiency on DS$_2$}: Relative $L_2$ error (\SI{}{\percent}) on the held-out test set as a function of training-set size. Entries are mean $\pm$ standard deviation over five random seeds. Percentages indicate the fraction of DS$_2$ used for training (uniform subsampling without replacement). All runs use identical model architecture and training recipe.}
    \vspace{.5em}
    \small
    \setlength{\tabcolsep}{3pt}
    \renewcommand{\arraystretch}{1}

    % Layout: 1 = percentage column, then V spacer, then single columns for u, v, p, alpha, Mean, each separated by V
    \begin{tabularx}{.95\textwidth}{
      c K
      c K
      c K
      c K
      c K
      c
    }
    \toprule
     \multirow[c]{2.5}{*}{\makecell[c]{\bfseries Percentage from \\ \bfseries total data [\SI{}{\percent}]}} & \multicolumn{10}{c}{\bfseries Relative $L_2$ Error [\SI{}{\percent}]} \\
    \cmidrule(lr){3-11}    
      &  % spacer
      & \textbf{$u$}
      &  % spacer
      & \textbf{$v$}
      &  % spacer
      & \textbf{$p$}
      &  % spacer
      & \textbf{$\alpha$}
      &  % spacer
      & \textbf{Mean} \\
    % underline only the variable columns, skipping V spacers
    \cmidrule(lr){1-1}\cmidrule(lr){3-3}\cmidrule(lr){5-5}\cmidrule(lr){7-7}\cmidrule(lr){9-9}\cmidrule(lr){11-11}
    % Data rows (leave all V spacer cells empty)
    $100$ & & $4.49$ & & $4.52$ & & $5.15$ & & $3.90$ & & $\mathbf{4.52}$ \\
    $90$ & & $6.37 \pm 0.06 $ & & $6.61  \pm 0.07 $ & & $6.66  \pm 0.06 $ & & $5.57  \pm 0.14 $ & & $\mathbf{6.30}  \pm \mathbf{0.04} $ \\
    $80$ & & $6.44  \pm 0.05 $ & & $6.66  \pm 0.05 $ & & $6.59  \pm 0.17 $ & & $5.67  \pm 0.08 $ & & $\mathbf{6.34}  \pm \mathbf{0.05} $ \\
    $70$ & & $6.50  \pm 0.03 $ & & $6.75  \pm 0.06 $ & & $6.75  \pm 0.11 $ & & $5.70  \pm 0.04 $ & & $\mathbf{6.43} \pm \mathbf{0.03} $ \\
    $60$ & & $6.51  \pm 0.03 $ & & $6.77  \pm 0.06 $ & & $6.75  \pm 0.16 $ & & $5.77  \pm 0.06 $ & & $\mathbf{6.45}  \pm \mathbf{0.05} $ \\
    $5$ & & $6.67  \pm 0.08 $ & & $6.93  \pm 0.10 $ & & $6.67  \pm 0.06 $ & & $5.96  \pm 0.12 $ & & $\mathbf{6.56}  \pm \mathbf{0.05} $ \\
    \bottomrule
    \end{tabularx}
    \label{tab:training_subsampling}
\end{table}

When considering \SI{100}{\percent} of the DS$_2$ dataset, our proposed model achieves a mean error of \SI{4.52}{\percent} on the test set. Reduction to \SI{90}{\percent} of the data leads to a mean error of about \SI[separate-uncertainty=true,multi-part-units=single]{6.30(0.04)}{\percent}. At \SI{80}{\percent} data, \SI[separate-uncertainty=true,multi-part-units=single]{6.34(0.05)}{\percent}. This trend continues until \SI{50}{\percent}, where the model accuracy achieves \SI[separate-uncertainty=true,multi-part-units=single]{6.56(0.05)}{\percent}. Several observations can be made about these outcomes. First, the overall performance degrades continuously and smoothly as the training set is reduced below \SI{90}{\percent}. The mean error increases roughly linear with decreasing data, and even with only half the data the error is still within $\approx2\SI{2}{\percent}$ of the full-data case. This smooth trend suggests that our model has not yet hit a data saturation point in the regime tested, and that it can still learn somewhat from each additional sample, but returns diminish as more data is added. In fact, the improvements beyond about \SIrange[]{70}{80}{\percent} of the data are very minor (the difference between \SI{80}{\percent} and \SI{100}{\percent} is only $0.18$ percentage points in mean error). This implies diminishing returns, and that beyond a certain dataset size, additional simulations contribute little new information, and the generalization of the model error flattens out. A related observation is the low variance across the five random subsets at each level, where the standard deviation of error is on the order of \SI{0.05}{\percent} or less in most cases, meaning the performance of the model is quite reproducible and not overly dependent on exactly which samples are picked. Together, these points indicate that the training process is stable and that the data samples in DS$_2$ are not highly redundant but also not vastly unique. In practical terms, this suggests that careful curation of the training set may be more valuable than throwing vast amounts of training data. For example, ensuring that the most physically distinct scenarios, e.g. different ingate positions, flow rates, etc., are included might achieve near-optimal accuracy without needing every possible simulation. Our findings support the view that targeted data selection or augmentation could outperform brute-force data generation in efficiency. Similar conclusions have been reached by other researchers. \citet{wen2022u} report that their U-\gls{FNO}-based surrogate for \ce{CO2} injection needed only one-third the data of a standard CNN to reach the same accuracy, and other work on multi-fidelity training has shown one can combine a few high-resolution simulations with many low-resolution ones to train effective models at reduced cost \citep{CUI20251736}. In our case, using \SIrange[]{50}{70}{\percent} of the dataset with smart sampling might be sufficient to achieve within $\approx0.1$ of the minimum attainable error. Beyond about \SI{75}{\percent} of the data, the effort spent generating more simulations might be better invested in diversifying conditions or improving the model rather than increasing the amount of training data.

In summary, the data-reduction study underscores that the proposed Fourier-Graph neural solver is data-efficient. Not only it does not require an extremely large simulation dataset to learn the underlying physics, but it also generalizes well even when trained on a subset of the full dataset. This is encouraging for real-world deployment, since obtaining simulation or experimental data is often a limiting factor. The small error increment under data removal suggests that the operator-learning approach is effectively extracting the key patterns from the flow physics. Once these are learned, additional examples yield diminishing returns. This aligns with the view that neural operators learn function-to-function mappings that interpolate smoothly between training scenarios. We emphasize that this does not imply that additional data are unnecessary. If parts of the parameter space are unsampled, the model will fail there. Rather, our results suggest that once the training distribution adequately covers the space, returns diminish. Future work could explore active learning strategies, where the model itself identifies which new simulations would most reduce its uncertainty, thereby optimizing data collection.

%% file: Plots/Results/colorbar_v0_test.pdf_tex
%% Creator: Inkscape 1.4.2 (f4327f4, 2025-05-13), www.inkscape.org
%% Accompanies image file 'colorbar_v0_test.pdf' (pdf, eps, ps)
\begingroup%
  \makeatletter%
  \providecommand\color[2][]{%
    \errmessage{(Inkscape) Color is used for the text in Inkscape, but the package 'color.sty' is not loaded}%
    \renewcommand\color[2][]{}%
  }%
  \providecommand\transparent[1]{%
    \errmessage{(Inkscape) Transparency is used (non-zero) for the text in Inkscape, but the package 'transparent.sty' is not loaded}%
    \renewcommand\transparent[1]{}%
  }%
  \providecommand\rotatebox[2]{#2}%
  \newcommand*\fsize{\dimexpr\f@size pt\relax}%
  \newcommand*\lineheight[1]{\fontsize{\fsize}{#1\fsize}\selectfont}%
  \ifx\svgwidth\undefined%
    \setlength{\unitlength}{123.71874619bp}%
    \ifx\svgscale\undefined%
      \relax%
    \else%
      \setlength{\unitlength}{\unitlength * \real{\svgscale}}%
    \fi%
  \else%
    \setlength{\unitlength}{\svgwidth}%
  \fi%
  \global\let\svgwidth\undefined%
  \global\let\svgscale\undefined%
  \makeatother%
  \begin{picture}(1,2.31783799)%
    \lineheight{1}%
    \setlength\tabcolsep{0pt}%
    \put(0,0){\includegraphics[width=\unitlength,page=1]{Plots/Results/colorbar_v0_test.pdf}}%
    \put(0.78209146,0.01679727){\makebox(0,0)[lt]{\lineheight{1.25}\smash{\begin{tabular}[t]{l}0.00\end{tabular}}}}%
    \put(0,0){\includegraphics[width=\unitlength,page=2]{Plots/Results/colorbar_v0_test.pdf}}%
    \put(0.78209146,0.21595866){\makebox(0,0)[lt]{\lineheight{1.25}\smash{\begin{tabular}[t]{l}0.18\end{tabular}}}}%
    \put(0,0){\includegraphics[width=\unitlength,page=3]{Plots/Results/colorbar_v0_test.pdf}}%
    \put(0.78209146,0.41512005){\makebox(0,0)[lt]{\lineheight{1.25}\smash{\begin{tabular}[t]{l}0.36\end{tabular}}}}%
    \put(0,0){\includegraphics[width=\unitlength,page=4]{Plots/Results/colorbar_v0_test.pdf}}%
    \put(0.78209146,0.6142815){\makebox(0,0)[lt]{\lineheight{1.25}\smash{\begin{tabular}[t]{l}0.55\end{tabular}}}}%
    \put(0,0){\includegraphics[width=\unitlength,page=5]{Plots/Results/colorbar_v0_test.pdf}}%
    \put(0.78209146,0.81344289){\makebox(0,0)[lt]{\lineheight{1.25}\smash{\begin{tabular}[t]{l}0.73\end{tabular}}}}%
    \put(0,0){\includegraphics[width=\unitlength,page=6]{Plots/Results/colorbar_v0_test.pdf}}%
    \put(0.78209146,1.01260428){\makebox(0,0)[lt]{\lineheight{1.25}\smash{\begin{tabular}[t]{l}0.91\end{tabular}}}}%
    \put(0,0){\includegraphics[width=\unitlength,page=7]{Plots/Results/colorbar_v0_test.pdf}}%
    \put(0.78209146,1.21176573){\makebox(0,0)[lt]{\lineheight{1.25}\smash{\begin{tabular}[t]{l}1.09\end{tabular}}}}%
    \put(0,0){\includegraphics[width=\unitlength,page=8]{Plots/Results/colorbar_v0_test.pdf}}%
    \put(0.78209146,1.41092712){\makebox(0,0)[lt]{\lineheight{1.25}\smash{\begin{tabular}[t]{l}1.28\end{tabular}}}}%
    \put(0,0){\includegraphics[width=\unitlength,page=9]{Plots/Results/colorbar_v0_test.pdf}}%
    \put(0.78209146,1.61008851){\makebox(0,0)[lt]{\lineheight{1.25}\smash{\begin{tabular}[t]{l}1.46\end{tabular}}}}%
    \put(0,0){\includegraphics[width=\unitlength,page=10]{Plots/Results/colorbar_v0_test.pdf}}%
    \put(0.78209146,1.80924995){\makebox(0,0)[lt]{\lineheight{1.25}\smash{\begin{tabular}[t]{l}1.64\end{tabular}}}}%
    \put(0.64,2.05){\makebox(0,0)[lt]{\lineheight{1.25}\smash{\begin{tabular}[t]{c}Velocity\\$[\si{\milli\metre\per\second}]$\end{tabular}}}}%
  \end{picture}%
\endgroup%

%% file: Plots/Results/ind1162_t0_error_rt.pdf_tex
%% Creator: Inkscape 1.4.2 (f4327f4, 2025-05-13), www.inkscape.org
%% Accompanies image file 'ind1162_t0_error.pdf' (pdf, eps, ps)
\begingroup%
  \makeatletter%
  \providecommand\color[2][]{%
    \errmessage{(Inkscape) Color is used for the text in Inkscape, but the package 'color.sty' is not loaded}%
    \renewcommand\color[2][]{}%
  }%
  \providecommand\transparent[1]{%
    \errmessage{(Inkscape) Transparency is used (non-zero) for the text in Inkscape, but the package 'transparent.sty' is not loaded}%
    \renewcommand\transparent[1]{}%
  }%
  \providecommand\rotatebox[2]{#2}%
  \newcommand*\fsize{\dimexpr\f@size pt\relax}%
  \newcommand*\lineheight[1]{\fontsize{\fsize}{#1\fsize}\selectfont}%
  \ifx\svgwidth\undefined%
    \setlength{\unitlength}{223.20000458bp}%
    \ifx\svgscale\undefined%
      \relax%
    \else%
      \setlength{\unitlength}{\unitlength * \real{\svgscale}}%
    \fi%
  \else%
    \setlength{\unitlength}{\svgwidth}%
  \fi%
  \global\let\svgwidth\undefined%
  \global\let\svgscale\undefined%
  \makeatother%
  \begin{picture}(1,1.49032251)%
    \lineheight{1}%
    \setlength\tabcolsep{0pt}%
    \put(0,0){\includegraphics[width=\unitlength,page=1]{Plots/Results/ind1162_t0_error_rt.pdf}}%
    %\put(0.15,0.33){\makebox(0,0)[lt]{\lineheight{1}\smash{\begin{tabular}[t]{l}\textcolor{white}{$r_t = 18.24\%$}\end{tabular}}}}%
  \end{picture}%
\endgroup%

%% file: Plots/Results/ind1162_t1_error_rt.pdf_tex
%% Creator: Inkscape 1.4.2 (f4327f4, 2025-05-13), www.inkscape.org
%% Accompanies image file 'ind1162_t1_error_rt.pdf' (pdf, eps, ps)
\begingroup%
  \makeatletter%
  \providecommand\color[2][]{%
    \errmessage{(Inkscape) Color is used for the text in Inkscape, but the package 'color.sty' is not loaded}%
    \renewcommand\color[2][]{}%
  }%
  \providecommand\transparent[1]{%
    \errmessage{(Inkscape) Transparency is used (non-zero) for the text in Inkscape, but the package 'transparent.sty' is not loaded}%
    \renewcommand\transparent[1]{}%
  }%
  \providecommand\rotatebox[2]{#2}%
  \newcommand*\fsize{\dimexpr\f@size pt\relax}%
  \newcommand*\lineheight[1]{\fontsize{\fsize}{#1\fsize}\selectfont}%
  \ifx\svgwidth\undefined%
    \setlength{\unitlength}{223.20000458bp}%
    \ifx\svgscale\undefined%
      \relax%
    \else%
      \setlength{\unitlength}{\unitlength * \real{\svgscale}}%
    \fi%
  \else%
    \setlength{\unitlength}{\svgwidth}%
  \fi%
  \global\let\svgwidth\undefined%
  \global\let\svgscale\undefined%
  \makeatother%
  \begin{picture}(1,1.49032251)%
    \lineheight{1}%
    \setlength\tabcolsep{0pt}%
    \put(0,0){\includegraphics[width=\unitlength,page=1]{Plots/Results/ind1162_t1_error_rt.pdf}}%
    %\put(0.15,0.33){\makebox(0,0)[lt]{\lineheight{1}\smash{\begin{tabular}[t]{l}\textcolor{white}{$r_t = 44.73\%$}\end{tabular}}}}%
  \end{picture}%
\endgroup%

%% file: Plots/Results/ind1162_t2_error_rt.pdf_tex
%% Creator: Inkscape 1.4.2 (f4327f4, 2025-05-13), www.inkscape.org
%% Accompanies image file 'ind1162_t2_error.pdf' (pdf, eps, ps)
\begingroup%
  \makeatletter%
  \providecommand\color[2][]{%
    \errmessage{(Inkscape) Color is used for the text in Inkscape, but the package 'color.sty' is not loaded}%
    \renewcommand\color[2][]{}%
  }%
  \providecommand\transparent[1]{%
    \errmessage{(Inkscape) Transparency is used (non-zero) for the text in Inkscape, but the package 'transparent.sty' is not loaded}%
    \renewcommand\transparent[1]{}%
  }%
  \providecommand\rotatebox[2]{#2}%
  \newcommand*\fsize{\dimexpr\f@size pt\relax}%
  \newcommand*\lineheight[1]{\fontsize{\fsize}{#1\fsize}\selectfont}%
  \ifx\svgwidth\undefined%
    \setlength{\unitlength}{223.20000458bp}%
    \ifx\svgscale\undefined%
      \relax%
    \else%
      \setlength{\unitlength}{\unitlength * \real{\svgscale}}%
    \fi%
  \else%
    \setlength{\unitlength}{\svgwidth}%
  \fi%
  \global\let\svgwidth\undefined%
  \global\let\svgscale\undefined%
  \makeatother%
  \begin{picture}(1,1.49032251)%
    \lineheight{1}%
    \setlength\tabcolsep{0pt}%
    \put(0,0){\includegraphics[width=\unitlength,page=1]{Plots/Results/ind1162_t2_error_rt.pdf}}%
    %\put(0.15,0.33){\makebox(0,0)[lt]{\lineheight{1}\smash{\begin{tabular}[t]{l}\textcolor{white}{$r_t = 34.16\%$}\end{tabular}}}}%
  \end{picture}%
\endgroup%

%% file: Plots/Results/ind1162_t3_error_rt.pdf_tex
%% Creator: Inkscape 1.4.2 (f4327f4, 2025-05-13), www.inkscape.org
%% Accompanies image file 'ind1162_t3_error.pdf' (pdf, eps, ps)
\begingroup%
  \makeatletter%
  \providecommand\color[2][]{%
    \errmessage{(Inkscape) Color is used for the text in Inkscape, but the package 'color.sty' is not loaded}%
    \renewcommand\color[2][]{}%
  }%
  \providecommand\transparent[1]{%
    \errmessage{(Inkscape) Transparency is used (non-zero) for the text in Inkscape, but the package 'transparent.sty' is not loaded}%
    \renewcommand\transparent[1]{}%
  }%
  \providecommand\rotatebox[2]{#2}%
  \newcommand*\fsize{\dimexpr\f@size pt\relax}%
  \newcommand*\lineheight[1]{\fontsize{\fsize}{#1\fsize}\selectfont}%
  \ifx\svgwidth\undefined%
    \setlength{\unitlength}{223.20000458bp}%
    \ifx\svgscale\undefined%
      \relax%
    \else%
      \setlength{\unitlength}{\unitlength * \real{\svgscale}}%
    \fi%
  \else%
    \setlength{\unitlength}{\svgwidth}%
  \fi%
  \global\let\svgwidth\undefined%
  \global\let\svgscale\undefined%
  \makeatother%
  \begin{picture}(1,1.49032251)%
    \lineheight{1}%
    \setlength\tabcolsep{0pt}%
    \put(0,0){\includegraphics[width=\unitlength,page=1]{Plots/Results/ind1162_t3_error_rt.pdf}}%
    %\put(0.15,0.33){\makebox(0,0)[lt]{\lineheight{1}\smash{\begin{tabular}[t]{l}\textcolor{white}{$r_t = 24.43\%$}\end{tabular}}}}%
  \end{picture}%
\endgroup%

%% file: Plots/Results/ind1162_t6_error_rt.pdf_tex
%% Creator: Inkscape 1.4.2 (f4327f4, 2025-05-13), www.inkscape.org
%% Accompanies image file 'ind1162_t6_error.pdf' (pdf, eps, ps)
\begingroup%
  \makeatletter%
  \providecommand\color[2][]{%
    \errmessage{(Inkscape) Color is used for the text in Inkscape, but the package 'color.sty' is not loaded}%
    \renewcommand\color[2][]{}%
  }%
  \providecommand\transparent[1]{%
    \errmessage{(Inkscape) Transparency is used (non-zero) for the text in Inkscape, but the package 'transparent.sty' is not loaded}%
    \renewcommand\transparent[1]{}%
  }%
  \providecommand\rotatebox[2]{#2}%
  \newcommand*\fsize{\dimexpr\f@size pt\relax}%
  \newcommand*\lineheight[1]{\fontsize{\fsize}{#1\fsize}\selectfont}%
  \ifx\svgwidth\undefined%
    \setlength{\unitlength}{223.20000458bp}%
    \ifx\svgscale\undefined%
      \relax%
    \else%
      \setlength{\unitlength}{\unitlength * \real{\svgscale}}%
    \fi%
  \else%
    \setlength{\unitlength}{\svgwidth}%
  \fi%
  \global\let\svgwidth\undefined%
  \global\let\svgscale\undefined%
  \makeatother%
  \begin{picture}(1,1.49032251)%
    \lineheight{1}%
    \setlength\tabcolsep{0pt}%
    \put(0,0){\includegraphics[width=\unitlength,page=1]{Plots/Results/ind1162_t6_error_rt.pdf}}%
    %\put(0.15,0.33){\makebox(0,0)[lt]{\lineheight{1}\smash{\begin{tabular}[t]{l}\textcolor{white}{$r_t = 28.25\%$}\end{tabular}}}}%
  \end{picture}%
\endgroup%

%% file: Plots/Results/ind1162_t9_error_rt.pdf_tex
%% Creator: Inkscape 1.4.2 (f4327f4, 2025-05-13), www.inkscape.org
%% Accompanies image file 'ind1162_t9_error.pdf' (pdf, eps, ps)
\begingroup%
  \makeatletter%
  \providecommand\color[2][]{%
    \errmessage{(Inkscape) Color is used for the text in Inkscape, but the package 'color.sty' is not loaded}%
    \renewcommand\color[2][]{}%
  }%
  \providecommand\transparent[1]{%
    \errmessage{(Inkscape) Transparency is used (non-zero) for the text in Inkscape, but the package 'transparent.sty' is not loaded}%
    \renewcommand\transparent[1]{}%
  }%
  \providecommand\rotatebox[2]{#2}%
  \newcommand*\fsize{\dimexpr\f@size pt\relax}%
  \newcommand*\lineheight[1]{\fontsize{\fsize}{#1\fsize}\selectfont}%
  \ifx\svgwidth\undefined%
    \setlength{\unitlength}{223.20000458bp}%
    \ifx\svgscale\undefined%
      \relax%
    \else%
      \setlength{\unitlength}{\unitlength * \real{\svgscale}}%
    \fi%
  \else%
    \setlength{\unitlength}{\svgwidth}%
  \fi%
  \global\let\svgwidth\undefined%
  \global\let\svgscale\undefined%
  \makeatother%
  \begin{picture}(1,1.49032251)%
    \lineheight{1}%
    \setlength\tabcolsep{0pt}%
    \put(0,0){\includegraphics[width=\unitlength,page=1]{Plots/Results/ind1162_t9_error_rt.pdf}}%
    %\put(0.15,0.33){\makebox(0,0)[lt]{\lineheight{1}\smash{\begin{tabular}[t]{l}\textcolor{white}{$r_t = 38.36\%$}\end{tabular}}}}%
  \end{picture}%
\endgroup%

%% file: Plots/Results/Prediction_overlay_15.pdf_tex
%% Creator: Inkscape 1.4.2 (f4327f4, 2025-05-13), www.inkscape.org
%% Accompanies image file 'Prediction_overlay_15.pdf' (pdf, eps, ps)
\begingroup%
  \makeatletter%
  \providecommand\color[2][]{%
    \errmessage{(Inkscape) Color is used for the text in Inkscape, but the package 'color.sty' is not loaded}%
    \renewcommand\color[2][]{}%
  }%
  \providecommand\transparent[1]{%
    \errmessage{(Inkscape) Transparency is used (non-zero) for the text in Inkscape, but the package 'transparent.sty' is not loaded}%
    \renewcommand\transparent[1]{}%
  }%
  \providecommand\rotatebox[2]{#2}%
  \newcommand*\fsize{\dimexpr\f@size pt\relax}%
  \newcommand*\lineheight[1]{\fontsize{\fsize}{#1\fsize}\selectfont}%
  \ifx\svgwidth\undefined%
    \setlength{\unitlength}{1\linewidth}%
    \ifx\svgscale\undefined%
      \relax%
    \else%
      \setlength{\unitlength}{\unitlength * \real{\svgscale}}%
    \fi%
  \else%
    \setlength{\unitlength}{\svgwidth}%
  \fi%
  \global\let\svgwidth\undefined%
  \global\let\svgscale\undefined%
  \makeatother%
  \begin{picture}(1,0.90916326)%
    \lineheight{1}%
    \setlength\tabcolsep{0pt}%
    \put(0,0){\includegraphics[width=\unitlength,page=1]{Plots/Results/Prediction_overlay_15.pdf}}%
    \put(0.42247429,0.71763677){\color[rgb]{1,1,1}\transparent{1}\makebox(0,0)[lt]{\lineheight{1.25}\smash{\begin{tabular}[t]{l}t=1\end{tabular}}}}%
    \put(0.44580176,0.63210264){\color[rgb]{1,1,1}\transparent{1}\makebox(0,0)[lt]{\lineheight{1.25}\smash{\begin{tabular}[t]{l}t=3\end{tabular}}}}%
    \put(0.46912926,0.52324102){\color[rgb]{1,1,1}\transparent{1}\makebox(0,0)[lt]{\lineheight{1.25}\smash{\begin{tabular}[t]{l}t=5\end{tabular}}}}%
    \put(0.48468093,0.4182673){\color[rgb]{1,1,1}\transparent{1}\makebox(0,0)[lt]{\lineheight{1.25}\smash{\begin{tabular}[t]{l}t=7\end{tabular}}}}%
    \put(0.48468093,0.31329359){\color[rgb]{1,1,1}\transparent{1}\makebox(0,0)[lt]{\lineheight{1.25}\smash{\begin{tabular}[t]{l}t=9\end{tabular}}}}%
    \put(0,0){\includegraphics[width=\unitlength,page=2]{Plots/Results/Prediction_overlay_15.pdf}}%
    \put(0.01876925,0.06142713){\makebox(0,0)[lt]{\lineheight{1.25}\smash{\begin{tabular}[t]{l}0.0\end{tabular}}}}%
    \put(0,0){\includegraphics[width=\unitlength,page=3]{Plots/Results/Prediction_overlay_15.pdf}}%
    \put(0.01876925,0.19963785){\makebox(0,0)[lt]{\lineheight{1.25}\smash{\begin{tabular}[t]{l}0.2\end{tabular}}}}%
    \put(0,0){\includegraphics[width=\unitlength,page=4]{Plots/Results/Prediction_overlay_15.pdf}}%
    \put(0.01876925,0.33784856){\makebox(0,0)[lt]{\lineheight{1.25}\smash{\begin{tabular}[t]{l}0.4\end{tabular}}}}%
    \put(0,0){\includegraphics[width=\unitlength,page=5]{Plots/Results/Prediction_overlay_15.pdf}}%
    \put(0.01876925,0.47605928){\makebox(0,0)[lt]{\lineheight{1.25}\smash{\begin{tabular}[t]{l}0.6\end{tabular}}}}%
    \put(0,0){\includegraphics[width=\unitlength,page=6]{Plots/Results/Prediction_overlay_15.pdf}}%
    \put(0.01876925,0.61426998){\makebox(0,0)[lt]{\lineheight{1.25}\smash{\begin{tabular}[t]{l}0.8\end{tabular}}}}%
    \put(0,0){\includegraphics[width=\unitlength,page=7]{Plots/Results/Prediction_overlay_15.pdf}}%
    \put(0.01876925,0.75248071){\makebox(0,0)[lt]{\lineheight{1.25}\smash{\begin{tabular}[t]{l}1.0\end{tabular}}}}%
    \put(0,0){\includegraphics[width=\unitlength,page=8]{Plots/Results/Prediction_overlay_15.pdf}}%
    \put(0.06965382,0.8866066){\makebox(0,0)[lt]{\lineheight{1.25}\smash{\begin{tabular}[t]{l}Volume\end{tabular}}}}%
    \put(0.03765894,0.84154461){\makebox(0,0)[lt]{\lineheight{1.25}\smash{\begin{tabular}[t]{l} Fraction [-]\end{tabular}}}}%
  \end{picture}%
\endgroup%

%% file: Plots/Results/Target_overlay_15.pdf_tex
%% Creator: Inkscape 1.4.2 (f4327f4, 2025-05-13), www.inkscape.org
%% Accompanies image file 'Target_overlay_15.pdf' (pdf, eps, ps)
\begingroup%
  \makeatletter%
  \providecommand\color[2][]{%
    \errmessage{(Inkscape) Color is used for the text in Inkscape, but the package 'color.sty' is not loaded}%
    \renewcommand\color[2][]{}%
  }%
  \providecommand\transparent[1]{%
    \errmessage{(Inkscape) Transparency is used (non-zero) for the text in Inkscape, but the package 'transparent.sty' is not loaded}%
    \renewcommand\transparent[1]{}%
  }%
  \providecommand\rotatebox[2]{#2}%
  \newcommand*\fsize{\dimexpr\f@size pt\relax}%
  \newcommand*\lineheight[1]{\fontsize{\fsize}{#1\fsize}\selectfont}%
  \ifx\svgwidth\undefined%
    \setlength{\unitlength}{1\linewidth}%
    \ifx\svgscale\undefined%
      \relax%
    \else%
      \setlength{\unitlength}{\unitlength * \real{\svgscale}}%
    \fi%
  \else%
    \setlength{\unitlength}{\svgwidth}%
  \fi%
  \global\let\svgwidth\undefined%
  \global\let\svgscale\undefined%
  \makeatother%
  \begin{picture}(1,1.04314776)%
    \lineheight{1}%
    \setlength\tabcolsep{0pt}%
    \put(0,0){\includegraphics[width=\unitlength,page=1]{Plots/Results/Target_overlay_15.pdf}}%
    \put(0.30418384,0.87353462){\color[rgb]{1,1,1}\transparent{1}\makebox(0,0)[lt]{\lineheight{1.25}\smash{\begin{tabular}[t]{l}t=1\end{tabular}}}}%
    \put(0.33265192,0.76915166){\color[rgb]{1,1,1}\transparent{1}\makebox(0,0)[lt]{\lineheight{1.25}\smash{\begin{tabular}[t]{l}t=3\end{tabular}}}}%
    \put(0.36112,0.63630063){\color[rgb]{1,1,1}\transparent{1}\makebox(0,0)[lt]{\lineheight{1.25}\smash{\begin{tabular}[t]{l}t=5\end{tabular}}}}%
    \put(0.38009872,0.50819428){\color[rgb]{1,1,1}\transparent{1}\makebox(0,0)[lt]{\lineheight{1.25}\smash{\begin{tabular}[t]{l}t=7\end{tabular}}}}%
    \put(0.38009872,0.38008792){\color[rgb]{1,1,1}\transparent{1}\makebox(0,0)[lt]{\lineheight{1.25}\smash{\begin{tabular}[t]{l}t=9\end{tabular}}}}%
    \put(0.27289339,0.99657675){\color[rgb]{1,1,1}\transparent{1}\makebox(0,0)[lt]{\lineheight{1.25}\smash{\begin{tabular}[t]{l}Target contours overlay\end{tabular}}}}%
  \end{picture}%
\endgroup%

%% file: Plots/Results/rmse_plot.pdf_tex
%% Creator: Inkscape 1.4.2 (f4327f4, 2025-05-13), www.inkscape.org
%% Accompanies image file 'rmse_plot.pdf' (pdf, eps, ps)
\begingroup%
  \makeatletter%
  \providecommand\color[2][]{%
    \errmessage{(Inkscape) Color is used for the text in Inkscape, but the package 'color.sty' is not loaded}%
    \renewcommand\color[2][]{}%
  }%
  \providecommand\transparent[1]{%
    \errmessage{(Inkscape) Transparency is used (non-zero) for the text in Inkscape, but the package 'transparent.sty' is not loaded}%
    \renewcommand\transparent[1]{}%
  }%
  \providecommand\rotatebox[2]{#2}%
  \newcommand*\fsize{\dimexpr\f@size pt\relax}%
  \newcommand*\lineheight[1]{\fontsize{\fsize}{#1\fsize}\selectfont}%
  \ifx\svgwidth\undefined%
    \setlength{\unitlength}{531.15522766bp}%
    \ifx\svgscale\undefined%
      \relax%
    \else%
      \setlength{\unitlength}{\unitlength * \real{\svgscale}}%
    \fi%
  \else%
    \setlength{\unitlength}{\svgwidth}%
  \fi%
  \global\let\svgwidth\undefined%
  \global\let\svgscale\undefined%
  \makeatother%
  \begin{picture}(1,0.98841093)%
    \lineheight{1}%
    \setlength\tabcolsep{0pt}%
    \put(0,0){\includegraphics[width=\unitlength,page=1]{Plots/Results/rmse_plot.pdf}}%
    \put(0.23947321,0.212466){\makebox(0,0)[lt]{\lineheight{1.25}\smash{\begin{tabular}[t]{l}0\end{tabular}}}}%
    \put(0,0){\includegraphics[width=\unitlength,page=2]{Plots/Results/rmse_plot.pdf}}%
    \put(0.3160497,0.212466){\makebox(0,0)[lt]{\lineheight{1.25}\smash{\begin{tabular}[t]{l}1\end{tabular}}}}%
    \put(0,0){\includegraphics[width=\unitlength,page=3]{Plots/Results/rmse_plot.pdf}}%
    \put(0.39262429,0.212466){\makebox(0,0)[lt]{\lineheight{1.25}\smash{\begin{tabular}[t]{l}2\end{tabular}}}}%
    \put(0,0){\includegraphics[width=\unitlength,page=4]{Plots/Results/rmse_plot.pdf}}%
    \put(0.46920078,0.212466){\makebox(0,0)[lt]{\lineheight{1.25}\smash{\begin{tabular}[t]{l}3\end{tabular}}}}%
    \put(0,0){\includegraphics[width=\unitlength,page=5]{Plots/Results/rmse_plot.pdf}}%
    \put(0.54577538,0.212466){\makebox(0,0)[lt]{\lineheight{1.25}\smash{\begin{tabular}[t]{l}4\end{tabular}}}}%
    \put(0,0){\includegraphics[width=\unitlength,page=6]{Plots/Results/rmse_plot.pdf}}%
    \put(0.62235187,0.212466){\makebox(0,0)[lt]{\lineheight{1.25}\smash{\begin{tabular}[t]{l}5\end{tabular}}}}%
    \put(0,0){\includegraphics[width=\unitlength,page=7]{Plots/Results/rmse_plot.pdf}}%
    \put(0.69893023,0.212466){\makebox(0,0)[lt]{\lineheight{1.25}\smash{\begin{tabular}[t]{l}6\end{tabular}}}}%
    \put(0,0){\includegraphics[width=\unitlength,page=8]{Plots/Results/rmse_plot.pdf}}%
    \put(0.77550484,0.212466){\makebox(0,0)[lt]{\lineheight{1.25}\smash{\begin{tabular}[t]{l}7\end{tabular}}}}%
    \put(0,0){\includegraphics[width=\unitlength,page=9]{Plots/Results/rmse_plot.pdf}}%
    \put(0.85208132,0.212466){\makebox(0,0)[lt]{\lineheight{1.25}\smash{\begin{tabular}[t]{l}8\end{tabular}}}}%
    \put(0,0){\includegraphics[width=\unitlength,page=10]{Plots/Results/rmse_plot.pdf}}%
    \put(0.92865593,0.212466){\makebox(0,0)[lt]{\lineheight{1.25}\smash{\begin{tabular}[t]{l}9\end{tabular}}}}%
    \put(0.42825613,0.14737164){\makebox(0,0)[lt]{\lineheight{1.25}\smash{\begin{tabular}[t]{l}\textbf{Prediction step $k$ [-]}\end{tabular}}}}%
    \put(0,0){\includegraphics[width=\unitlength,page=11]{Plots/Results/rmse_plot.pdf}}%
    \put(0.1432806,0.27124958){\makebox(0,0)[lt]{\lineheight{1.25}\smash{\begin{tabular}[t]{l}0\end{tabular}}}}%
    \put(0,0){\includegraphics[width=\unitlength,page=12]{Plots/Results/rmse_plot.pdf}}%
    \put(0.1432806,0.38113836){\makebox(0,0)[lt]{\lineheight{1.25}\smash{\begin{tabular}[t]{l}5\end{tabular}}}}%
    \put(0,0){\includegraphics[width=\unitlength,page=13]{Plots/Results/rmse_plot.pdf}}%
    \put(0.13386904,0.49102526){\makebox(0,0)[lt]{\lineheight{1.25}\smash{\begin{tabular}[t]{l}10\end{tabular}}}}%
    \put(0,0){\includegraphics[width=\unitlength,page=14]{Plots/Results/rmse_plot.pdf}}%
    \put(0.13386904,0.60091216){\makebox(0,0)[lt]{\lineheight{1.25}\smash{\begin{tabular}[t]{l}15\end{tabular}}}}%
    \put(0,0){\includegraphics[width=\unitlength,page=15]{Plots/Results/rmse_plot.pdf}}%
    \put(0.13386904,0.71079905){\makebox(0,0)[lt]{\lineheight{1.25}\smash{\begin{tabular}[t]{l}20\end{tabular}}}}%
    \put(0,0){\includegraphics[width=\unitlength,page=16]{Plots/Results/rmse_plot.pdf}}%
    \put(0.13386904,0.82068783){\makebox(0,0)[lt]{\lineheight{1.25}\smash{\begin{tabular}[t]{l}25\end{tabular}}}}%
    \put(0.10170122,0.43045267){\rotatebox{90}{\makebox(0,0)[lt]{\lineheight{1.25}\smash{\begin{tabular}[t]{l}\textbf{Error $\mathbf{r_{\alpha}^k}$ [\%]}\end{tabular}}}}}%
    \put(0,0){\includegraphics[width=\unitlength,page=17]{Plots/Results/rmse_plot.pdf}}%
    \put(0.275,0.52){\makebox(0,0)[lt]{\smash{{\fontsize{7}{8}\selectfont \begin{tabular}[t]{@{}l@{}}\colorbox{white}{\textcolor{black}{$9.38\%$}}\end{tabular}}}}}
    \put(0.42,0.35){\makebox(0,0)[lt]{\smash{{\fontsize{7}{8}\selectfont \begin{tabular}[t]{@{}l@{}}\colorbox{white}{\textcolor{black}{$6.20\%$}}\end{tabular}}}}}
    \put(0.525,0.50){\makebox(0,0)[lt]{\smash{{\fontsize{7}{8}\selectfont \begin{tabular}[t]{@{}l@{}}\colorbox{white}{\textcolor{black}{$7.36\%$}}\end{tabular}}}}}
    \put(0.73,0.47){\makebox(0,0)[lt]{\smash{{\fontsize{7}{8}\selectfont \begin{tabular}[t]{@{}l@{}}\colorbox{white}{\textcolor{black}{$12.05\%$}}\end{tabular}}}}}
    \put(0.80,0.72){\makebox(0,0)[lt]{\smash{{\fontsize{7}{8}\selectfont \begin{tabular}[t]{@{}l@{}}\colorbox{white}{\textcolor{black}{$17.79\%$}}\end{tabular}}}}}

  \end{picture}%
\endgroup%

%% file: Plots/Results/colorbar3.pdf_tex
%% Creator: Inkscape 1.4.2 (f4327f4, 2025-05-13), www.inkscape.org
%% Accompanies image file 'colorbar3.pdf' (pdf, eps, ps)
\begingroup%
  \makeatletter%
  \providecommand\color[2][]{%
    \errmessage{(Inkscape) Color is used for the text in Inkscape, but the package 'color.sty' is not loaded}%
    \renewcommand\color[2][]{}%
  }%
  \providecommand\transparent[1]{%
    \errmessage{(Inkscape) Transparency is used (non-zero) for the text in Inkscape, but the package 'transparent.sty' is not loaded}%
    \renewcommand\transparent[1]{}%
  }%
  \providecommand\rotatebox[2]{#2}%
  \newcommand*\fsize{\dimexpr\f@size pt\relax}%
  \newcommand*\lineheight[1]{\fontsize{\fsize}{#1\fsize}\selectfont}%
  \ifx\svgwidth\undefined%
    \setlength{\unitlength}{1\linewidth}%
    \ifx\svgscale\undefined%
      \relax%
    \else%
      \setlength{\unitlength}{\unitlength * \real{\svgscale}}%
    \fi%
  \else%
    \setlength{\unitlength}{\svgwidth}%
  \fi%
  \global\let\svgwidth\undefined%
  \global\let\svgscale\undefined%
  \makeatother%
  \begin{picture}(1,0.33123988)%
    \lineheight{1}%
    \setlength\tabcolsep{0pt}%
    \put(0,0){\includegraphics[width=\unitlength,page=1]{Plots/Results/colorbar3.pdf}}%
    \put(0.025,0.19){\makebox(0,0)[lt]{\lineheight{1.25}\smash{\begin{tabular}[t]{l}-24\end{tabular}}}}%
    \put(0,0){\includegraphics[width=\unitlength,page=2]{Plots/Results/colorbar3.pdf}}%
    \put(0.135,0.19){\makebox(0,0)[lt]{\lineheight{1.25}\smash{\begin{tabular}[t]{l}15\end{tabular}}}}%
    \put(0,0){\includegraphics[width=\unitlength,page=3]{Plots/Results/colorbar3.pdf}}%
    \put(0.235,0.19){\makebox(0,0)[lt]{\lineheight{1.25}\smash{\begin{tabular}[t]{l}55\end{tabular}}}}%
    \put(0,0){\includegraphics[width=\unitlength,page=4]{Plots/Results/colorbar3.pdf}}%
    \put(0.336,0.19){\makebox(0,0)[lt]{\lineheight{1.25}\smash{\begin{tabular}[t]{l}94\end{tabular}}}}%
    \put(0,0){\includegraphics[width=\unitlength,page=5]{Plots/Results/colorbar3.pdf}}%
    \put(0.435,0.19){\makebox(0,0)[lt]{\lineheight{1.25}\smash{\begin{tabular}[t]{l}133\end{tabular}}}}%
    \put(0,0){\includegraphics[width=\unitlength,page=6]{Plots/Results/colorbar3.pdf}}%
    \put(0.535,0.19){\makebox(0,0)[lt]{\lineheight{1.25}\smash{\begin{tabular}[t]{l}172\end{tabular}}}}%
    \put(0.45,0.25){\makebox(0,0)[lt]{\lineheight{1}\smash{\begin{tabular}[t]{l}Pressure [Pa]\end{tabular}}}}%
    \put(0,0){\includegraphics[width=\unitlength,page=7]{Plots/Results/colorbar3.pdf}}%
    \put(0.635,0.19){\makebox(0,0)[lt]{\lineheight{1.25}\smash{\begin{tabular}[t]{l}212\end{tabular}}}}%
    \put(0,0){\includegraphics[width=\unitlength,page=8]{Plots/Results/colorbar3.pdf}}%
    \put(0.735,0.19){\makebox(0,0)[lt]{\lineheight{1.25}\smash{\begin{tabular}[t]{l}251\end{tabular}}}}%
    \put(0,0){\includegraphics[width=\unitlength,page=9]{Plots/Results/colorbar3.pdf}}%
    \put(0.835,0.19){\makebox(0,0)[lt]{\lineheight{1.25}\smash{\begin{tabular}[t]{l}290\end{tabular}}}}%
    \put(0,0){\includegraphics[width=\unitlength,page=10]{Plots/Results/colorbar3.pdf}}%
    \put(0.935,0.19){\makebox(0,0)[lt]{\lineheight{1.25}\smash{\begin{tabular}[t]{l}330\end{tabular}}}}%
    \put(0,0){\includegraphics[width=\unitlength,page=11]{Plots/Results/colorbar3.pdf}}%
    \put(0.26707868,0.135){\makebox(0,0)[lt]{\lineheight{1.25}\smash{\begin{tabular}[t]{l}10\end{tabular}}}}%
    \put(0,0){\includegraphics[width=\unitlength,page=12]{Plots/Results/colorbar3.pdf}}%
    \put(0.43189193,0.135){\makebox(0,0)[lt]{\lineheight{1.25}\smash{\begin{tabular}[t]{l}25\end{tabular}}}}%
    \put(0,0){\includegraphics[width=\unitlength,page=13]{Plots/Results/colorbar3.pdf}}%
    \put(0.59670524,0.135){\makebox(0,0)[lt]{\lineheight{1.25}\smash{\begin{tabular}[t]{l}50\end{tabular}}}}%
    \put(0,0){\includegraphics[width=\unitlength,page=14]{Plots/Results/colorbar3.pdf}}%
    \put(0.76151846,0.135){\makebox(0,0)[lt]{\lineheight{1.25}\smash{\begin{tabular}[t]{l}100\end{tabular}}}}%
    \put(0.05,0.135){\makebox(0,0)[lt]{\lineheight{1}\smash{\begin{tabular}[t]{l}Isocontour\end{tabular}}}}%
    \put(0.85,0.135){\makebox(0,0)[lt]{\lineheight{1}\smash{\begin{tabular}[t]{l}[Pa]\end{tabular}}}}%
  \end{picture}%
\endgroup%

%% file: Plots/Results/colorbar_pressure.pdf_tex
%% Creator: Inkscape 1.4.2 (f4327f4, 2025-05-13), www.inkscape.org
%% Accompanies image file 'colorbar_pressure.pdf' (pdf, eps, ps)
\begingroup%
  \makeatletter%
  \providecommand\color[2][]{%
    \errmessage{(Inkscape) Color is used for the text in Inkscape, but the package 'color.sty' is not loaded}%
    \renewcommand\color[2][]{}%
  }%
  \providecommand\transparent[1]{%
    \errmessage{(Inkscape) Transparency is used (non-zero) for the text in Inkscape, but the package 'transparent.sty' is not loaded}%
    \renewcommand\transparent[1]{}%
  }%
  \providecommand\rotatebox[2]{#2}%
  \newcommand*\fsize{\dimexpr\f@size pt\relax}%
  \newcommand*\lineheight[1]{\fontsize{\fsize}{#1\fsize}\selectfont}%
  \ifx\svgwidth\undefined%
    \setlength{\unitlength}{237.234375bp}%
    \ifx\svgscale\undefined%
      \relax%
    \else%
      \setlength{\unitlength}{\unitlength * \real{\svgscale}}%
    \fi%
  \else%
    \setlength{\unitlength}{\svgwidth}%
  \fi%
  \global\let\svgwidth\undefined%
  \global\let\svgscale\undefined%
  \makeatother%
  \begin{picture}(1,2.02009874)%
    \lineheight{1}%
    \setlength\tabcolsep{0pt}%
    \put(0,0){\includegraphics[width=\unitlength,page=1]{Plots/Results/colorbar_pressure.pdf}}%
    \put(0.65090561,1.05075409){\makebox(0,0)[lt]{\lineheight{1.25}\smash{\begin{tabular}[t]{l}-23\end{tabular}}}}%
    \put(0,0){\includegraphics[width=\unitlength,page=2]{Plots/Results/colorbar_pressure.pdf}}%
    \put(0.65090561,1.12898894){\makebox(0,0)[lt]{\lineheight{1.25}\smash{\begin{tabular}[t]{l}15\end{tabular}}}}%
    \put(0,0){\includegraphics[width=\unitlength,page=3]{Plots/Results/colorbar_pressure.pdf}}%
    \put(0.65090561,1.20722381){\makebox(0,0)[lt]{\lineheight{1.25}\smash{\begin{tabular}[t]{l}55\end{tabular}}}}%
    \put(0,0){\includegraphics[width=\unitlength,page=4]{Plots/Results/colorbar_pressure.pdf}}%
    \put(0.65090561,1.28545869){\makebox(0,0)[lt]{\lineheight{1.25}\smash{\begin{tabular}[t]{l}94\end{tabular}}}}%
    \put(0,0){\includegraphics[width=\unitlength,page=5]{Plots/Results/colorbar_pressure.pdf}}%
    \put(0.65090561,1.36369354){\makebox(0,0)[lt]{\lineheight{1.25}\smash{\begin{tabular}[t]{l}133\end{tabular}}}}%
    \put(0,0){\includegraphics[width=\unitlength,page=6]{Plots/Results/colorbar_pressure.pdf}}%
    \put(0.65090561,1.44192842){\makebox(0,0)[lt]{\lineheight{1.25}\smash{\begin{tabular}[t]{l}172\end{tabular}}}}%
    \put(0,0){\includegraphics[width=\unitlength,page=7]{Plots/Results/colorbar_pressure.pdf}}%
    \put(0.65090561,1.5201633){\makebox(0,0)[lt]{\lineheight{1.25}\smash{\begin{tabular}[t]{l}212\end{tabular}}}}%
    \put(0,0){\includegraphics[width=\unitlength,page=8]{Plots/Results/colorbar_pressure.pdf}}%
    \put(0.65090561,1.59839814){\makebox(0,0)[lt]{\lineheight{1.25}\smash{\begin{tabular}[t]{l}251\end{tabular}}}}%
    \put(0,0){\includegraphics[width=\unitlength,page=9]{Plots/Results/colorbar_pressure.pdf}}%
    \put(0.65090561,1.67663302){\makebox(0,0)[lt]{\lineheight{1.25}\smash{\begin{tabular}[t]{l}290\end{tabular}}}}%
    \put(0,0){\includegraphics[width=\unitlength,page=10]{Plots/Results/colorbar_pressure.pdf}}%
    \put(0.65090561,1.75486789){\makebox(0,0)[lt]{\lineheight{1.25}\smash{\begin{tabular}[t]{l}329\end{tabular}}}}%
    \put(0.58,1.87){\makebox(0,0)[lt]{\lineheight{0.9}\smash{\begin{tabular}[t]{c}Pressure\\$[\si{\pascal}]$\end{tabular}}}}%
    \put(0,0){\includegraphics[width=\unitlength,page=11]{Plots/Results/colorbar_pressure.pdf}}%
  \end{picture}%
\endgroup%

%% file: Plots/Results/absolute_error_map_r2.pdf_tex
%% Creator: Inkscape 1.4.2 (f4327f4, 2025-05-13), www.inkscape.org
%% Accompanies image file 'absolute_error_map_r2.pdf' (pdf, eps, ps)
\begingroup%
  \makeatletter%
  \providecommand\color[2][]{%
    \errmessage{(Inkscape) Color is used for the text in Inkscape, but the package 'color.sty' is not loaded}%
    \renewcommand\color[2][]{}%
  }%
  \providecommand\transparent[1]{%
    \errmessage{(Inkscape) Transparency is used (non-zero) for the text in Inkscape, but the package 'transparent.sty' is not loaded}%
    \renewcommand\transparent[1]{}%
  }%
  \providecommand\rotatebox[2]{#2}%
  \newcommand*\fsize{\dimexpr\f@size pt\relax}%
  \newcommand*\lineheight[1]{\fontsize{\fsize}{#1\fsize}\selectfont}%
  \ifx\svgwidth\undefined%
    \setlength{\unitlength}{352.79999542bp}%
    \ifx\svgscale\undefined%
      \relax%
    \else%
      \setlength{\unitlength}{\unitlength * \real{\svgscale}}%
    \fi%
  \else%
    \setlength{\unitlength}{\svgwidth}%
  \fi%
  \global\let\svgwidth\undefined%
  \global\let\svgscale\undefined%
  \makeatother%
  \begin{picture}(1,1)%
    \lineheight{1}%
    \setlength\tabcolsep{0pt}%
    \put(0,0){\includegraphics[width=\unitlength,page=1]{Plots/Results/absolute_error_map_r2.pdf}}%
    %\put(0.14,0.815){\makebox(0,0)[lt]{\lineheight{1}\smash{\begin{tabular}[t]{l}\textcolor{white}{$r_t = 72.93\%$}\end{tabular}}}}%
  \end{picture}%
\endgroup%

%% file: Plots/Results/absolute_error_map_r4.pdf_tex
%% Creator: Inkscape 1.4.2 (f4327f4, 2025-05-13), www.inkscape.org
%% Accompanies image file 'absolute_error_map_r4.pdf' (pdf, eps, ps)
\begingroup%
  \makeatletter%
  \providecommand\color[2][]{%
    \errmessage{(Inkscape) Color is used for the text in Inkscape, but the package 'color.sty' is not loaded}%
    \renewcommand\color[2][]{}%
  }%
  \providecommand\transparent[1]{%
    \errmessage{(Inkscape) Transparency is used (non-zero) for the text in Inkscape, but the package 'transparent.sty' is not loaded}%
    \renewcommand\transparent[1]{}%
  }%
  \providecommand\rotatebox[2]{#2}%
  \newcommand*\fsize{\dimexpr\f@size pt\relax}%
  \newcommand*\lineheight[1]{\fontsize{\fsize}{#1\fsize}\selectfont}%
  \ifx\svgwidth\undefined%
    \setlength{\unitlength}{352.79999542bp}%
    \ifx\svgscale\undefined%
      \relax%
    \else%
      \setlength{\unitlength}{\unitlength * \real{\svgscale}}%
    \fi%
  \else%
    \setlength{\unitlength}{\svgwidth}%
  \fi%
  \global\let\svgwidth\undefined%
  \global\let\svgscale\undefined%
  \makeatother%
  \begin{picture}(1,1)%
    \lineheight{1}%
    \setlength\tabcolsep{0pt}%
    \put(0,0){\includegraphics[width=\unitlength,page=1]{Plots/Results/absolute_error_map_r4.pdf}}%
    %\put(0.15,0.85){\makebox(0,0)[lt]{\lineheight{1}\smash{\begin{tabular}[t]{l}\textcolor{white}{$r_t = 36.10\%$}\end{tabular}}}}%
  \end{picture}%
\endgroup%

%% file: Plots/Results/absolute_error_map_r6.pdf_tex
%% Creator: Inkscape 1.4.2 (f4327f4, 2025-05-13), www.inkscape.org
%% Accompanies image file 'absolute_error_map_r6.pdf' (pdf, eps, ps)
\begingroup%
  \makeatletter%
  \providecommand\color[2][]{%
    \errmessage{(Inkscape) Color is used for the text in Inkscape, but the package 'color.sty' is not loaded}%
    \renewcommand\color[2][]{}%
  }%
  \providecommand\transparent[1]{%
    \errmessage{(Inkscape) Transparency is used (non-zero) for the text in Inkscape, but the package 'transparent.sty' is not loaded}%
    \renewcommand\transparent[1]{}%
  }%
  \providecommand\rotatebox[2]{#2}%
  \newcommand*\fsize{\dimexpr\f@size pt\relax}%
  \newcommand*\lineheight[1]{\fontsize{\fsize}{#1\fsize}\selectfont}%
  \ifx\svgwidth\undefined%
    \setlength{\unitlength}{352.79999542bp}%
    \ifx\svgscale\undefined%
      \relax%
    \else%
      \setlength{\unitlength}{\unitlength * \real{\svgscale}}%
    \fi%
  \else%
    \setlength{\unitlength}{\svgwidth}%
  \fi%
  \global\let\svgwidth\undefined%
  \global\let\svgscale\undefined%
  \makeatother%
  \begin{picture}(1,1)%
    \lineheight{1}%
    \setlength\tabcolsep{0pt}%
    \put(0,0){\includegraphics[width=\unitlength,page=1]{Plots/Results/absolute_error_map_r6.pdf}}%
    %\put(0.15,0.85){\makebox(0,0)[lt]{\lineheight{1}\smash{\begin{tabular}[t]{l}\textcolor{white}{$r_t = 15.96\%$}\end{tabular}}}}%
  \end{picture}%
\endgroup%

%% file: Plots/Results/absolute_error_map_r7.pdf_tex
%% Creator: Inkscape 1.4.2 (f4327f4, 2025-05-13), www.inkscape.org
%% Accompanies image file 'absolute_error_map_r7.pdf' (pdf, eps, ps)
\begingroup%
  \makeatletter%
  \providecommand\color[2][]{%
    \errmessage{(Inkscape) Color is used for the text in Inkscape, but the package 'color.sty' is not loaded}%
    \renewcommand\color[2][]{}%
  }%
  \providecommand\transparent[1]{%
    \errmessage{(Inkscape) Transparency is used (non-zero) for the text in Inkscape, but the package 'transparent.sty' is not loaded}%
    \renewcommand\transparent[1]{}%
  }%
  \providecommand\rotatebox[2]{#2}%
  \newcommand*\fsize{\dimexpr\f@size pt\relax}%
  \newcommand*\lineheight[1]{\fontsize{\fsize}{#1\fsize}\selectfont}%
  \ifx\svgwidth\undefined%
    \setlength{\unitlength}{352.79999542bp}%
    \ifx\svgscale\undefined%
      \relax%
    \else%
      \setlength{\unitlength}{\unitlength * \real{\svgscale}}%
    \fi%
  \else%
    \setlength{\unitlength}{\svgwidth}%
  \fi%
  \global\let\svgwidth\undefined%
  \global\let\svgscale\undefined%
  \makeatother%
  \begin{picture}(1,1)%
    \lineheight{1}%
    \setlength\tabcolsep{0pt}%
    \put(0,0){\includegraphics[width=\unitlength,page=1]{Plots/Results/absolute_error_map_r7.pdf}}%
    %\put(0.15,0.85){\makebox(0,0)[lt]{\lineheight{1}\smash{\begin{tabular}[t]{l}\textcolor{white}{$r_t = 18.20\%$}\end{tabular}}}}%
  \end{picture}%
\endgroup%

%% file: Plots/Results/absolute_error_map_r8.pdf_tex
%% Creator: Inkscape 1.4.2 (f4327f4, 2025-05-13), www.inkscape.org
%% Accompanies image file 'absolute_error_map_rt.pdf' (pdf, eps, ps)
\begingroup%
  \makeatletter%
  \providecommand\color[2][]{%
    \errmessage{(Inkscape) Color is used for the text in Inkscape, but the package 'color.sty' is not loaded}%
    \renewcommand\color[2][]{}%
  }%
  \providecommand\transparent[1]{%
    \errmessage{(Inkscape) Transparency is used (non-zero) for the text in Inkscape, but the package 'transparent.sty' is not loaded}%
    \renewcommand\transparent[1]{}%
  }%
  \providecommand\rotatebox[2]{#2}%
  \newcommand*\fsize{\dimexpr\f@size pt\relax}%
  \newcommand*\lineheight[1]{\fontsize{\fsize}{#1\fsize}\selectfont}%
  \ifx\svgwidth\undefined%
    \setlength{\unitlength}{352.79999542bp}%
    \ifx\svgscale\undefined%
      \relax%
    \else%
      \setlength{\unitlength}{\unitlength * \real{\svgscale}}%
    \fi%
  \else%
    \setlength{\unitlength}{\svgwidth}%
  \fi%
  \global\let\svgwidth\undefined%
  \global\let\svgscale\undefined%
  \makeatother%
  \begin{picture}(1,1)%
    \lineheight{1}%
    \setlength\tabcolsep{0pt}%
    \put(0,0){\includegraphics[width=\unitlength,page=1]{Plots/Results/absolute_error_map_r8.pdf}}%
    %\put(0.15,0.85){\makebox(0,0)[lt]{\lineheight{1}\smash{\begin{tabular}[t]{l}\textcolor{white}{$r_t = 19.02\%$}\end{tabular}}}}%
  \end{picture}%
\endgroup%

%% file: Plots/Results/absolute_error_map_r9.pdf_tex
%% Creator: Inkscape 1.4.2 (f4327f4, 2025-05-13), www.inkscape.org
%% Accompanies image file 'absolute_error_map.pdf' (pdf, eps, ps)
\begingroup%
  \makeatletter%
  \providecommand\color[2][]{%
    \errmessage{(Inkscape) Color is used for the text in Inkscape, but the package 'color.sty' is not loaded}%
    \renewcommand\color[2][]{}%
  }%
  \providecommand\transparent[1]{%
    \errmessage{(Inkscape) Transparency is used (non-zero) for the text in Inkscape, but the package 'transparent.sty' is not loaded}%
    \renewcommand\transparent[1]{}%
  }%
  \providecommand\rotatebox[2]{#2}%
  \newcommand*\fsize{\dimexpr\f@size pt\relax}%
  \newcommand*\lineheight[1]{\fontsize{\fsize}{#1\fsize}\selectfont}%
  \ifx\svgwidth\undefined%
    \setlength{\unitlength}{352.79999542bp}%
    \ifx\svgscale\undefined%
      \relax%
    \else%
      \setlength{\unitlength}{\unitlength * \real{\svgscale}}%
    \fi%
  \else%
    \setlength{\unitlength}{\svgwidth}%
  \fi%
  \global\let\svgwidth\undefined%
  \global\let\svgscale\undefined%
  \makeatother%
  \begin{picture}(1,1)%
    \lineheight{1}%
    \setlength\tabcolsep{0pt}%
    \put(0,0){\includegraphics[width=\unitlength,page=1]{Plots/Results/absolute_error_map_r9.pdf}}%
    %\put(0.15,0.85){\makebox(0,0)[lt]{\normalsize\smash{\begin{tabular}[t]{l}\textcolor{white}{$r_t = 18.44\%$}\end{tabular}}}}%
  \end{picture}%
\endgroup%

%% file: Plots/Results/ind742_t5_error_sp1.pdf_tex
%% Creator: Inkscape 1.4.2 (f4327f4, 2025-05-13), www.inkscape.org
%% Accompanies image file 'ind742_t5_error.pdf' (pdf, eps, ps)
\begingroup%
  \makeatletter%
  \providecommand\color[2][]{%
    \errmessage{(Inkscape) Color is used for the text in Inkscape, but the package 'color.sty' is not loaded}%
    \renewcommand\color[2][]{}%
  }%
  \providecommand\transparent[1]{%
    \errmessage{(Inkscape) Transparency is used (non-zero) for the text in Inkscape, but the package 'transparent.sty' is not loaded}%
    \renewcommand\transparent[1]{}%
  }%
  \providecommand\rotatebox[2]{#2}%
  \newcommand*\fsize{\dimexpr\f@size pt\relax}%
  \newcommand*\lineheight[1]{\fontsize{\fsize}{#1\fsize}\selectfont}%
  \ifx\svgwidth\undefined%
    \setlength{\unitlength}{223.20000458bp}%
    \ifx\svgscale\undefined%
      \relax%
    \else%
      \setlength{\unitlength}{\unitlength * \real{\svgscale}}%
    \fi%
  \else%
    \setlength{\unitlength}{\svgwidth}%
  \fi%
  \global\let\svgwidth\undefined%
  \global\let\svgscale\undefined%
  \makeatother%
  \begin{picture}(1,1.49032251)%
    \lineheight{1}%
    \setlength\tabcolsep{0pt}%
    \put(0,0){\includegraphics[width=\unitlength,page=1]{Plots/Results/ind742_t5_error_sp1.pdf}}%
    %\put(0.15,0.33){\makebox(0,0)[lt]{\lineheight{1}\smash{\begin{tabular}[t]{l}\textcolor{white}{$r_t = 34.84\%$}\end{tabular}}}}%
    \put(0.01,0.32){\makebox(0,0)[lt]{\smash{{\fontsize{7}{8}\selectfont \begin{tabular}[t]{@{}l@{}}\colorbox{white}{\textcolor{black}{$s_s = 1$}}\end{tabular}}}}}
  \end{picture}%
\endgroup%

%% file: Plots/Results/ind742_t5_error_sp2.pdf_tex
%% Creator: Inkscape 1.4.2 (f4327f4, 2025-05-13), www.inkscape.org
%% Accompanies image file 'ind742_t5_error_sp2.pdf' (pdf, eps, ps)
\begingroup%
  \makeatletter%
  \providecommand\color[2][]{%
    \errmessage{(Inkscape) Color is used for the text in Inkscape, but the package 'color.sty' is not loaded}%
    \renewcommand\color[2][]{}%
  }%
  \providecommand\transparent[1]{%
    \errmessage{(Inkscape) Transparency is used (non-zero) for the text in Inkscape, but the package 'transparent.sty' is not loaded}%
    \renewcommand\transparent[1]{}%
  }%
  \providecommand\rotatebox[2]{#2}%
  \newcommand*\fsize{\dimexpr\f@size pt\relax}%
  \newcommand*\lineheight[1]{\fontsize{\fsize}{#1\fsize}\selectfont}%
  \ifx\svgwidth\undefined%
    \setlength{\unitlength}{223.20000458bp}%
    \ifx\svgscale\undefined%
      \relax%
    \else%
      \setlength{\unitlength}{\unitlength * \real{\svgscale}}%
    \fi%
  \else%
    \setlength{\unitlength}{\svgwidth}%
  \fi%
  \global\let\svgwidth\undefined%
  \global\let\svgscale\undefined%
  \makeatother%
  \begin{picture}(1,1.49032251)%
    \lineheight{1}%
    \setlength\tabcolsep{0pt}%
    \put(0,0){\includegraphics[width=\unitlength,page=1]{Plots/Results/ind742_t5_error_sp2.pdf}}%
    %\put(0.15,0.33){\makebox(0,0)[lt]{\lineheight{1}\smash{\begin{tabular}[t]{l}\textcolor{white}{$r_t = 37.93\%$}\end{tabular}}}}%
    \put(0.01,0.32){\makebox(0,0)[lt]{\smash{{\fontsize{7}{8}\selectfont \begin{tabular}[t]{@{}l@{}}\colorbox{white}{\textcolor{black}{$s_s = 2$}}\end{tabular}}}}}
  \end{picture}%
\endgroup%

%% file: Plots/Results/ind742_t5_error_sp3.pdf_tex
%% Creator: Inkscape 1.4.2 (f4327f4, 2025-05-13), www.inkscape.org
%% Accompanies image file 'ind742_t5_error_sp3.pdf' (pdf, eps, ps)
\begingroup%
  \makeatletter%
  \providecommand\color[2][]{%
    \errmessage{(Inkscape) Color is used for the text in Inkscape, but the package 'color.sty' is not loaded}%
    \renewcommand\color[2][]{}%
  }%
  \providecommand\transparent[1]{%
    \errmessage{(Inkscape) Transparency is used (non-zero) for the text in Inkscape, but the package 'transparent.sty' is not loaded}%
    \renewcommand\transparent[1]{}%
  }%
  \providecommand\rotatebox[2]{#2}%
  \newcommand*\fsize{\dimexpr\f@size pt\relax}%
  \newcommand*\lineheight[1]{\fontsize{\fsize}{#1\fsize}\selectfont}%
  \ifx\svgwidth\undefined%
    \setlength{\unitlength}{223.20000458bp}%
    \ifx\svgscale\undefined%
      \relax%
    \else%
      \setlength{\unitlength}{\unitlength * \real{\svgscale}}%
    \fi%
  \else%
    \setlength{\unitlength}{\svgwidth}%
  \fi%
  \global\let\svgwidth\undefined%
  \global\let\svgscale\undefined%
  \makeatother%
  \begin{picture}(1,1.49032251)%
    \lineheight{1}%
    \setlength\tabcolsep{0pt}%
    \put(0,0){\includegraphics[width=\unitlength,page=1]{Plots/Results/ind742_t5_error_sp3.pdf}}%
    %\put(0.15,0.33){\makebox(0,0)[lt]{\lineheight{1}\smash{\begin{tabular}[t]{l}\textcolor{white}{$r_t = 37.27\%$}\end{tabular}}}}%
    \put(0.01,0.32){\makebox(0,0)[lt]{\smash{{\fontsize{7}{8}\selectfont \begin{tabular}[t]{@{}l@{}}\colorbox{white}{\textcolor{black}{$s_s = 3$}}\end{tabular}}}}}
  \end{picture}%
\endgroup%

%% file: Plots/Results/ind742_t5_error_sp4.pdf_tex
%% Creator: Inkscape 1.4.2 (f4327f4, 2025-05-13), www.inkscape.org
%% Accompanies image file 'ind742_t5_error_sp4.pdf' (pdf, eps, ps)
\begingroup%
  \makeatletter%
  \providecommand\color[2][]{%
    \errmessage{(Inkscape) Color is used for the text in Inkscape, but the package 'color.sty' is not loaded}%
    \renewcommand\color[2][]{}%
  }%
  \providecommand\transparent[1]{%
    \errmessage{(Inkscape) Transparency is used (non-zero) for the text in Inkscape, but the package 'transparent.sty' is not loaded}%
    \renewcommand\transparent[1]{}%
  }%
  \providecommand\rotatebox[2]{#2}%
  \newcommand*\fsize{\dimexpr\f@size pt\relax}%
  \newcommand*\lineheight[1]{\fontsize{\fsize}{#1\fsize}\selectfont}%
  \ifx\svgwidth\undefined%
    \setlength{\unitlength}{223.20000458bp}%
    \ifx\svgscale\undefined%
      \relax%
    \else%
      \setlength{\unitlength}{\unitlength * \real{\svgscale}}%
    \fi%
  \else%
    \setlength{\unitlength}{\svgwidth}%
  \fi%
  \global\let\svgwidth\undefined%
  \global\let\svgscale\undefined%
  \makeatother%
  \begin{picture}(1,1.49032251)%
    \lineheight{1}%
    \setlength\tabcolsep{0pt}%
    \put(0,0){\includegraphics[width=\unitlength,page=1]{Plots/Results/ind742_t5_error_sp4.pdf}}%
    %\put(0.15,0.33){\makebox(0,0)[lt]{\lineheight{1}\smash{\begin{tabular}[t]{l}\textcolor{white}{$r_t = 38.17\%$}\end{tabular}}}}%
    \put(0.01,0.32){\makebox(0,0)[lt]{\smash{{\fontsize{7}{8}\selectfont \begin{tabular}[t]{@{}l@{}}\colorbox{white}{\textcolor{black}{$s_s = 4$}}\end{tabular}}}}}
  \end{picture}%
\endgroup%

%% file: Plots/Results/ind742_t5_error_sp5.pdf_tex
%% Creator: Inkscape 1.4.2 (f4327f4, 2025-05-13), www.inkscape.org
%% Accompanies image file 'ind742_t5_error_sp5.pdf' (pdf, eps, ps)
\begingroup%
  \makeatletter%
  \providecommand\color[2][]{%
    \errmessage{(Inkscape) Color is used for the text in Inkscape, but the package 'color.sty' is not loaded}%
    \renewcommand\color[2][]{}%
  }%
  \providecommand\transparent[1]{%
    \errmessage{(Inkscape) Transparency is used (non-zero) for the text in Inkscape, but the package 'transparent.sty' is not loaded}%
    \renewcommand\transparent[1]{}%
  }%
  \providecommand\rotatebox[2]{#2}%
  \newcommand*\fsize{\dimexpr\f@size pt\relax}%
  \newcommand*\lineheight[1]{\fontsize{\fsize}{#1\fsize}\selectfont}%
  \ifx\svgwidth\undefined%
    \setlength{\unitlength}{223.20000458bp}%
    \ifx\svgscale\undefined%
      \relax%
    \else%
      \setlength{\unitlength}{\unitlength * \real{\svgscale}}%
    \fi%
  \else%
    \setlength{\unitlength}{\svgwidth}%
  \fi%
  \global\let\svgwidth\undefined%
  \global\let\svgscale\undefined%
  \makeatother%
  \begin{picture}(1,1.49032251)%
    \lineheight{1}%
    \setlength\tabcolsep{0pt}%
    \put(0,0){\includegraphics[width=\unitlength,page=1]{Plots/Results/ind742_t5_error_sp5.pdf}}%
    %\put(0.15,0.33){\makebox(0,0)[lt]{\lineheight{1}\smash{\begin{tabular}[t]{l}\textcolor{white}{$r_t = 38.46\%$}\end{tabular}}}}%
    \put(0.01,0.32){\makebox(0,0)[lt]{\smash{{\fontsize{7}{8}\selectfont \begin{tabular}[t]{@{}l@{}}\colorbox{white}{\textcolor{black}{$s_s = 5$}}\end{tabular}}}}}
  \end{picture}%
\endgroup%

%% file: Plots/Results/colorbar_v0_0.98.pdf_tex
%% Creator: Inkscape 1.4.2 (f4327f4, 2025-05-13), www.inkscape.org
%% Accompanies image file 'colorbar_v0_0.98.pdf' (pdf, eps, ps)
\begingroup%
  \makeatletter%
  \providecommand\color[2][]{%
    \errmessage{(Inkscape) Color is used for the text in Inkscape, but the package 'color.sty' is not loaded}%
    \renewcommand\color[2][]{}%
  }%
  \providecommand\transparent[1]{%
    \errmessage{(Inkscape) Transparency is used (non-zero) for the text in Inkscape, but the package 'transparent.sty' is not loaded}%
    \renewcommand\transparent[1]{}%
  }%
  \providecommand\rotatebox[2]{#2}%
  \newcommand*\fsize{\dimexpr\f@size pt\relax}%
  \newcommand*\lineheight[1]{\fontsize{\fsize}{#1\fsize}\selectfont}%
  \ifx\svgwidth\undefined%
    \setlength{\unitlength}{123.71874619bp}%
    \ifx\svgscale\undefined%
      \relax%
    \else%
      \setlength{\unitlength}{\unitlength * \real{\svgscale}}%
    \fi%
  \else%
    \setlength{\unitlength}{\svgwidth}%
  \fi%
  \global\let\svgwidth\undefined%
  \global\let\svgscale\undefined%
  \makeatother%
  \begin{picture}(1,2.31783799)%
    \lineheight{1}%
    \setlength\tabcolsep{0pt}%
    \put(0,0){\includegraphics[width=\unitlength,page=1]{Plots/Results/colorbar_v0_0.98.pdf}}%
    \put(0.78209146,0.01679727){\makebox(0,0)[lt]{\lineheight{1.25}\smash{\begin{tabular}[t]{l}0.00\end{tabular}}}}%
    \put(0,0){\includegraphics[width=\unitlength,page=2]{Plots/Results/colorbar_v0_0.98.pdf}}%
    \put(0.78209146,0.21595866){\makebox(0,0)[lt]{\lineheight{1.25}\smash{\begin{tabular}[t]{l}0.11\end{tabular}}}}%
    \put(0,0){\includegraphics[width=\unitlength,page=3]{Plots/Results/colorbar_v0_0.98.pdf}}%
    \put(0.78209146,0.41512005){\makebox(0,0)[lt]{\lineheight{1.25}\smash{\begin{tabular}[t]{l}0.22\end{tabular}}}}%
    \put(0,0){\includegraphics[width=\unitlength,page=4]{Plots/Results/colorbar_v0_0.98.pdf}}%
    \put(0.78209146,0.6142815){\makebox(0,0)[lt]{\lineheight{1.25}\smash{\begin{tabular}[t]{l}0.33\end{tabular}}}}%
    \put(0,0){\includegraphics[width=\unitlength,page=5]{Plots/Results/colorbar_v0_0.98.pdf}}%
    \put(0.78209146,0.81344289){\makebox(0,0)[lt]{\lineheight{1.25}\smash{\begin{tabular}[t]{l}0.44\end{tabular}}}}%
    \put(0,0){\includegraphics[width=\unitlength,page=6]{Plots/Results/colorbar_v0_0.98.pdf}}%
    \put(0.78209146,1.01260428){\makebox(0,0)[lt]{\lineheight{1.25}\smash{\begin{tabular}[t]{l}0.54\end{tabular}}}}%
    \put(0,0){\includegraphics[width=\unitlength,page=7]{Plots/Results/colorbar_v0_0.98.pdf}}%
    \put(0.78209146,1.21176573){\makebox(0,0)[lt]{\lineheight{1.25}\smash{\begin{tabular}[t]{l}0.65\end{tabular}}}}%
    \put(0,0){\includegraphics[width=\unitlength,page=8]{Plots/Results/colorbar_v0_0.98.pdf}}%
    \put(0.78209146,1.41092712){\makebox(0,0)[lt]{\lineheight{1.25}\smash{\begin{tabular}[t]{l}0.76\end{tabular}}}}%
    \put(0,0){\includegraphics[width=\unitlength,page=9]{Plots/Results/colorbar_v0_0.98.pdf}}%
    \put(0.78209146,1.61008851){\makebox(0,0)[lt]{\lineheight{1.25}\smash{\begin{tabular}[t]{l}0.87\end{tabular}}}}%
    \put(0,0){\includegraphics[width=\unitlength,page=10]{Plots/Results/colorbar_v0_0.98.pdf}}%
    \put(0.78209146,1.80924995){\makebox(0,0)[lt]{\lineheight{1.25}\smash{\begin{tabular}[t]{l}0.98\end{tabular}}}}%
    \put(0.585,2.21808198){\makebox(0,0)[lt]{\lineheight{1.25}\smash{\begin{tabular}[t]{c}Velocity\\$[\si{\milli\metre\per\second}]$\end{tabular}}}}%
  \end{picture}%
\endgroup%

%% file: Plots/Results/Target_overlay_sp_60_14.pdf_tex
%% Creator: Inkscape 1.4.2 (f4327f4, 2025-05-13), www.inkscape.org
%% Accompanies image file 'Target_overlay_sp_60_14.pdf' (pdf, eps, ps)
\begingroup%
  \makeatletter%
  \providecommand\color[2][]{%
    \errmessage{(Inkscape) Color is used for the text in Inkscape, but the package 'color.sty' is not loaded}%
    \renewcommand\color[2][]{}%
  }%
  \providecommand\transparent[1]{%
    \errmessage{(Inkscape) Transparency is used (non-zero) for the text in Inkscape, but the package 'transparent.sty' is not loaded}%
    \renewcommand\transparent[1]{}%
  }%
  \providecommand\rotatebox[2]{#2}%
  \newcommand*\fsize{\dimexpr\f@size pt\relax}%
  \newcommand*\lineheight[1]{\fontsize{\fsize}{#1\fsize}\selectfont}%
  \ifx\svgwidth\undefined%
    \setlength{\unitlength}{1\linewidth}%
    \ifx\svgscale\undefined%
      \relax%
    \else%
      \setlength{\unitlength}{\unitlength * \real{\svgscale}}%
    \fi%
  \else%
    \setlength{\unitlength}{\svgwidth}%
  \fi%
  \global\let\svgwidth\undefined%
  \global\let\svgscale\undefined%
  \makeatother%
  \begin{picture}(1,0.87526366)%
    \lineheight{1}%
    \setlength\tabcolsep{0pt}%
    \put(0,0){\includegraphics[width=\unitlength,page=1]{Plots/Results/Target_overlay_sp_60_14.pdf}}%
    \put(0.01,0.12152639){\makebox(0,0)[lt]{\lineheight{1.25}\smash{\begin{tabular}[t]{l}0.0\end{tabular}}}}%
    \put(0,0){\includegraphics[width=\unitlength,page=2]{Plots/Results/Target_overlay_sp_60_14.pdf}}%
    \put(0.01,0.22877289){\makebox(0,0)[lt]{\lineheight{1.25}\smash{\begin{tabular}[t]{l}0.2\end{tabular}}}}%
    \put(0,0){\includegraphics[width=\unitlength,page=3]{Plots/Results/Target_overlay_sp_60_14.pdf}}%
    \put(0.01,0.3360194){\makebox(0,0)[lt]{\lineheight{1.25}\smash{\begin{tabular}[t]{l}0.4\end{tabular}}}}%
    \put(0,0){\includegraphics[width=\unitlength,page=4]{Plots/Results/Target_overlay_sp_60_14.pdf}}%
    \put(0.01,0.44326587){\makebox(0,0)[lt]{\lineheight{1.25}\smash{\begin{tabular}[t]{l}0.6\end{tabular}}}}%
    \put(0,0){\includegraphics[width=\unitlength,page=5]{Plots/Results/Target_overlay_sp_60_14.pdf}}%
    \put(0.01,0.55051235){\makebox(0,0)[lt]{\lineheight{1.25}\smash{\begin{tabular}[t]{l}0.8\end{tabular}}}}%
    \put(0,0){\includegraphics[width=\unitlength,page=6]{Plots/Results/Target_overlay_sp_60_14.pdf}}%
    \put(0.01,0.65775885){\makebox(0,0)[lt]{\lineheight{1.25}\smash{\begin{tabular}[t]{l}1.0\end{tabular}}}}%
    \put(0,0){\includegraphics[width=\unitlength,page=7]{Plots/Results/Target_overlay_sp_60_14.pdf}}%
    \put(0.04694466,0.87529386){\makebox(0,0)[lt]{\lineheight{1.25}\smash{\begin{tabular}[t]{l}Volume\end{tabular}}}}%
    \put(0.0192405,0.81236312){\makebox(0,0)[lt]{\lineheight{1.25}\smash{\begin{tabular}[t]{l}Fraction [-]\end{tabular}}}}%
  \end{picture}%
\endgroup%

%% file: 6_Conclusion.tex
\section{Conclusion}
\label{conclusion}

This work presents a Fourier-Graph neural solver for modeling $2$D two-phase mold filling. The proposed model maps geometry, \gls{IC}/\gls{BC} data, and inlet settings to transient $({\gls{u},\gls{v},\gls{p},\gls{alpha}})$ fields on unstructured meshes. Results show that the surrogate model is able to reproduce the large scale advection and the advance of the fluid-air interface across unseen gating locations and inlet parameters while maintaining a stable rollout over the studied horizons. The aggregate error remains within a few percent across velocity, pressure, and volume fraction, and qualitative comparisons indicate that errors concentrate near steep gradients and high-curvature portions of the interface. Overall, velocity and volume fraction predictions are robust, whereas pressure is more sensitive.

The ablations quantify how training resolution and data volume shape performance. Spatial subsampling of the training mesh causes a monotonic increase in error, with performance degradation concentrating around sharp, gradient-rich fronts. Temporal subsampling yields gentler deterioration and the model remains tolerant to missing frames within the studied horizon. Finally, data reduction produces a smooth and moderate error increase with diminishing gains beyond roughly \SI{75}{\percent} of the dataset. These findings offer practical guidance for dataset design, with spatial resolution being prioritized when accuracy is more important. Regarding dataset size and diversity, our investigations suggest that additional data should be curated to expand diversity rather than at quantity once the core regimes are covered. Taken together, these results support neural operators as fast surrogates for design-in-the-loop studies of gating systems.

The present study also reveals clear limitations, with the pressure field remaining the most challenging quantity for modeling. In general, the proposed neural solver delivers under resolved flows around localized gradients and along the fluid-air interface. In future works, we plan to improve pressure fidelity and physics consistency across steps. Possible research avenues include complementing the optimization loss with relative $L_2$ that punishes interface contour errors, divergence statistics, and global mass errors. We also plan to test longer horizons and spatial and temporal sampling strategies to study stability at extended times. Another promising research direction is the exploration of multi fidelity data schedules that combine a small set of high resolution cases with many coarse cases to improve sample efficiency and flow accuracy. Finally, expansions to $3$D and thermal-solidification coupling are natural and necessary modelings to realistically surrogate casting physics.

%% file: 7_Acknowledgement.tex
\section*{Acknowledgement}
% Which one should we use?
% For "Tier3 Grundversorgung" resources (without a specific project or external funding), please use:
The authors gratefully acknowledge the scientific support and HPC resources provided by the Erlangen National High Performance Computing Center (NHR@FAU) of the Friedrich-Alexander-Universität Erlangen-Nürnberg (FAU). The hardware is funded by the German Research Foundation (DFG).

%% file: 8_Contributions.tex
\section*{Author contributions statement}

\textbf{Edgard M. Minete}: Software; Conceptualization; Methodology; Validation; Formal analysis; Investigation; Data Curation; Writing – Original Draft; Visualization. \textbf{Mathis Immertreu}: Software; Validation; Investigation. \textbf{Fabian Teichmann}: Conceptualization; Methodology; Writing – Original Draft; Writing – Review \& Editing; Supervision; Project administration; Funding acquisition. \textbf{Sebastian Müller}: Methodology; Resources; Writing – Original Draft; Writing – Review \& Editing; Supervision; Funding acquisition.